\documentclass[fleqn,usenatbib]{mnras}

\usepackage{newtxtext,newtxmath}

\usepackage[T1]{fontenc}
\usepackage{ae,aecompl}

\usepackage{graphicx}	
\usepackage{amsmath}	
\usepackage{amssymb}	
\usepackage{booktabs}
\usepackage{caption}
\usepackage{subcaption}
\usepackage{float}
\usepackage{multirow}
\usepackage{dsfont}
\usepackage{relsize}
\usepackage{xargs}                      		
\usepackage{xfrac}

\newcommandx{\revised}{\color{black}} 
\newcommandx{\revisedtwo}{\color{black}} 

\newcommand{\ra}[1] {\renewcommand{\arraystretch}{#1}}
\newcommand\Nz{{N_{\mathsmaller{0}}}}
\newcommand\nn{\overline{N}}
\newcommand\ff{\boldsymbol{\mathsf{\overline{F}}}}
\newcommand\zz{\boldsymbol{\mathsf{\overline{Z}}}}
\newcommand\Sigmaz{{\boldsymbol{\Sigma}}_{\mathsmaller{0}}}
\newcommand\Uz{{\boldsymbol{\mathsf{U}}_{\mathsmaller{0}}}}
\newcommand\Vz{{\boldsymbol{\mathsf{V}}_{\mathsmaller{0}}}}
\newcommand\nl{\boldsymbol{n}'}
\newcommand\PPhi{\boldsymbol{\Phi}}
\newcommand\PPsi{\boldsymbol{\Psi}}
\newcommand\R{\boldsymbol{\mathsf{R}}}
\newcommand\G{\boldsymbol{\mathsf{G}}}
\newcommand\F{\boldsymbol{\mathsf{F}}}
\newcommand\Z{\boldsymbol{\mathsf{Z}}}
\newcommand\Ii{\boldsymbol{\mathsf{I}}}
\newcommand\U{\boldsymbol{\mathsf{U}}}
\newcommand\V{\boldsymbol{\mathsf{V}}}
\newcommand\SSigma{\boldsymbol{\Sigma}}
\newcommand\Ss{\boldsymbol{\mathsf{S}}}
\newcommand\W{\boldsymbol{\mathsf{W}}}
\newcommand\RGt{{\boldsymbol{\mathsf{R}}_{\mathrm{grid}}}}
\newcommand\RFPhit{{\boldsymbol{\mathsf{R}}_{\mathrm{sing}}}}
\newcommand\RPhit{{\boldsymbol{\mathsf{R}}_{\mathrm{dirt}}}}
\DeclareMathOperator{\Diag}{Diag}
\DeclareMathOperator{\Null}{Null}

\title[Dimensionality reduction for interferometric imaging]{A Fourier dimensionality reduction model for big data interferometric imaging}
\author[S. Vijay Kartik et al.]{
S. Vijay Kartik,$^{1}$\thanks{E-mail: vijay.kartik@epfl.ch}
Rafael E. Carrillo,$^{1}$
Jean-Philippe Thiran$^{1}$
and Yves Wiaux$^{2}$\thanks{E-mail: y.wiaux@hw.ac.uk}
\\
$^{1}$ Signal Processing Laboratory (LTS5), Ecole polytechnique f{\'e}d{\'e}rale de Lausanne (EPFL),
      CH-1015 Lausanne, Switzerland\\
$^{2}$Institute of Sensors, Signals, and Systems, Heriot-Watt University, Edinburgh EH14 4AS, UK
}

\date{Accepted 2017 February 27. Received 2017 February 21; in original form 2016 September 5}

\pubyear{2017}

\begin{document}
\label{firstpage}
\pagerange{\pageref{firstpage}--\pageref{lastpage}}
\maketitle

\begin{abstract}
Data dimensionality reduction in radio interferometry can provide {\revised savings} of computational resources for image reconstruction {\revised through reduced memory footprints and lighter computations per iteration}, which is {\revised important} for the scalability of imaging methods to the big data setting of the next-generation telescopes. 
This article sheds new light on dimensionality reduction from the perspective of the compressed sensing theory and studies its interplay with imaging algorithms designed in the context of convex optimization.
We propose a post-gridding linear data embedding to the space spanned by the left singular vectors of the measurement operator, providing a dimensionality reduction below image size. 
This embedding preserves the null space of the measurement operator and hence its sampling properties are also preserved in light of the compressed sensing theory. 
We show that this can be approximated by first computing the dirty image and then applying a weighted subsampled discrete Fourier transform to obtain the final reduced data vector. 
This Fourier dimensionality reduction model ensures a fast implementation of the full measurement operator, essential for any iterative image reconstruction method. 
The proposed reduction also preserves the i.i.d. Gaussian properties of the original measurement noise. 
For convex optimization-based imaging algorithms, this is key to justify the use of the standard $\ell_2$-norm as the data fidelity term. 
Our simulations confirm that this dimensionality reduction approach can be leveraged by convex optimization algorithms with no loss in imaging quality relative to reconstructing the image from the complete visibility data set. 
{\revised Reconstruction results in simulation settings with no direction dependent effects or calibration errors show promising performance of the proposed dimensionality reduction. Further tests on real data are planned as an extension of the current work.}
\textsc{matlab} code implementing the proposed reduction method is available on GitHub.
\end{abstract}




\section{Introduction}
Image reconstruction in radio interferometry is intrinsically an ill-posed inverse problem due to the fact that the visibilities essentially identify an incomplete Fourier coverage of the image of interest. 
The theory of compressed sensing demonstrates that a signal admitting a sparse representation in some adequate basis can be recovered from incomplete sampling~\citep{donoho_compressed_2006, candes_robust_2006, candes_stable_2006}. 
As a consequence, the development of compressed sensing-based imaging methods in radio interferometry is an active research area, and novel work applying compressed sensing theory was reported soon after its establishment~\citep{wiaux_compressed_2009, wiaux_spread_2009}. 
More involved work to handle specific problems has also been performed, primarily on wide field-of-view observations~\citep{mcewen_compressed_2011} and non-coplanar effects~\citep{wolz_revisiting_2013}.
Approaches implementing sparse reconstruction for radio interferometry continue to be investigated, and report consistently increasing reconstruction performance, as recent work described in~\citet{li_application_2011, carrillo_sparsity_2012, carrillo_purify:_2014, garsden_lofar_2015, dabbech_moresane:_2015, ferrari_multi-frequency_2015} demonstrates.

Large radio telescopes -- like the Low Frequency Array (LOFAR), the upgraded Karl G. Jansky Very Large Array (VLA) and the future Square Kilometre Array (SKA) -- are expected to produce data at an extremely high rate. For instance, estimates for data rates in the first phase of SKA operation are around five terabits per second~\citep{broekema_square_2015}. 
The resulting images are designed to be in the gigapixels range, and with a high dynamic range of six/seven orders of magnitude~\citep{cornwell_wide_2012, wijnholds_signal_2014}. 
This high data rate requirement presents a great signal processing challenge and has led to new research in the design and development of scalable image recovery and data-handling methods in radio interferometry. 
Due to the very demanding requirements for image reconstruction algorithms, big data methods need to be studied and employed; in particular, High Performance Computing (HPC)-ready solutions are needed to scale with the memory and CPU requirements of these decidedly compute-intensive processing tasks.

From a data acquisition standpoint, radio-interferometric imaging is strictly a compressed sensing problem only when the amount of continuous visibilities measurements is lower than the image size. 
In the big data regime of next-generation telescopes, the data volumes would actually be much larger than the image size, to reach high dynamic ranges. 
However, this does not change the ill-posed nature of the problem due to the intrinsic incomplete Fourier coverage. Therefore, from a reconstruction standpoint, the properties that the measurement operator needs to satisfy to enable accurate recovery from sparsity promoting convex optimization algorithms are those prescribed by the compressed sensing theory. 
The question raised is thus: ``Can image reconstruction methods scale to the big data regime?''

One approach to this problem is to study parallelized and distributed optimization algorithms that can split the data as needed~\citep{ferrari_distributed_2014, carrillo_scalable_2015, onose_scalable_2016}. 
In this article we contemplate the idea of linearly embedding the data in a lower-dimensional space before feeding them to the image reconstruction algorithms. 
This linear data dimensionality reduction approach has recently been considered in the numerical linear algebra literature under the name of `sketching', which, in general terms, addresses the question of solving a high-dimensional optimization problem by embedding it into a lower-dimensional space~\citep{mahoney_randomized_2011, woodruff_sketching_2014}. 
Sketching and similar random projection methods for dimensionality reduction of manifold-modelled data~\citep{baraniuk_random_2007, hegde_random_2007} were introduced and studied at the same time as the emergence of compressed sensing theory. 
Recent work drawing parallels between dimensionality reduction and compressed sensing~\citep{baraniuk_simple_2008, krahmer_new_2010} can be traced back to the seminal work on the Johnson-Lindenstrauss lemma~\citep{johnson_extensions_1984}.

Assuming an original linear measurement operator $\PPhi$ from the radio interferometry measurement equation, our task is to design a sketching/dimensionality reduction operator $\R$  leading to a full measurement operator $\PPhi'=\R\PPhi$ with the same properties as $\PPhi$ as dictated by the compressed sensing theory. In the context of sketching, dimensionality reduction is performed by random projections. We do not follow that approach here, but instead leverage the singular value decomposition of the original measurement operator.

This article is organized as follows. Section~\ref{sec:csimaging} gives a brief overview of the ill-posed inverse problem of radio-interferometric imaging, and discusses convex optimization-based imaging methods. Additionally, we highlight key properties and guarantees that the compressed sensing theory requires and provides, respectively, for signal reconstruction.
In Section~\ref{sec:dimreduction} we study dimensionality reduction techniques in the particular setting of using compressed sensing-based reconstruction algorithms. 
We present preliminary studies undertaken in this context, and derive the theoretically optimal dimensionality reduction operator from a singular value decomposition perspective of the measurement operator. 
We then introduce a novel post-gridding dimensionality reduction consisting of first mapping gridded visibilities back to image space, i.e., computing the dirty image, and then performing a weighted subsampled discrete Fourier transform to obtain the final reduced data vector with dimension below image size. 
We advocate that this procedure is optimal in reducing the dimension of the data vector, while preserving the compressed sensing properties and fast implementation of the final measurement operator $\PPhi'$. 
It also ensures that the independent and identically distributed (i.i.d.) Gaussian properties of the measurement noise are preserved, thus setting the image reconstruction problem appropriately for convex optimization-based reconstruction. 
We highlight alternative dimensionality reduction methods that lead to gridded visibilities of dimension above image size, and dirty images of dimension equal to image size, and make a comparative study on the reconstruction quality using these methods.

We show that the image reconstruction quality with visibilities `reduced' using the proposed method is comparable to that using the complete visibilities set and that with gridded visibilities. The performance associated with the use of the dirty image as reduced data vector is suboptimal. These results are reported in Section~\ref{sec:plots} through simulations with data sets containing up to 26 million continuous visibilities using SKA-like simulated $uv$ coverages.
We present our conclusions on this proposed dimensionality reduction method in Section~\ref{sec:conclusions}, and outline possible directions to extend this work and combine it with other concurrent approaches in scalable image recovery in radio interferometry.
\textsc{matlab} code containing this work is available on GitHub\footnote{\url{http://basp-group.github.io/fourierdimredn}} and is expected to be integrated into \textsc{`purify'}\footnote{\url{http://basp-group.github.io/purify}}, a radio-interferometric imaging software proposed and described in \citet{carrillo_purify:_2014}.

\section{Compressed sensing perspective on radio-interferometric imaging}
\label{sec:csimaging}
\subsection{Radio interferometer measurement equation}
\label{subsec:measurementeqn}
A radio interferometer contains an array of antennas, and each pair of antennas constitutes a `baseline'. 
A baseline measures, at each instant of observation, the correlation of the electric fields coming from the source under scrutiny.
The baseline (measured in units of the wavelength at the centre frequency of the observation band) has components $u$,\,$v$\,and\,$w$; ($u$,$v$) form the coordinates of the visibilities ($\boldsymbol{y}$) plane, the corresponding coordinates for the source intensity distribution ($\boldsymbol{x}$) being $l$ and $m$.
On a small field of view, and assuming narrow-band intensity incoherent signals ($l$,\,$m$ small, {\revised with negligible $w$-term effects}), the general measurement equation (described and explained in detail in \citet{thompson_interferometry_2001, smirnov_revisiting_2011}) boils down -- through the van Cittert-Zernike theorem -- to an incomplete Fourier coverage of sky brightness. 
The measurement equation is then given by
\begin{equation}
\label{eq:contrimeintegral}
\boldsymbol{y}(u,v) = \iint \limits_{\Omega} \boldsymbol{A}(l,m,u,v)\,\boldsymbol{x}(l,m)\operatorname{e}^{-2\pi\!\operatorname{i} (ul+vm)}\,\operatorname{d}\!l\operatorname{d}\!m,
\end{equation}
where $\boldsymbol{y}(u,v)$ denotes the obtained visibilities from an underlying source intensity ``image'' $\boldsymbol{x}(l,m)$, and $\boldsymbol{A}(l,m,u,v)$ represents all antenna properties including collecting area, beam pattern and other possible direction dependent effects {\revised and $w$-term corrections}, and the integral is computed over $\Omega = \{(l,m):l^2+m^2<1\}$.
The discretized form of the associated linear measurement model reads in matrix form as
\begin{equation}
\label{eq:rime}
\boldsymbol{y}=\PPhi \boldsymbol{x} + \boldsymbol{n},
\end{equation} where $\boldsymbol{x} \in \mathbb{C}^{N}$ is the (vectorized) image to be recovered and $\boldsymbol{y} \in \mathbb{C}^{M}$ the visibilities vector, $\boldsymbol{n} \in \mathbb{C}^{M}$ being the noise in the measurements; the `measurement operator' $\PPhi \in \mathbb{C}^{M\times N}$ covers the linear relation between the signal and the continuous visibilities, and is given by $\PPhi=\G\ff\boldsymbol{\mathsf{D}}_{\mathsmaller{\mathrm{R}}}\Z\boldsymbol{\mathsf{B}}$, where $\boldsymbol{\mathsf{B}}$ denotes the primary beam, $\Z$ {\revised the zero-padding of the image} needed to compute the {\revised discrete Fourier transform} of $\boldsymbol{x}$ on a finer sampling grid in the Fourier domain, $\ff$ the {\revised discrete Fourier transform} operator in the oversampled case, $\G$ a convolution interpolation operator to map from the discrete frequency grid to the continuous $uv$ plane (each row containing the interpolation kernel), and $\boldsymbol{\mathsf{D}}_{\mathsmaller{\mathrm{R}}}$ a diagonal matrix to implement the reciprocal of the inverse Fourier transform of the interpolation kernel used in $\G$, to undo the effects of the convolution by the interpolation kernel in the frequency domain. 
In the work presented in this article, we assume $\boldsymbol{\mathsf{B}}=\Ii$, where $\Ii$ is identity.
Note that direction-dependent effects can be accounted for in this approach by allowing general interpolation kernels in each row of the matrix $\G$.
We define the combined operator $\zz=\boldsymbol{\mathsf{D}}_{\mathsmaller{\mathrm{R}}}\Z$ for brevity in further discussions.
With these assumptions and notations we now have the measurement operator given by
\begin{equation}
\label{eq:phidefn}
\PPhi=\G\ff\zz\quad \in \mathbb{C}^{M\times N}.
\end{equation}

The superscript `$^{\dagger}$' is used throughout this article to mean the adjoint operator/matrix.
It may be noted here that $\G^{\dagger}$ is the `gridding' operator mapping continuous visibilities to the oversampled discrete Fourier grid, $\ff^{\dagger}$ is the inverse Fourier transform and $\zz^{\dagger}$ is a (scaled) cropping to image size. 
More details and motivations for the components of the measurement operator $\PPhi $ listed above can be found in \citet{rau_advances_2009, carrillo_purify:_2014}.
In this article we assume complete knowledge of $\PPhi$ and therefore pre-calibrated continuous visibilities $\boldsymbol{y}$.

\subsection{Sparse reconstruction methods for imaging}
\label{subsec:sparsereconstruction}
Radio-interferometric image recovery poses an inverse problem which is ill-posed since $\PPhi $ models an \emph{incomplete} measurement of the visibility/$uv$ space. 
The most widely used image reconstruction method in radio interferometry is the \textsc{clean} algorithm~\citep{hogbom_aperture_1974} and its several variants like {\revised Cotton-Schwab-\textsc{clean}~\citep{schwab_relaxing_1984}, Multi-frequency-\textsc{clean}~\citep{sault_mfclean_1994} }and Multiscale-\textsc{clean}~\citep{cornwell_multiscale_2008}.
In its most basic form, \textsc{clean} forms a `model' of the image through an iterative method with the following sequential steps: (i) gridding the residual visibilities as $\G^{\dagger}(\boldsymbol{y}-\PPhi\boldsymbol{x})$, (ii) forming the residual dirty image $\zz^{\dagger}\ff^{\dagger}\G^{\dagger}(\boldsymbol{y}-\PPhi\boldsymbol{x}) = \PPhi^{\dagger}(\boldsymbol{y}-\PPhi\boldsymbol{x})$, and finally (iii) selecting the peak of the residual image as a model component.
In essence, \textsc{clean} follows a `matching pursuit'-type algorithmic structure~\citep{mallat_matching_1993}, using a gradient descent to minimize the residual norm $\|\boldsymbol{y}-\PPhi \boldsymbol{x}\|_2$, with the implicit use of the sparsity of the underlying signal to regularize the inverse problem. Sparsity is reflected in the assumption that the sky model contains a few point sources, selected one at a time with \textsc{clean}.

A signal $\boldsymbol{x}\in\mathbb{C}^N$ has a $k$-sparse representation if there exists a basis $\PPsi$ so that $\boldsymbol{x}=\PPsi \boldsymbol{\alpha}$ where $\boldsymbol{\alpha}$ has at most $k$ non-zero entries.
 $\boldsymbol{x}$ is called \emph{compressible} if the remaining $N-k$ entries are non-zero but negligible entries.
Compressed sensing-based approaches explicitly leverage the sparsity of the signal, typically using an $\ell_1$-norm minimization scheme.
These approaches have been presented as robust recovery methods for radio interferometry images in the literature, e.g.,~\citet{wiaux_compressed_2009, li_application_2011}.
We take a so-called `analysis'-based approach to recover the image directly, as opposed to recovering a sparse representation and subsequently `synthesizing' the image from that.
We apply the sparsity restrictions while maintaining reasonable fidelity with the initial data, leading to the solution to the inverse problem of the form
\begin{equation}
\label{eq:minimizationprob}
\min\limits_{\boldsymbol{x}\in \mathbb{R}_{+}^N}\|\PPsi^{\dagger}\boldsymbol{x}\|_1 \text{ s.t. } \|\boldsymbol{y}-\PPhi \boldsymbol{x}\|_2^2 \leq \varepsilon,
\end{equation}
where $\varepsilon$ is an upper bound on the energy of the noise $\boldsymbol{n}$ {\revised in visibility space}, given by $\|\boldsymbol{n}\|_2^2$ (see \citet{candes_compressed_2011, carrillo_sparsity_2012} for more details on the analysis-based approach). 
In Eq.~(\ref{eq:minimizationprob}) the condition imposed on the signal $\boldsymbol{x}$ is that it be sparse in a dictionary $\PPsi \in  \mathbb{C}^{N\times D}$, where $D$ is the number of elements in the dictionary.
The sparsity prior is expressed as a convex $\ell_1$-norm approximation of the less tractable non-convex $\ell_0$-norm that reflects the actual sparsity.
The data fidelity term is expressed as an $\ell_2$-norm. 
In the general case, the data fidelity term is actually a negative log-likelihood term given by $\boldsymbol{n}^{\dagger}{\boldsymbol{\mathsf{C}}_n}^{-1}\boldsymbol{n}$, where $\boldsymbol{\mathsf{C}}_n$ is the noise covariance matrix. 
This simplifies to $\|\boldsymbol{n}\|^2_2/\sigma_n^2$ under the assumption of i.i.d. Gaussian noise.
We also note the importance of having a reliable estimate of the upper bound $\varepsilon$ for the data fidelity term and the physical relevance of limiting this to the total energy in the noise; in the case of continuous visibilities, this is analytically computed since the noise energy, given by $\|\boldsymbol{n}\|_2^2$, has a $\chi^2$ distribution if $\boldsymbol{n}$ has i.i.d. Gaussian entries. 
For reasonably high degrees of freedom -- corresponding, in this setting, to high dimensional noise vectors -- a $\chi^2$ distribution exhibits concentration of measure, thus providing a sharp upper bound that is directly computable from the mean.
In our simulations for image reconstruction we compute this upper bound as two standard deviations beyond the mean, which includes a large percentile of the distribution.
If the assumption of i.i.d. Gaussian entries of the noise fails to hold true, then determining $\varepsilon$ is not a direct analytical computation since the noise energy no longer follows a $\chi^2$ distribution.

Applying a stricter sparsity constraint by using the concept of `average sparsity' over a concatenation of orthonormal bases $\PPsi_i \in  \mathbb{C}^{N}$ to form the dictionary $\PPsi$, the \textsc{sara} algorithm presented in \citet{carrillo_sparsity_2012} shows improved reconstruction results over {\revisedtwo other convex optimization-based imaging methods that use sparse reconstruction.}
Average sparsity is imposed over a dictionary that consists of the Dirac basis and the first eight Daubechies wavelets. This is different from other imaging methods that may impose sparsity in a single orthonormal basis.
\citet{garsden_lofar_2015, dabbech_moresane:_2015, ferrari_multi-frequency_2015} and others have also reported improved signal reconstruction using different sparse regularization techniques.
With the help of proximal splitting methods (see \citet{combettes_proximal_2011} for a comprehensive summary of these methods), a complex minimization problem can be split into multiple tractable subproblems, and this is the approach taken in \citet{carrillo_sparsity_2012, carrillo_purify:_2014} for image reconstruction in radio interferometry. 
In the work presented here, we build on these proposed image reconstruction methods; the inverse problem is regularized using \textsc{sara} and solved using an Alternating Direction Method of Multipliers (ADMM)-based proximal splitting algorithm, which was shown to be potentially scalable to medium/large-sized data in \citet{carrillo_scalable_2015, onose_scalable_2016}.

\subsection{Compressed sensing properties and guarantees}
\label{subsec:csprops}
Compressed sensing theory provides guarantees on stable signal reconstruction from ill-posed inverse problems provided the underlying signal and the measurement operator satisfy certain properties. 
The concepts of sparsity and incoherence, and the availability of tractable reconstruction methods are the pillars of compressed sensing theory.
For a more detailed justification of sparsity and incoherence in compressed sensing, see \citet{candes_sparsity_2007}. 
In this article, we focus on two properties of the measurement operator $\PPhi$, viz., the Null Space Property (NSP) and the Restricted Isometry Property (RIP). 
These properties, if satisfied by $\PPhi$, guarantee stable recovery of the signal $\boldsymbol{x}$ from the measurements $\boldsymbol{y}$ as they appear in Eq.~(\ref{eq:rime}).
A concise explanation follows:

The NSP is essential to guarantee exact signal recovery; $\PPhi$ satisfies the NSP of order $k$ with constant $\gamma\in (0,1)$ if \begin{equation}
\label{eq:nsp}
\|\boldsymbol{\eta}_T\|_1 \leq \gamma\|\boldsymbol{\eta}_{T^{\text{\tiny c}}}\|_1
\end{equation}
for all sets $T\subset \{1,\dots,N\}$, $|T| \leq k$ and for all $\boldsymbol{\eta}\in \Null(\PPhi)$. 
Here $\boldsymbol{\eta}_T$ is obtained by setting entries of $\boldsymbol{\eta}$ to zero for indices which are not in $T$. $T^{\text{\tiny c}}$ is the complement of $T$.
Put differently, the NSP means that no $k$-sparse signals are contained in the null space of the operator $\PPhi$.
It can then be shown that for a $k$-sparse signal $\boldsymbol{x}$ the reconstruction achieved using $\ell_1$-minimization is exact, and for a more general $\boldsymbol{x}$, the reconstruction error is bounded (Theorem 1 and its proof in \citet[pp.~199]{fornasier_compressive_2011}).

The RIP characterizes stable signal recovery in the presence of noise, and ensures that two different $k$-sparse signals remain well separated even after the application of the measurement operator. $\PPhi$ satisfies the RIP of order $k$ with constant $\delta\in (0,1)$ if
\begin{equation}
\label{eq:rip}
(1-\delta)\|\boldsymbol{x}\|^2_2 \leq \|\PPhi \boldsymbol{x}\|^2_2 \leq (1+\delta)\|\boldsymbol{x}\|^2_2
\end{equation}
for all $k$-sparse signals $\boldsymbol{x}$. 
The RIP implies the NSP (Lemma 2 and its proof in \citet[pp.~200]{fornasier_compressive_2011}). 
The fact that the RIP also implies robustness to measurement noise has been proven~\citep{candes_stable_2006}, and bounds on the reconstruction error have also been deduced~\citep{foucart_mathematical_2013}.
In our case with the requirement for the measurement operator to satisfy the RIP of order $2k$, it is worth noting that an operator $\PPhi$ satisfying the RIP of order $2k$ is in fact the Johnson-Lindenstrauss embedding for the case where $\boldsymbol{x}$ would be the difference between two $k$-sparse signals~\citep{krahmer_new_2010}.

Some examples of $\PPhi$ satisfying the RIP and NSP are Gaussian, Bernoulli, and partial random Fourier measurement matrices. 
Additionally, a spread-spectrum operator has been studied and applied to radio-interferometric imaging by \citet{wiaux_spread_2009}, where its universality relative to the sparsity dictionary $\PPsi$ is demonstrated through comparable image recovery results for simulated radio interferometry data containing non-negligible $w$-component. 
In theory, they all ensure robust sparse reconstruction. 
Compressed sensing-based reconstruction methods lend themselves readily to radio-interferometric imaging as the measurement operator is always a modified version of the partial Fourier matrix.

However, since the convex optimization algorithms that are employed for compressed sensing-based imaging amount to a non-linear iterative reconstruction involving repeated application of $\PPhi, \PPsi$ and their adjoint operators (\citet{carrillo_purify:_2014, yang_alternating_2011} contain more detailed overviews of the convex optimization algorithms used in this work), the image recovery method that leads to the solution in Eq.~(\ref{eq:minimizationprob}) relies on fast implementations of $\PPhi$ and $\PPsi$. 
In the context of our imaging techniques with a predefined concatenation of bases $\PPsi$ whose dimensionality is proportional to the image size, we can see that as the data size increases, so does the memory and computing requirement to manipulate and perform operations with $\PPhi$, whose dimensionality is proportional to data size.

Thus, dimensionality reduction of the visibilities while maintaining the desired image size is a much needed step to be able to handle big data and build scalable imaging algorithms based on convex optimization, and more so when viewed in the context of compressed sensing-based imaging methods with iterative computations directly dependent on data size.

\section{Data dimensionality reduction}
\label{sec:dimreduction}
\subsection{Preliminary studies and tests}
\label{subsec:dimredcomplexity}
As mentioned in Section~\ref{subsec:csprops}, a fast and scalable implementation of the measurement operator $\PPhi$ is critical for the viability of convex optimization-based image recovery methods. 
The complexity of computing the intermediate step involving $\PPhi$ ($=\G\ff\zz$) in the optimization algorithm can increase rapidly with increasing data size. 
Given an image of $N$ pixels, and $M$ visibilities obtained with non-uniform Fourier transform involving an interpolation kernel of size $k\times k$, the asymptotic complexity of applying $\PPhi$ is seen to be $\mathcal{O}(Mk^2+NlogN)$, since it is the complexity of matrix operations involving the matrix $\G$ with $M$ rows of $k^2$ non-zero entries, added with the complexity of an $N-$sized FFT.

As seen from the complexity, this image recovery solution is very demanding, both in terms of computing time and memory. 
The desirability of dimension reduction is apparent at this point, to enable us to (i) reduce the data \emph{size}, thereby decreasing memory requirements, (ii) keep the measurement operator \emph{fast}, thus reducing the computing time, (iii) preserve compressed sensing properties  (notably the NSP) of the measurement operator to guarantee accurate signal reconstruction, and (iv) preserve the i.i.d. Gaussian properties of the original measurement noise, in order to facilitate an easy computation of the data fidelity term of the objective function in our convex optimization algorithms through an $\ell_2$-norm.

We understand dimensionality reduction as the process of linearly mapping a higher-dimensional vector $\boldsymbol{y} \in \mathbb{C}^{M}$to a lower-dimensional vector $\boldsymbol{y}' \in \mathbb{C}^{M'}$ such that $M' \ll M$. 
This is typically achieved by applying an `embedding' operator $\R \in \mathbb{C}^{M'\times M}$, so that $\boldsymbol{y}' = \R\boldsymbol{y}$ is of dimension $M'$. 
Applying such an embedding operator to the radio interferometry measurement equation defined in Eq.~(\ref{eq:rime}), we obtain the full measurement operator
\begin{equation}
\PPhi'=\R\PPhi,
\end{equation}
and the reduced inverse problem is then given as
\begin{equation}
\boldsymbol{y}'=\R\PPhi \boldsymbol{x}+\R\boldsymbol{n}.
\end{equation}
The choice of $\R$ is critical as it affects not just the distortion of $\boldsymbol{y}$ but also the properties of $\PPhi$ that originally led to guaranteed image recovery through compressed sensing-based reconstruction methods. 
Additionally, $\R$ modifies the original noise vector $\boldsymbol{n}$. In our setting for radio-interferometric imaging, $\boldsymbol{n}$ is assumed to be uncorrelated, having i.i.d. zero-mean Gaussian components. 
After applying an embedding operator $\R$, the `embedded' noise $\nl=\R\boldsymbol{n}$ has a covariance matrix 
\begin{equation}
\label{eq:generalcovmat}
\boldsymbol{\mathsf{C}}_{n'} = \nl(\nl)^{\dagger}=\sigma_n^2\R\R^{\dagger},
\end{equation}
which is not necessarily diagonal; i.e., the embedded noise $\nl$ is, in general, correlated.

As a preliminary step in choosing an appropriate embedding operator, some of the options for $\R$ as mentioned in Section~\ref{subsec:csprops} were studied. 
A Gaussian random matrix provides the ideal embedding, since the full measurement operator $\PPhi'=\R\PPhi$ would retain any original NSP of $\PPhi$ and thus guarantee signal recovery as per compressed sensing theory. 
Additionally, on average the `embedded' noise is decorrelated even without the assumptions of i.i.d. Gaussian noise in original measurements.
This is also ideal since, the data fidelity term and its upper bound are easily computed, as explained in Section~\ref{subsec:sparsereconstruction}.
Unfortunately, the Gaussian random matrix renders the measurement operator very slow, and the asymptotic complexity of applying $\PPhi'$ can then be seen to be $\mathcal{O}(M'N)$ since the Gaussian random matrix is a fully dense matrix of size $M'\times N$. 
As $M'$ is typically some proportion $p$ of the image size $N$, the asymptotic complexity for the Gaussian random matrix is then $\mathcal{O}(p N^2)$, which is clearly much worse than the original asymptotic value which was dominated by $\mathcal{O}(Mk^2)$, since $M\ll N^2$.
Recent work involving small-scale simulations \citep{kartik_dimension_2015} suggests that Gaussian random embedding could perform better than other embeddings (see e.g., gridding, in Section~\ref{subsec:otherdimreds}). 
But the time and memory required to run these small-scale simulations are prohibitively high; thus, for lack of a scalable and fast implementation of the Gaussian random matrix/operator, other dimensionality reduction schemes need to be chosen.

Another possibility studied for the embedding matrix $\R$ was a `spread-spectrum' like operator $\R=\boldsymbol{\mathsf{M\ff D}}_{\pm 1}$, where $\boldsymbol{\mathsf{D}}_{\pm 1}$ is a diagonal random sign matrix with entries $\pm 1$, $\ff$ the {\revised discrete Fourier transform} operator, and $\boldsymbol{\mathsf{M}}$ an $M'\times N$ random selection matrix which embeds the data vector to the final size $M'$. 
We note also that this embedding is similar to the subsampled randomized Hadamard transform operator as described in \citet{tropp_improved_2011}. Although the individual sub-operators in $\R$ have fast implementations, the modified measurement operator $\R\PPhi$ has several issues which make this embedding unsuitable. 
Firstly, it can no longer be directly proven that the measurement operator satisfies the RIP, as required by compressed sensing theory for exact signal recovery. 
Secondly, the application of this measurement operator $\R\PPhi$ is very slow, since the sub-operators $\boldsymbol{\mathsf{M}}$, $\ff$, $\boldsymbol{\mathsf{D}}_{\pm 1}$ and $\PPhi$ need to be combined together and pre-computed as a dense $M'\times N$-sized matrix, and the asymptotic complexity in applying $\R\PPhi$ in this case would be $\mathcal{O}(M'N)$, as for the case with a Gaussian random matrix.

Other embedding matrices, including random projection matrices either used within or without sketching techniques~\citep{woodruff_sketching_2014, li_very_2006, bingham_random_2001} provide dimensionality reduction, and may also have fast implementations of $\R$.
But since most suffer from a lack of sparsity, the modified measurement operator $\R\PPhi$ is not sparse - which precludes a fast implementation (especially for matrix-vector multiplications) and causes a large memory footprint.
They are thus unsuitable for repeated application in the iterative methods used for image reconstruction in our case.

Our attempts at approaching dimensionality reduction from a compressed sensing perspective lead us to consider the NSP and the RIP of the full measurement operator. 
With the exception of random matrices, constructing fast matrices satisfying the RIP is known to be non-trivial -- although recent attempts towards addressing this are presented in \citet{nelson_new_2014}.
Also, verifying the RIP for deterministic matrices is NP-hard, as shown in \citet{bandeira_certifying_2012}. 
So the idea is to devise an embedding operator that reduces the dimensionality of the measurements while preserving the NSP of the original measurement operator, thus maintaining the same compressed sensing-based guarantees on recovering the image.

We note that many state-of-the-art imaging techniques in radio interferometry include a `gridding'-like subroutine, mapping continuous visibilities to the discrete Fourier grid with an operator similar to $\R = \PPhi^{\dagger}$~\citep{dabbech_moresane:_2015, li_application_2011}. 
Typically, this is a gridding to the discrete Fourier grid through $\G^{\dagger}$, or a mapping back to image space through $\PPhi^{\dagger}$.
Gridding has been studied and developed further by \citet{sullivan_fast_2012} with the introduction of the `Fast Holographic Deconvolution' technique; this technique leverages the lossless information property~\citep{tegmark_how_1997} that is being used to reduce Cosmic Microwave Background data, and introduces the \emph{Holographic Mapping} function $\boldsymbol{\mathsf{H}}=\G^{\dagger}\G$. $\boldsymbol{\mathsf{H}}$ models the mapping between a continuous visibility and the corresponding equivalent in the gridded, `holographic' map without having to go through separate interpolation and gridding steps -- which are the most time consuming parts of standard imaging techniques. 
A pre-computed holographic matrix $\boldsymbol{\mathsf{H}}$ is stored before image reconstruction starts, and therefore provides a way to quicken the imaging process. 
Additionally, the compact support of the interpolation kernel present in each row of the matrix $\G$ ensures that $\boldsymbol{\mathsf{H}}$ remains sparse, so its repeated application is also not a hindrance to the imaging technique.
We see that gridding is an appropriate technique to reduce data dimensionality while maintaining information content, and with the use of a holographic mapping it can be incorporated in imaging techniques without incurring a large cost in terms of image reconstruction time. 
In the next section we present a post-gridding dimensionality reduction technique that introduces an additional step to achieve reducing dimensionality.

\subsection{Singular vector space embedding}
\label{subsec:svd}
\subsubsection{Optimal dimensionality reduction model}
An ideal dimensionality reduction method would result in a final data dimension as small as possible while simultaneously guaranteeing accurate image reconstruction by retaining the NSP of the original measurement operator $\PPhi$.
{\revised The null space of $\PPhi$ arises from the incomplete Fourier coverage that forms all visibilities.}
Observing the singular value decomposition (SVD) of $\PPhi$ given by 
\begin{equation}
\PPhi=\U\SSigma\V^{\dagger},
\end{equation}
where $\U\in \mathbb{C}^{M\times M}$, $\V\in \mathbb{C}^{N\times N}$ are unitary matrices and $\SSigma\in \mathbb{C}^{M\times N}$ is a rectangular diagonal matrix containing the singular values of $\PPhi$, {\revised we note that the existence of the null space of $\PPhi$ implies that some singular values are necessarily zero-valued.
In fact, the singular values $\SSigma_{i}$ occupy a continuous spectrum of values, with large values corresponding to Fourier grid points with contribution to multiple interpolation kernels present in $\G$, and gradually decreasing to the minimum value of zero corresponding to Fourier grid points with no such contribution, thus leading to an incomplete $uv$ coverage}.
We can see that retaining the non-zero singular values of $\PPhi$ effectively retains the null space of $\PPhi$.
Following this, we rewrite the SVD as 
\begin{equation}
\PPhi=\Uz\Sigmaz\Vz^{\dagger},
\end{equation}
where $\Uz\in \mathbb{C}^{M\times \Nz}$, $\Sigmaz\in \mathbb{C}^{\Nz\times \Nz}$ and $\Vz\in \mathbb{C}^{N\times \Nz}$ are truncated versions of $\U$, $\SSigma$ by only retaining columns (rows for $\V$) corresponding to non-zero singular values of $\PPhi$. 
Clearly, the number of non-zero singular values is $\Nz \leq min(N,M)$ since $\SSigma\in \mathbb{C}^{M\times N}$.

An optimal dimensionality reduction operator on $\PPhi$ would then be a projection on its left singular vectors {\revised that correspond to non-zero singular values, since the null space of $\Phi$ is retained through these left singular vectors and thus no information is lost. The projection is given by:}
\begin{align}
\R_{\mathrm{sing-o}}&=\Uz^{\dagger} \notag \\
&=\Sigmaz^{-1}\Vz^{\dagger}\PPhi^{\dagger}.
\label{eq:idealdimred}
\end{align}
The full measurement operator therefore reads as a weighted subsampling in the right singular vector basis:
\begin{equation}
\PPhi'_{\mathrm{sing-o}}=\Sigmaz\Vz^{\dagger}.
\label{eq:phisingo}
\end{equation}
The corresponding `embedded' noise $\Uz^{\dagger}\boldsymbol{n}$ has a covariance matrix  
\begin{align}
\boldsymbol{\mathsf{C}}_{n'} &= \sigma_n^2\Uz^{\dagger}\Uz \nonumber \\
&=\sigma_n^2\Ii.
\label{eq:covmatsingo}
\end{align}
This follows from Eq.~(\ref{eq:generalcovmat}), since columns of $\U$ are orthonormal by definition.
The noise thus remains fully decorrelated after dimensionality reduction, which allows us to continue using an $\ell_2$-norm as the data fidelity term in the minimization algorithm, as explained in Section~\ref{subsec:sparsereconstruction}.

Put differently, the ideal dimensionality reduction involves a gridding-like operation performed in radio interferometry to obtain the dirty image (shown here by $\PPhi^{\dagger}$), followed by a projection on the right singular vectors of $\PPhi$ corresponding to non-zero singular values, and finally followed by a weighting operation with the inverse of the non-zero singular values of $\PPhi$.

In theory, therefore, the ideal dimensionality reduction $\R_{\mathrm{sing-o}}$ reduces data to a dimension $\Nz\leq N$. It ensures that the full measurement operator $\PPhi'_{\mathrm{sing-o}}$ preserves the null space of $\PPhi$, therefore retaining any original NSP of $\PPhi$. 
It also induces a decorrelated noise in the reduced dimension, thus enabling the minimization algorithm to use an $\ell_2$-norm of the noise for the data fidelity term. 
In reality, however, this operator $\R$ is difficult to implement since the SVD is computationally expensive, with an asymptotic complexity of $\mathcal{O}(N^3)$~\citep{golub_matrix_1996}. 
Additionally, since $\R_{\mathrm{sing-o}}$ may not have a guaranteed fast implementation, applying it iteratively in our minimization algorithms would also be prohibitively expensive.
This renders the optimal reduction method impractical.
We propose to get around this limitation by building an approximate version $\RFPhit$ of $\R_{\mathrm{sing-o}}$ that can be readily computed and applied.

\subsubsection{Approximate Fourier reduction model}
\label{subsubsec:diagapprox}

In finding a valid approximation of the ideal dimensionality reduction given by Eq.~(\ref{eq:idealdimred}), we attempt to approximate the unitary matrix $\V$.
We can note that $\V$ is in fact the eigenbasis of $\PPhi^{\dagger}\PPhi$, since it contains the right singular vectors of $\PPhi$ as defined in Section~\ref{subsec:svd}.
To understand the eigenbasis of $\PPhi^{\dagger}\PPhi$, we probe its structure, expanding it to its constituent operators as defined in Eq.~(\ref{eq:phidefn}).
This gives us 
\begin{align}
\PPhi^{\dagger}\PPhi&=(\G\ff\zz)^{\dagger}(\G\ff\zz) \nonumber \\
&=(\zz^{\dagger}\ff^{\dagger})(\G^{\dagger}\G)(\ff\zz).
\label{eq:rearrangedsvd}
\end{align}

\begin{figure}
	\begin{subfigure}{1\linewidth}
		\centering
  		\includegraphics[trim={0px 0px 0px 0px}, clip, width=\linewidth]{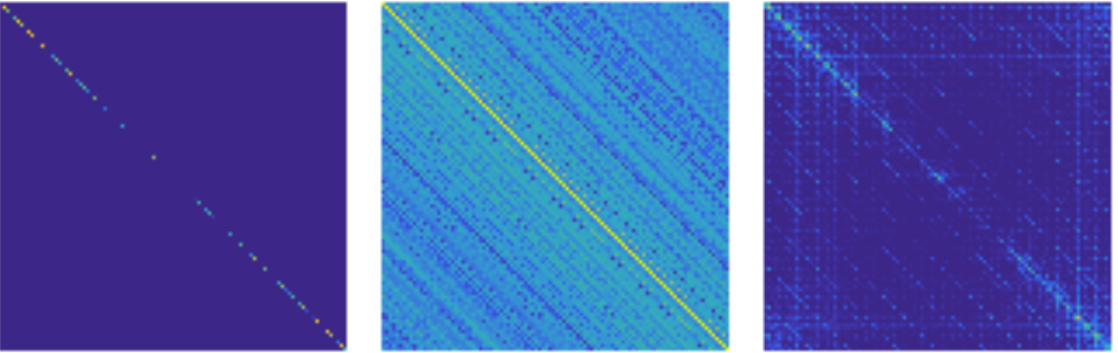}
		\vspace{1pt}
	\end{subfigure}
	\begin{subfigure}{1\linewidth}
  		\centering
  		\includegraphics[trim={0px 0px 0px 0px}, clip, width=\linewidth]{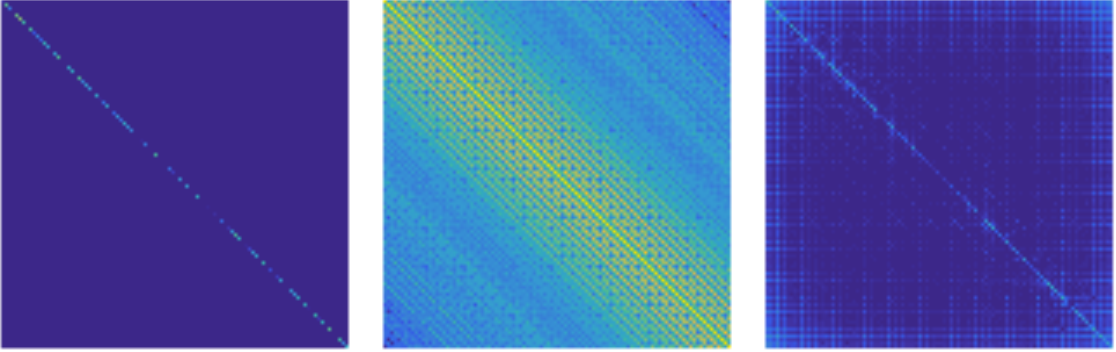}
	\end{subfigure}
	\caption{Numerical results (shown here in $\log_{10}$ scale) illustrating correlation of noise in the reduced dimension through varying degrees of `diagonality' of the initial noise covariance matrix. Left: $\G^{\dagger}\G$, for reduction through $\RGt$; Centre: $\PPhi^{\dagger}\PPhi$, for reduction through $\RPhit$; Right: $\F\PPhi^{\dagger}\PPhi\F^{\dagger}$, for reduction through $\RFPhit$. Top row: Gaussian random coverage. Bottom row: SKA-like coverage.}
	\label{fig:covmats}
\end{figure}

The central term in Eq.~(\ref{eq:rearrangedsvd}) is the holographic map $\boldsymbol{\mathsf{H}}=\G^{\dagger}\G$, comprising individual elements $(\G^{\dagger}\G)_{ij}$ that denote the simultaneous contributions of different interpolation kernels that would map continuous visibilities onto the pixel $(i,j)$ on the discrete Fourier grid.
Since each visibility is obtained by integrating a small region of the $uv$ plane, the interpolation kernels have compact support (e.g., the $8\times 8$ Kaiser-Bessel kernels used to calculate the non-uniform Fourier transform in our simulations).
Thus, the simultaneous contributions of different interpolation kernels are largely limited to small areas of overlapping support, and consequently limited to having significant  contribution only for pixels $(i,j)$ where $i,j$ are of similar value.
In other words, the largest values of $\G^{\dagger}\G$ are on and immediately around its diagonal.
This is also seen through numerical results as shown in Fig.~\ref{fig:covmats}, where the illustrations for $\G^{\dagger}\G$ can be seen to be extremely close to a diagonal matrix. It should be noted that here we implicitly assume that there are no DDEs and that the antenna array is coplanar ($w=0$). If these assumptions become invalid, the interpolation kernels present as rows of $\G$ can no longer be simply represented with compact support, and $\G^{\dagger}\G$ no longer remains overwhelmingly diagonal.

Now we prepend and append $\F^{\dagger}\F$ to Eq.~(\ref{eq:rearrangedsvd}) -- the crucial observation being that $\F$ is an image-sized {\revised Fourier transform} as opposed to the oversampled {\revised Fourier transform} $\ff$. $\F$ can then be expressed as $\F=\Z^{\dagger}\ff\zz$.
Eq.~(\ref{eq:rearrangedsvd}) can then be rewritten as 
\begin{equation}
\PPhi^{\dagger}\PPhi=\F^{\dagger} \left [ (\F\zz^{\dagger}\ff^{\dagger})(\G^{\dagger}\G)(\ff\zz\F^{\dagger})\right ] \F.
\label{eq:rewrittenFPhit}
\end{equation}
We note that the term $\F\zz^{\dagger}\ff^{\dagger}=\Z^{\dagger}\ff\zz\zz^{\dagger}\ff^{\dagger}$ as a whole performs a convolution with the inverse Fourier transform of $\zz\zz^{\dagger}$. Since $\zz\zz^{\dagger}$ is --  within limits of the scaling introduced in $\zz$ by $\boldsymbol{\mathsf{D}}_{\mathsmaller{\mathrm{R}}}$ (see Section~\ref{subsec:measurementeqn}) -- a partially distorted version of a two-dimensional pulse function of width equal to half of the field of view of the observations, its inverse Fourier transform is given by a sinc function with non-zero values at integer-indices, and a two-pixel wide main lobe.
The convolution with such a sinc function, when performed on $\G^{\dagger}\G$, results in a `smearing' of its diagonal character, with more non-zero values now appearing at off-diagonal locations. 
This smearing effect is compounded since $\F\zz^{\dagger}\ff^{\dagger}$ occurs as a pre- and post-operation on $\G^{\dagger}\G$. 
However, the smearing does not radically affect the diagonal {\revised character} since the main lobe of the sinc function has a small width.
Numerical results seen in Fig.~\ref{fig:covmats} show that $\F\PPhi^{\dagger}\PPhi\F^{\dagger}$ indeed regains much of the diagonal character from $\G^{\dagger}\G$, remaining close to a fully diagonal matrix. The ringing effect observed around the diagonal may be attributed to the side lobes of the sinc function.

{\revised
As a quantitative measure of the diagonal character of the matrices shown in Fig.~\ref{fig:covmats}, we define, for a matrix $\boldsymbol{\mathsf{C}}$, the ratio
$\mathlarger{\beta}_{\boldsymbol{\mathsf{C}}} = {\| \Diag_{\mathrm{band}}(\boldsymbol{\mathsf{C}}) \|_{\mathsmaller{F}}}/{\| \boldsymbol{\mathsf{C}} \|_{\mathsmaller{F}}}$, where $\Diag_{\mathrm{band}}(\boldsymbol{\mathsf{C}})$ is a band diagonal matrix formed from a thin band around the main diagonal of $\boldsymbol{\mathsf{C}}$ given by $\Diag(\boldsymbol{\mathsf{C}})$, and $\| \cdot \|_{\mathsmaller{F}}$ is the Frobenius norm.
Numerical results using test $uv$ coverages included in this work show the following typical values: $\mathlarger{\beta}_{\G^{\dagger}\G} \approx 0.95$, $\mathlarger{\beta}_{\PPhi^{\dagger}\PPhi} \approx 0.50$, and $\mathlarger{\beta}_{\F\PPhi^{\dagger}\PPhi\F^{\dagger}} \approx 0.90$, illustrating that the overwhelming majority of significant elements of the matrix $\F\PPhi^{\dagger}\PPhi\F^{\dagger}$ are on and around the diagonal - in a thin band corresponding to $2\%$ of the matrix size.

}

We thus see that $\PPhi^{\dagger}\PPhi \approx \F^{\dagger}[\Diag(\F\PPhi^{\dagger}\PPhi\F^{\dagger})]\F$, which is very close to the eigendecomposition of $\PPhi^{\dagger}\PPhi$ given by $\PPhi^{\dagger}\PPhi=\V\SSigma^2\V^{\dagger}$. This motivates the approximation of
\begin{equation}
\label{eq:approxf}
\V^{\dagger}\approx \F\quad \mathrm{and}\quad \SSigma^2\approx \Diag(\F\PPhi^{\dagger}\PPhi\F^{\dagger}).
\end{equation}
The approximation of the eigenbasis of $\PPhi^{\dagger}\PPhi$ by the orthogonal columns of the {\revised discrete Fourier transform} operator immediately renders our ideal dimensionality reduction operator feasible. 
The computationally expensive SVD of $\PPhi$ no longer needs to be explicitly calculated, as the {\revised discrete Fourier transform} operator is known without any knowledge of $\PPhi$. 
Moreover, fast implementations of $\F$ exist in the form of the FFT algorithm.
$\Vz$ is then given by $\Ss\F$, $\Ss$ being a subsampling matrix to select the dimensions corresponding to the $\Nz \leq N$ non-zero singular values present in $\SSigma$, thus producing a dimensionality reduction below image size.
The diagonal matrix $\Sigma$ is obtained by simply computing the square root of $\Diag(\F\PPhi^{\dagger}\PPhi\F^{\dagger})$. A similar selection of $\Nz$ dimensions leads to $\Sigmaz=\Ss\SSigma$, thus giving the  approximation
\begin{equation}
\label{eq:approxfzero}
\Vz^{\dagger}\approx \Ss\F\quad \mathrm{and}\quad \Sigmaz^2\approx \Ss\,\Diag(\F\PPhi^{\dagger}\PPhi\F^{\dagger}).
\end{equation}
This leads us to propose a Fourier model of the dimensionality reduction operator consisting of mapping gridded visibilities back to image space, i.e., computing a dirty image, and then performing a weighted subsampled {\revised discrete Fourier transform}, given by
\begin{equation}
\label{eq:rfphit}
\R_{\mathrm{sing}}=\Sigmaz^{-1}\Ss\F\PPhi^{\dagger} \quad \in \mathbb{C}^{\Nz\times M}.
\end{equation}
The full measurement operator is then given by
\begin{equation}
\PPhi'_{\mathrm{sing}}=\Sigmaz^{-1}\Ss\F\PPhi^{\dagger}\PPhi \quad \in \mathbb{C}^{\Nz\times N},
\label{eq:phising}
\end{equation}
where $\PPhi$ is given by Eq.~(\ref{eq:phidefn}).
Finally, the full measurement operator is tangible, and is suitably fast for repeated application in minimization algorithms.

Here we review the properties of our Fourier dimensionality reduction. 
$\PPhi'_{\mathrm{sing}}$ being an approximation of $\PPhi'_{\mathrm{sing-o}}$, we assume that it approximately preserves the null space of $\PPhi$. 
We {\revised have also seen} that $\F\PPhi^{\dagger}\PPhi\F^{\dagger}$ is {\revised largely diagonal - this diagonal character is maintained as long as the interpolation/de-gridding kernels used to compute the continuous visibilities have compact support}.
This {\revised in turn} implies that the covariance matrix $\sigma_n^2\RFPhit\RFPhit^{\dagger}$ of the embedded noise is {\revised largely diagonal as well}.
The weighting by $\Sigmaz^{-1}$ explicitly normalizes all the diagonal values of the noise covariance matrix to the original noise variance $\sigma_n^2$.
$\PPhi'_{\mathrm{sing}}$ also achieves the same dimensionality reduction to $\Nz$ as $\PPhi'_{\mathrm{sing-o}}$. 
Crucially, it exhibits a fast implementation since its constituent operators are diagonal, sparse and Fourier matrices.
A summary of properties of $\PPhi'_{\mathrm{sing-o}}$ and $\PPhi'_{\mathrm{sing}}$ is shown in Table~\ref{tab:props}. 

We extend the idea of approximating the initial noise covariance matrix to a further degree by assuming $\F\PPhi^{\dagger}\PPhi\F^{\dagger} \propto \Ii$ without explicit computation of $\F\PPhi^{\dagger}\PPhi\F^{\dagger}$, thus leading to $\R=\F\PPhi^{\dagger}$. 
This variant of our approach is also investigated in our simulations.

\subsection{Standard gridding-based dimensionality reductions}
\label{subsec:otherdimreds}
\subsubsection{Embedding visibilities to the dirty image}
Embedding visibilities to the dirty image is a standard way to reduce dimensionality, and is performed in many image reconstruction methods in radio interferometry, {\revised essentially through an image-based deconvolution with appropriate weighting}. 
Setting it in the terminology presented here, it amounts to using an embedding operator $\R=\PPhi^{\dagger}$, as previously explained in Section~\ref{subsec:sparsereconstruction}.
The corresponding noise is highly correlated, and this can be seen from the covariance matrix $\PPhi^{\dagger}\PPhi$ shown in Fig.~\ref{fig:covmats}, which contains significant off-diagonal elements. 

In order to be able to use this dimensionality reduction in our minimization problem formulation (Eq.~(\ref{eq:minimizationprob})), the embedded noise would need to have i.i.d. Gaussian entries. 
This is achieved by assuming, as done previously,  an approximation of the noise covariance matrix by its diagonal
\begin{equation}
\W^{2}=\Diag(\PPhi^{\dagger}\PPhi).
\end{equation}
$\W$ is invertible since $\PPhi^{\dagger}\PPhi$ is the dirty beam and hence typically non-zero along its main diagonal -- this also implies that there is no potential for further subsampling based on zero values.
We subsequently apply a weighting $\W^{-1}$ to obtain the dimension embedding operator
\begin{equation}
\label{eq:rphit}
\R_{\mathrm{dirt}}=\W^{-1}\PPhi^{\dagger} \quad \in \mathbb{C}^{N\times M}.
\end{equation}
The full measurement operator is therefore given as
\begin{equation}
\PPhi'_{\mathrm{dirt}}=\W^{-1}\PPhi^{\dagger}\PPhi \quad \in \mathbb{C}^{N\times N}.
\label{eq:phidirt}
\end{equation}
It preserves the null space of $\PPhi$, thus retaining any original NSP.
Indeed the SVD or eigendecomposition of $\PPhi^{\dagger}\PPhi$ reads as $\PPhi^{\dagger}\PPhi=\V\SSigma^2\V^{\dagger}$.
Applying $\RPhit$ is also fast as the individual suboperators in $\PPhi$ have fast implementations.
However, the embedded noise covariance matrix $\sigma^2_n\RPhit\RPhit^{\dagger}$ is far from diagonal, as seen in Fig.~\ref{fig:covmats}, though with diagonal entries all equal to the original noise variance $\sigma^2_n$. 
For completeness in our comparison of results, we extend our approximation of the initial noise covariance matrix by nevertheless assuming $\PPhi^{\dagger}\PPhi \propto \Ii$, resulting in the variant of $\RPhit$ given by $\R=\PPhi^{\dagger}$.

\subsubsection{Gridding visibilities}
Gridding continuous visibilities to discrete Fourier points reduces the data dimension to the size of the oversampled discrete Fourier grid, and is routinely performed in radio-interferometric imaging as a first step. 
Gridding can be seen as applying the embedding $\R=\G^{\dagger}$ to continuous visibilities.
As noted earlier in Section~\ref{subsubsec:diagapprox}, in the general case with DDEs and non-coplanar antenna arrays, the corresponding noise covariance matrix $\sigma_n^2\G^{\dagger}\G$ is non-diagonal, but under our initial assumptions of the absence of DDEs and $w=0$, we note that the noise covariance matrix is {\revised largely diagonal}.
This is also seen in Fig.~\ref{fig:covmats} in the form of a highly diagonal structure of $\G^{\dagger}\G$.

For the benefit of the minimization problem as posed in Eq.~(\ref{eq:minimizationprob}), an approximate i.i.d. Gaussian nature of the embedded noise is ensured by weighting the embedding operator with $\overline{\W}^{-1}$, where
\begin{equation}
\overline{\W}^{2}=\Diag(\G^{\dagger}\G).
\label{eq:wgrid}
\end{equation}
More precisely, this weighting $\overline{\W}^{-1}$ simply normalizes all the diagonal values of the noise covariance matrix to the original noise variance $\sigma^2_n$.

Here we note a natural further dimensionality reduction by discarding those discrete Fourier grid points that are not covered by any interpolation kernel support over the $uv$ plane. 
As contributions of the different interpolation kernels over a given discrete Fourier grid point correspond to individual columns of the matrix $\G$, discrete grid points that are not thus covered manifest as all-zero columns of $\G$ and consequently, zeros on the diagonal of $\G^{\dagger}\G$. 
A subsampling operator $\overline{\Ss}$ can then be applied to the embedding operator to only select dimensions corresponding to non-zero diagonal values of $\G^{\dagger}\G$. A similar selection of dimensions on $\overline{\W}$ gives $\overline{\W}_{\mathsmaller{0}}$ of size $\nn$ below the dimension of the oversampled discrete Fourier grid, thus leading to the dimensionality reduction operator
\begin{equation}
\label{eq:rgt}
\RGt=\overline{\W}_{\mathsmaller{0}}^{-1}\overline{\Ss}\G^{\dagger} \quad\in \mathbb{C}^{\nn\times M}.
\end{equation}
We see that the full measurement operator
\begin{equation}
\PPhi'_{\mathrm{grid}}=\overline{\W}_{\mathsmaller{0}}^{-1}\overline{\Ss}\G^{\dagger}\PPhi \quad\in \mathbb{C}^{\nn\times N}
\label{eq:phigrid}
\end{equation}
preserves the null space of $\PPhi$, following from $\Null(\PPhi) \subseteq \Null(\G^{\dagger}\PPhi) \subseteq \Null(\PPhi^{\dagger}\PPhi)$, thus retaining any original NSP.
The diagonal dominated nature of $\sigma^2_n\RGt\RGt^{\dagger}$ has already been shown, which leads to an appropriate modelling of the noise.
Also, applying $\RGt$ is fast owing to the sparsity of $\G^{\dagger}$.

Note that in the context of this dimensionality reduction with $\RGt$, the weighting matrix $\overline{\W}_{\mathsmaller{0}}^{-1}$, in fact designed for {\revised optimal} weighting of the embedded visibilities, also operates as uniform weighting. Indeed, the diagonal values of $\G^{\dagger}\G$ are a measure of the density of continuous visibilities at each discrete grid point.

As with $\RFPhit$, we also approximate the initial noise covariance matrix in this case by assuming $\G^{\dagger}\G\propto \Ii$, leading to the dimensionality reduction $\R=\G^{\dagger}$. This variant is included in comparisons of reconstruction quality.

\begin{table*}
	\centering
	\caption{Summary of the different dimensionality reduction methods with their advantages and disadvantages with respect to compressed sensing-based imaging.}
	\ra{1.3}
	\begin{tabular}{@{}lcccc@{}}
		\toprule
		Properties of $\PPhi'=\R\PPhi$ (Full meas. operator) & $\R=\R_{\mathrm{sing-o}}$ & $\R=\RFPhit$ & $\R=\RPhit$ & $\R=\RGt$ \\
		\hline
		Approximate null space preservation& Yes & Yes & Yes & Yes \\
		Fast implementation& No & Yes & Yes & Yes \\ 
		{\revised Largely diagonal} noise covariance matrix& Yes & Yes & No & Yes \\
   		Final dimension& $N_0 \leq N$ & $N_0 \leq N$ & $N$ & $\nn \leq 4N$ \\
		\bottomrule
	\end{tabular}
	\label{tab:props}
\end{table*}

\subsection{Feature comparison}
\label{subsec:comparisontable}
A comparison of the different dimensionality reduction methods $\RFPhit$, $\RPhit$ and $\RGt$ is shown in Table~\ref{tab:props}.
We note in the listing that $\RFPhit$ provides a good combination of the desired final dimension, the guarantees for compressed sensing-based imaging to reconstruct images, and a {\revised largely diagonal} noise covariance matrix that enables us to embed this technique in the convex optimization algorithm we employ for imaging. 
$\RPhit$ embeds the data to image size while maintaining any original NSP of $\PPhi$, and has a fast implementation; however, it fails to appropriately model the noise and is therefore less suitable for the minimization problem which requires i.i.d. Gaussian embedded noise to enable a simple $\ell_{2}$-norm data fidelity term.
$\RGt$ continues to maintain the NSP of $\PPhi$ and the i.i.d. Gaussian properties of the noise, which are essential for image reconstruction using our convex optimization algorithms. However, the data reduction is limited to a size $\nn \leq 4N$.

As mentioned in Sections~\ref{subsec:svd} and \ref{subsec:otherdimreds}, for each of $\RFPhit$, $\RPhit$ and $\RGt$, an attempt is also made to further approximate the initial noise covariance matrix by the identity matrix in order to render the application of the respective dimensionality reduction methods even faster. 
Eqs.~(\ref{eq:rfphit}), (\ref{eq:rphit}) and (\ref{eq:rgt}) are then simplified to $\R=\F\PPhi^{\dagger}$, $\R=\PPhi^{\dagger}$ and $\R=\G^{\dagger}$ respectively. 
However, this approximation is seen to be inappropriate, leading to poorer modelling of the noise and consequently lowering image reconstruction quality. 
$\RFPhit$, $\RPhit$, $\RGt$ and their respective variants are used to reduce dimensionality before performing image reconstruction in different settings and with varying data sizes. 

\subsection{Further reduction by thresholding}
\label{subsec:holeycoverage}

As a conservative dimensionality reduction method, $\RFPhit$ would embed to a final data size $\Nz \approx N$ under the assumption that there are very few zero-valued singular values of $\PPhi$, within limits of numerical precision, and all corresponding singular vectors are thus necessary to retain the information content of $\PPhi$. 
Similarly, $\RGt$ embeds to a size $\nn\approx 4N$ under the assumption of having contributions to the continuous visibilities from most discrete Fourier grid points. 
However, further dimensionality reduction may be obtained in both cases by a thresholding strategy.

We first consider the Fourier dimensionality reduction model based on $\RFPhit$. 
The approach described below to further reduce the final embedding dimension consists of discarding the data dimensions associated with singular values $\SSigma_i$ below a threshold, rather than conservatively discarding those equal to zero only {\revised -- this is made possible due to the fact that the singular values occupy a range of values going down to zero, as discussed in Section~\ref{subsec:svd}.}
In other words, {\revised through such a thresholding operation,} we will attempt a low-rank approximation of the original singular value matrix $\SSigma$, and consequently reduce our final data dimension to the corresponding low-rank.
For the sake of the simplicity of this argument, we consider $\RFPhit$ to be equal to $\R_{\mathrm{sing-o}}$.
From Eq.~(\ref{eq:phisingo}), we see that the full measurement operator $\PPhi'_{\mathrm{sing-o}}$ reads as a weighted subsampling in the orthonormal basis $\V^{\dagger}$, with the weights given by the singular values of $\PPhi$. 
From Eq.~(\ref{eq:covmatsingo}), the noise covariance matrix reads as $\sigma^2_n\Ii$.
In order to safely discard a given singular value $\SSigma_i$ for the dimension $i$ without losing information, its effect on the corresponding embedded visibility $\boldsymbol{y}'_i$ would need to be negligible relative to the embedded noise level $\sigma_n$:
\begin{equation}
\label{eq:ynoisethreshold}
\lvert\boldsymbol{y}'_i\rvert < \gamma\sigma_n, \quad \mathrm{with~} \gamma=\mathcal{O}(1).
\end{equation}
In general, one has $\lvert\boldsymbol{y}'_i\rvert \leq \Sigma_i\|\boldsymbol{x}\|_2$, which is saturated only in the case where $\boldsymbol{x}$ is fully aligned with the right singular vector $\V_i$.
This means that the condition
\begin{equation}
\SSigma_i\|\boldsymbol{x}\|_2 < \gamma\sigma_n
\end{equation}
is sufficient to ensure the requirement imposed by Eq.~(\ref{eq:ynoisethreshold}) to disregard dimension $i$.
In other words, the data dimension $i$ can be discarded with no adverse effect on signal reconstruction if the corresponding singular value, computed as given in Eq.~(\ref{eq:approxfzero}), is below a noise-based threshold:
\begin{equation}
\label{eq:diagthreshold}
\SSigma_i<\frac{\gamma\sigma_n}{\|\boldsymbol{x}\|_2}.
\end{equation}

Secondly, we consider the gridding-based dimensionality reduction $\RGt$. The approach to further reducing the final embedding dimension will again consist of discarding the data dimensions associated with the weights $\overline{\W}_i$ below a threshold, rather than conservatively discarding those equal to zero only. A similar bound as Eq.~(\ref{eq:diagthreshold}) can be deduced as follows for thresholding out data dimensions. Again, for the sake of this very argument only, the full measurement operator $\PPhi'_{\mathrm{grid}}$  in Eq.~(\ref{eq:phigrid}) can be approximated as weighted subsampling in the Fourier basis, with weights $\overline{\W}$ computed from Eq.~(\ref{eq:wgrid}). The noise covariance matrix exhibits diagonal values all equal to the original noise variance $\sigma^2_n$. The same reasoning as for $\RFPhit$ now applies and the data dimension $i$ can be discarded with no adverse effect on signal reconstruction if the corresponding weight is below the following noise-based threshold:
\begin{equation}
\label{eq:diagthresholdgt}
\overline{\W}_i<\frac{\gamma\sigma_n}{\|\boldsymbol{x}\|_2}.
\end{equation}

The threshold computation in Eqs.~(\ref{eq:diagthreshold}) and (\ref{eq:diagthresholdgt}) needs knowledge of $\|\boldsymbol{x}\|_2$, which is a priori not available from interferometric data.
One would naturally want to estimate $\|\boldsymbol{x}\|_2$ from the dirty image.
This is supported by recent work showing that $\boldsymbol{x}$ can be bounded by the dirty image~\citep{wijnholds_data_2011, sardarabadi_radio_2016}. 

{\revised 
\subsection{Reduced computational requirements}
Current radio-interferometric imaging techniques involve processing in the data space (of dimension $M$) and a lower dimensional sparsity space (of dimension $N$ or $\nn$). 
For \textsc{clean}-based algorithms this can be seen in the move between `major' cycles in the data dimension $M$, `minor' cycles with gridded visibilities of dimension $\nn$, and the image space of dimension $N$ with an implicit sparsity assumption. 
For convex optimization-based algorithms like the one used in this work, this is typically seen in the concurrent computation of a data fidelity term with vectors of dimension $M$, and a sparsity prior of lower dimension $N$.

The goal of dimensionality reduction as described here is to reduce the computational load of imaging methods for next-generation radio interferometers where $M$ is very large, {\revisedtwo on the order of $10^{10}$}. 
The proposed dimensionality reduction method using $\RFPhit$ reduces data size by significant amounts to $\Nz \leq N \ll M$, and these lower-dimensional data cause a smaller memory footprint in imaging pipelines. 
The existence of fast sub-operator implementations and a low-sized full measurement operator translate to faster computations per iteration of the convex optimization algorithms. 
The properties of the full measurement operator $\PPhi'$ as listed in Table.~\ref{tab:props} are a good indicator of the computational savings afforded by $\RFPhit$ as compared to other reduction methods. 
The applicability of the proposed dimensionality reduction method using $\RFPhit$ has the advantage of resulting in a reduced data dimension that is independent of the initial data size (this is also true for $\RPhit$ and $\RGt$, which result in reduced data sizes of $N$ and $\nn$ respectively). 

The initial set-up of the imagers shall indeed be affected by an increase in the initial dimension -- in particular, the pre-computation of the holographic matrix and the appropriate weights to be used in the imaging algorithm. 
However, these pre-computations have a one-time cost, and subsequent imaging is unaffected, depending only on the embedded data and thus using fewer resources in terms of memory and computing time.
}

\section{Simulations and results}
\label{sec:plots}
\subsection{Simulation settings}
\label{subsec:simulations}
The effectiveness of the proposed dimensionality reduction method was demonstrated through simulations.
Image quality comparisons were made between reconstructions through the dimensionality reduction methods $\RFPhit$, $\RPhit$ and $\RGt$, and their respective variants $\R=\F\PPhi^{\dagger}$, $\R=\PPhi^{\dagger}$ and $\R=\G^{\dagger}$.
In a first setting, a conservative dimensionality reduction was performed for each case -- $\RFPhit$ accounting for dimensions corresponding to all non-zero singular values of $\PPhi$, and $\RGt$ accounting for dimensions corresponding to all discrete Fourier grid points that have non-zero contribution to the continuous visibilities through interpolation kernels. 
The final data dimension after reduction in this setting was seen to be $\Nz \approx N$ for $\RFPhit$,  $\Nz=N$ for $\RPhit$, and $\nn \approx 4N$ for $\RGt$.

\begin{figure}
	\centering
	\begin{minipage}{.48\linewidth}
	        \centering
  		\includegraphics[trim={0px 0px 0px 0px}, clip, height=3.80cm]{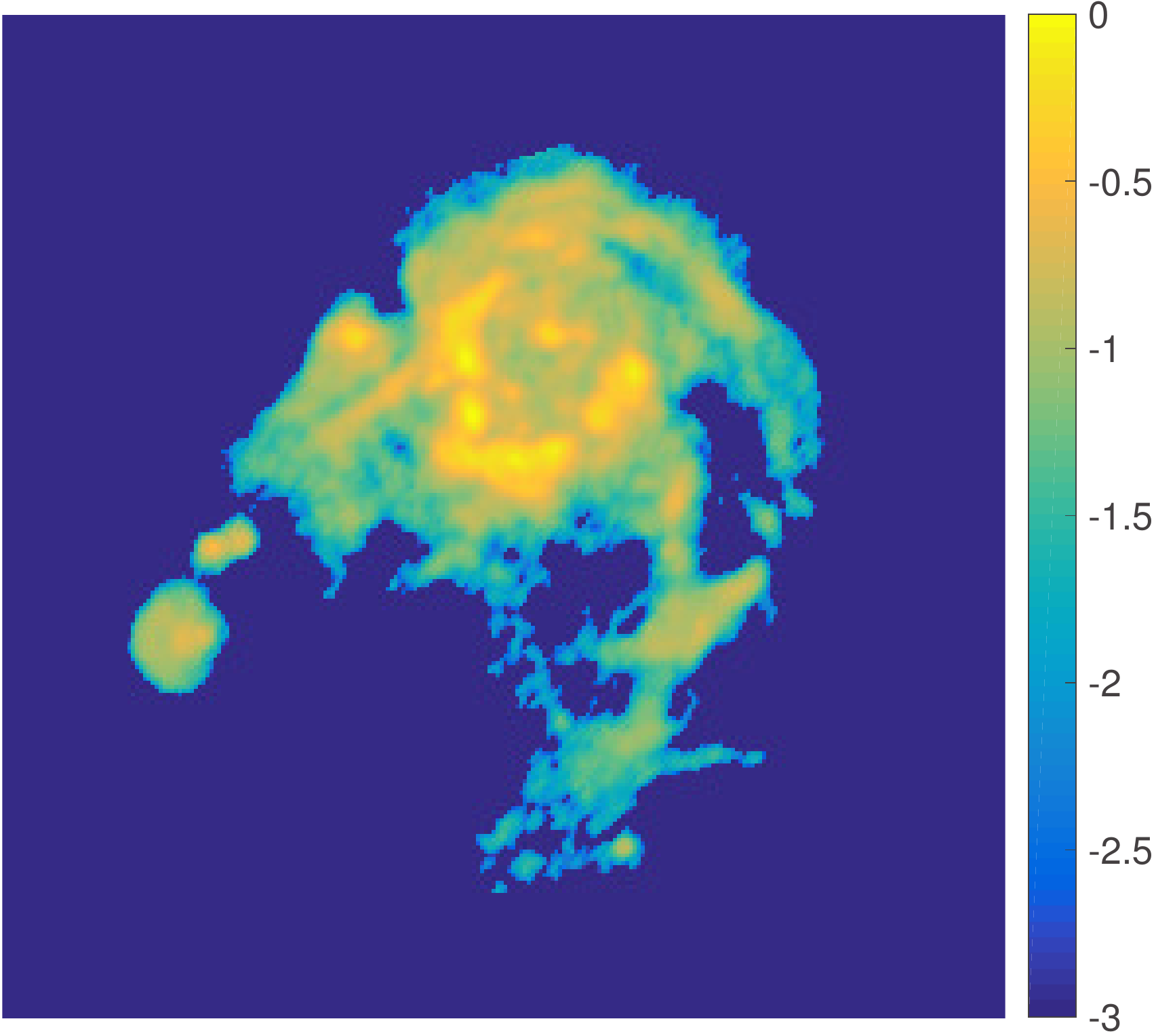}
	\end{minipage}
	\hspace{2pt}
	\begin{minipage}{.48\linewidth}
	        \centering
  		\includegraphics[trim={0px 0px 0px 0px}, clip, height=3.80cm]{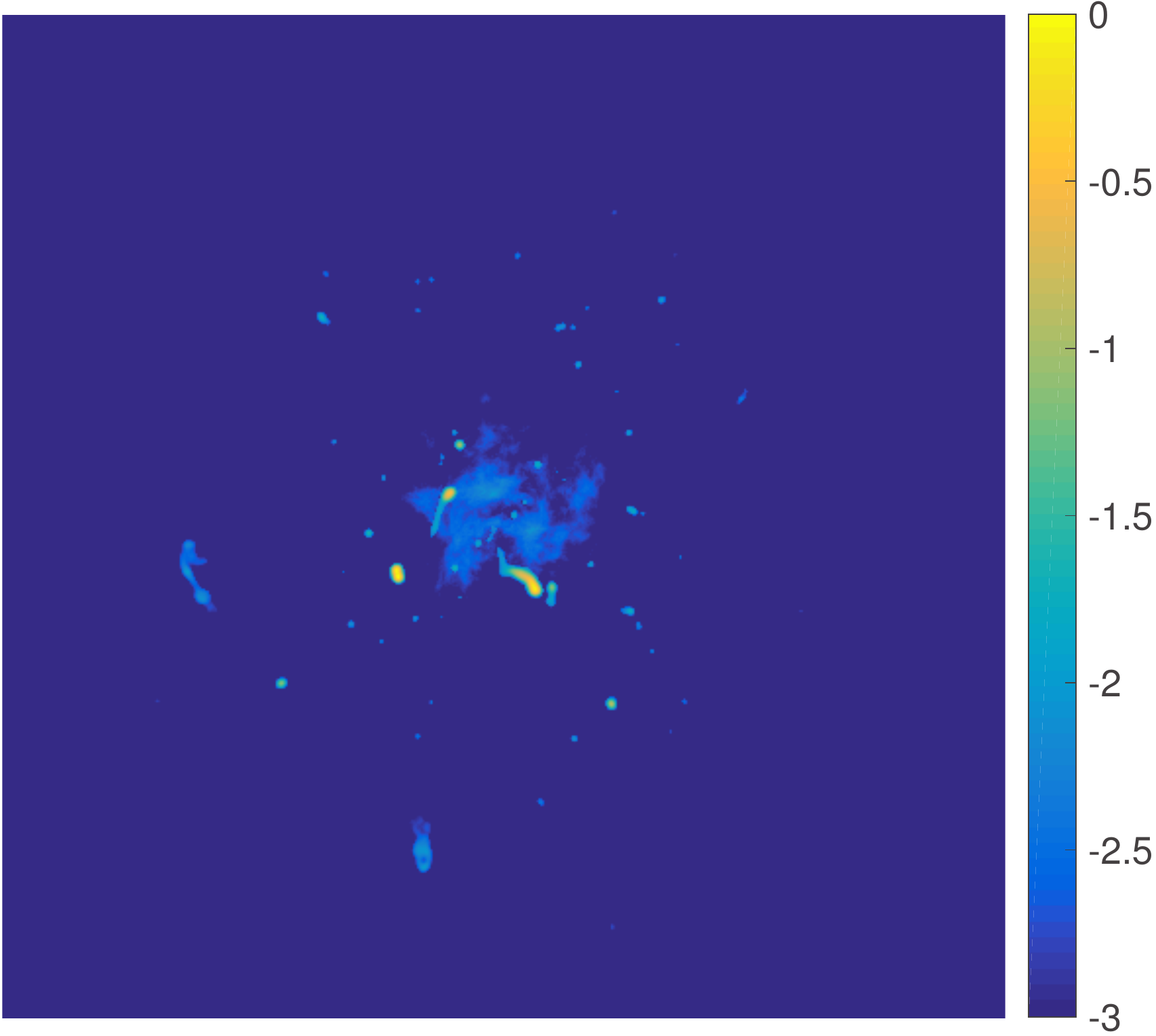}
	\end{minipage}
	
	\hspace{-6pt}
	\begin{minipage}{.98\linewidth}
  		\centering
  		\includegraphics[trim={0px 0px 0px 0px}, clip, height=3.84cm]{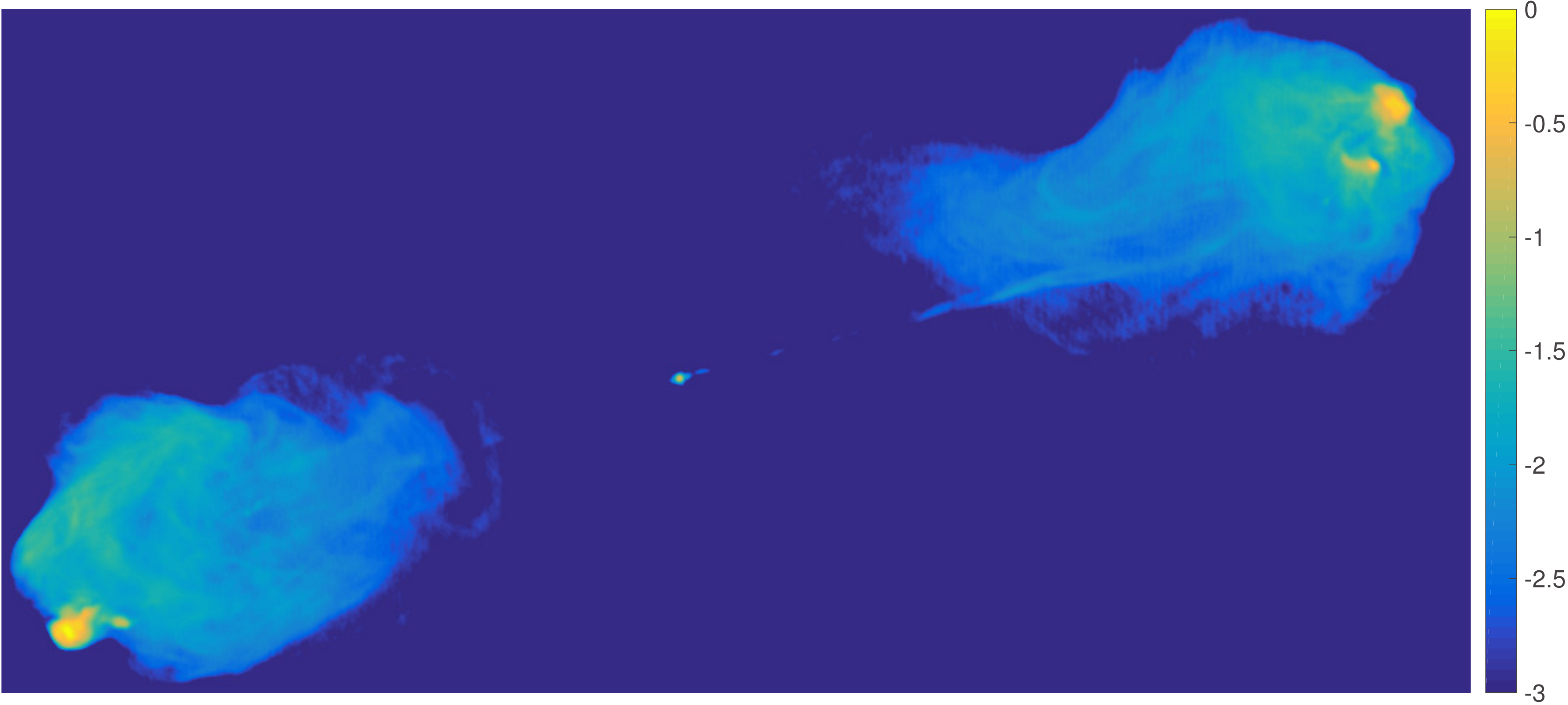}
	\end{minipage}

	\caption{The test images in $\log_{10}$ scale, clockwise from top left: M31 ($256\times 256$ pixels), a simulated galaxy cluster ($512\times 512$ pixels) and Cygnus A ($477\times 1025$ pixels).}
	\label{fig:testimages}
\end{figure}

Simulations were performed on different test images chosen for their varied characteristics: (i) the classic `M31' image has a compact structure showing an H\textsc{ii} region of the M31 galaxy ($256\times 256$ pixels); (ii) an image of a galaxy cluster ($512\times 512$ pixels) simulated using the `\textsc{faraday}' tool \citep{Murgia2004}, has high dynamic range by design; (iii) a partial image of the Cygnus A radio galaxy ($477\times 1025$ pixels) includes a strong central core, two strong jets and lobes of diffuse structure with bright hotspots. These test images are shown in Fig.~\ref{fig:testimages}.

\begin{figure}
	\begin{minipage}{0.49\columnwidth}
		\hspace{-5pt}
		\centering
  		\includegraphics[trim={0px 0px 0px 0px}, clip, height=4.1cm]{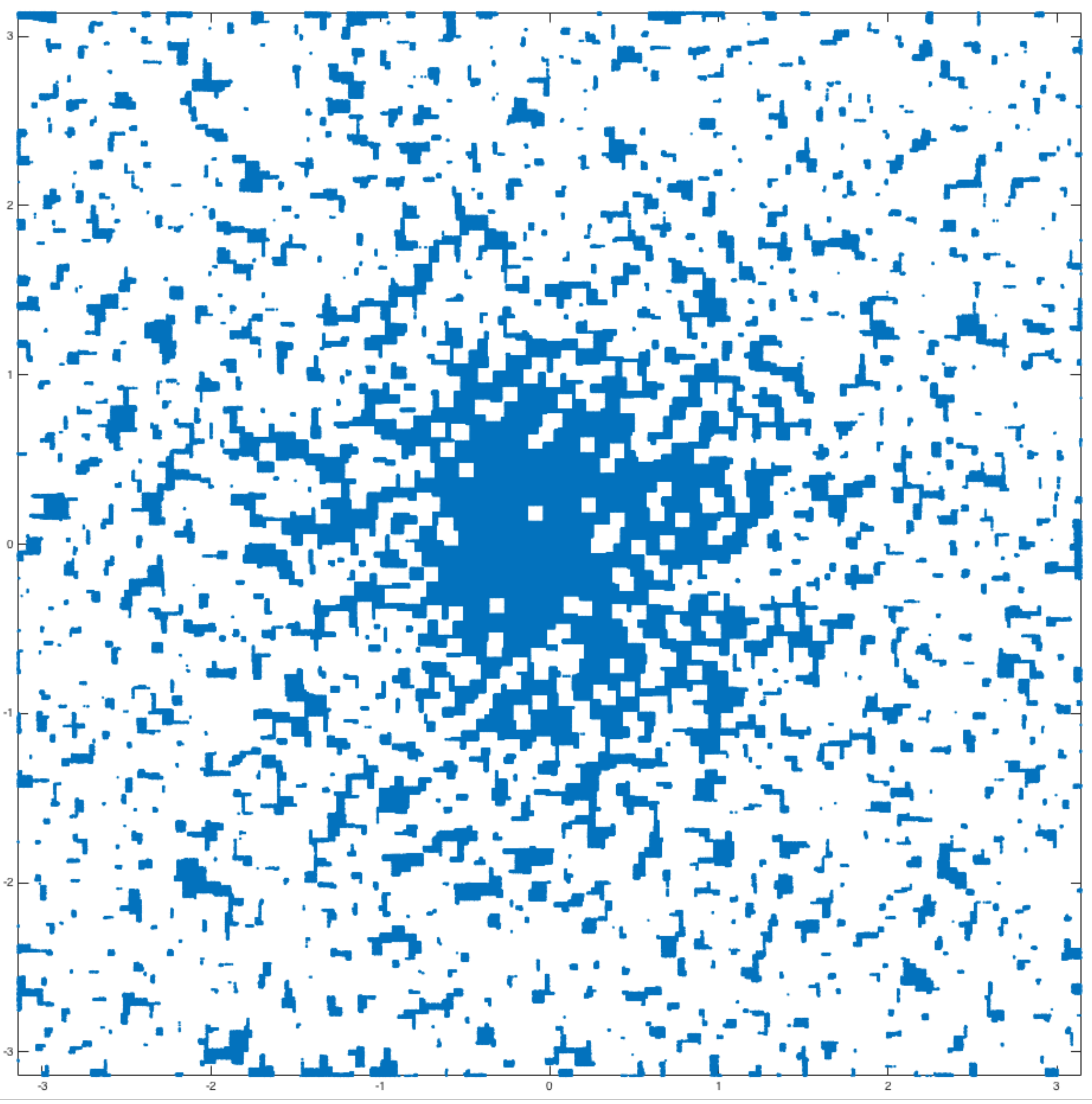}
	\end{minipage}
	\hspace{2pt}
	\begin{minipage}{0.49\columnwidth}
  		\centering
  		\includegraphics[trim={0px 0px 0px 0px}, clip, height=4.15cm]{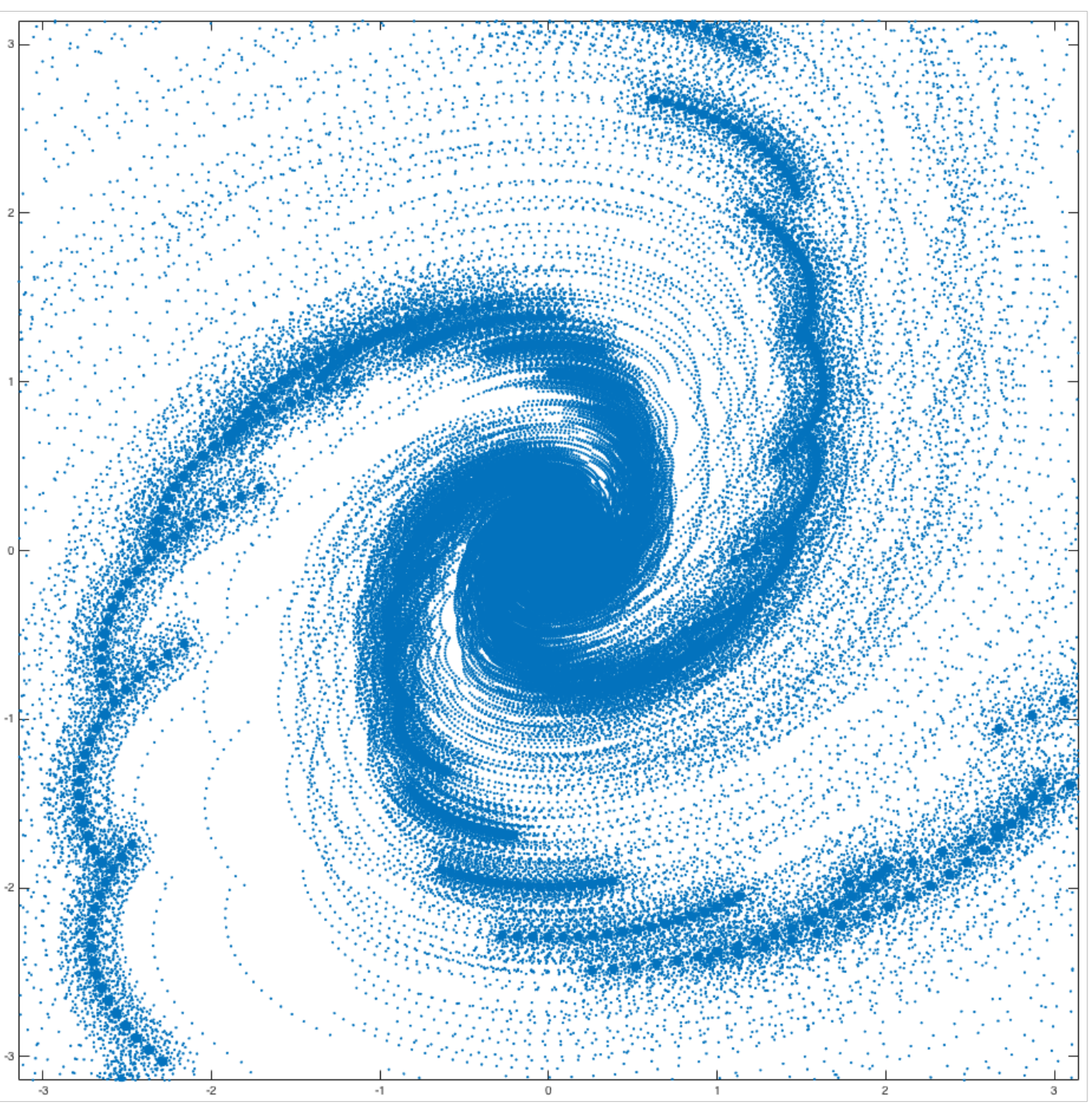}
	\end{minipage}
	\caption{Illustration of simulated $uv$ coverages over normalized frequency axes ${[{-\pi}, \pi]}$. Left: Random Gaussian density profile with the $uv$ plane increasingly sparsely covered at higher frequencies (approximately 3,000,000 $uv$ points); Right: Profile generated with SKA-like baselines, with partial ellipses simulating dense coverage at lower frequencies and sparser coverage at higher frequencies. A telescope configuration of 254 antennas was used to obtain approximately 650,000 $uv$ points.}
	\label{fig:coverages}
\end{figure}

Two categories of $uv$ coverages were used to simulate telescope measurements. 
One with synthetic coverages with a random Gaussian sampling profile with missing frequency regions, and another with more realistic SKA-like coverages generated with a simulated telescope configuration of 254 antennas (Fig.~\ref{fig:coverages}), using the `lwimager' tool made available as part of the `Casacore' software suite\footnote{The lwimager implementation is available in the `casarest' package in Casacore at \url{https://github.com/casacore/casarest}}. The SKA-like coverages correspond to observation times ranging from 30 minutes to 8 hours, depending on the image and the initial data dimension.
The frequencies were normalized to lie in the interval ${[{-\pi}, \pi]}$. The $(0, 0)$ component of the $uv$ plane was not included in generating visibilities, reflecting the real situation of the absence of zero-length baselines.

For each coverage, complex visibilities were generated {\revised over the continuous $uv$ plane using the measurement model described in Eq.~\ref{eq:rime}}, and these were perturbed with complex additive white Gaussian noise to give an input signal-to-noise ratio (SNR) of 30~dB. 
The input SNR was defined as  $\text{SNR}_{\text{i}}=20\log_{10}(\|\PPhi\boldsymbol{x}\|_2/\|\boldsymbol{n}\|_2)$.
The continuous visibilities dimension was varied over a wide range, in multiples of image size, from $10$ to $100$.
This corresponds to an approximate range of 650,000 to 26 million visibilities over different test images.
Continuous visibilities were generated with a non-uniform oversampled Fourier transform ($2\times$ oversampling in each image dimension) using $8\times 8$ Kaiser-Bessel interpolation kernels as described and implemented by \citet{fessler_nonuniform_2003}.
$\ell_1$-minimization was performed using the \textsc{sara} algorithm implemented in \textsc{purify}. \textsc{sara} regularizes the inverse problem by imposing `average sparsity' of the signal over a set of bases, as explained in Section~\ref{subsec:sparsereconstruction}. The resulting minimization problem was solved using an ADMM-based proximal splitting method -- see \citet{carrillo_sparsity_2012, carrillo_purify:_2014} for further details on \textsc{sara} and \citet{carrillo_scalable_2015, onose_scalable_2016} for implementation details of the ADMM-based minimizer.

Additionally, a more aggressive dimensionality reduction was performed for $\RFPhit$ and $\RGt$ by only retaining dimensions corresponding to significant values of embedded data, as described in Section~\ref{subsec:holeycoverage}.
This led to a final data dimension size of $\Nz \ll N$ and $\nn \ll 4N$ respectively.
Image reconstruction was performed for SKA-like coverages using these further reduced data.

As a comparison baseline, images were also recovered using the complete visibilities set. 
This corresponds to the trivial `embedding' $\R=\Ii$. 
Finally, the reconstruction was compared with model images obtained using the Multiscale~\textsc{clean} (\textsc{ms-clean}) algorithm~\citep{cornwell_multiscale_2008} as provided in the `WSClean' program~\citep{offringa_wsclean_2014}. \textsc{ms-clean} was run on continuous visibilities simulated from SKA-like $uv$ coverages.
{\revised The synthetic Gaussian-profile $uv$ coverages were generated on the fly during simulations, and since WSClean takes measurement sets as input, \textsc{ms-clean} was not run on data simulated using these coverages.}

We used SNR and dynamic range (DR) of the reconstructed image as measures of image quality, and compared these values across different methods in our simulations. We define `Reconstruction SNR' as
\begin{equation}
\text{SNR}_{\hat{\boldsymbol{x}}} = 20\log_{10}\left(\frac{\|\boldsymbol{x}\|_2}{\|\boldsymbol{x}-\hat{\boldsymbol{x}}\|_2}\right),	
\end{equation}
and `Reconstruction DR' as
\begin{equation}
\text{DR}_{\hat{\boldsymbol{x}}} = \sqrt N\|\PPhi\|^2 \frac{max(\hat{\boldsymbol{x}})}{\|\PPhi^{\dagger}(\boldsymbol{y} - \PPhi\hat{\boldsymbol{x}})\|_2},
\end{equation}
where $\boldsymbol{x}$ is the underlying test image measured through the operator $\PPhi$ (with spectral norm $\|\PPhi\|$) to give visibilities $\boldsymbol{y}$, and $\hat{\boldsymbol{x}}$ is the reconstructed image.
For the methods $\RFPhit$, $\RPhit$, $\RGt$ and their respective variants, the DR was computed using the corresponding measurement operators as given in Eqs.~(\ref{eq:phising}), (\ref{eq:phidirt}) and (\ref{eq:phigrid}), along with the respective embedded visibilities.

{\revised
Since reconstruction results from \textsc{ms-clean} are either in the form of a restored image containing added residual, or a model image containing extended components, direct SNR and DR computations are not readily apparent and no longer remain a valid way to compare reconstruction performance between \textsc{clean} and the proposed compressed sensing-based imaging methods. Therefore, the \textsc{ms-clean} reconstructions are {\revisedtwo presented here as model images (without being convolved with the beam and adding the residual)} for visual comparison with the other methods described in this work.
}

\subsection{Image reconstruction results}
\label{subsec:results}
\subsubsection{Accounting for all non-zero singular values}
\begin{figure*}
	\begin{subfigure}{1\linewidth}
		\centering
  		\begin{subfigure}{.5\linewidth}
			\centering
  			\includegraphics[trim={0 0 0 0}, clip, width=\columnwidth]{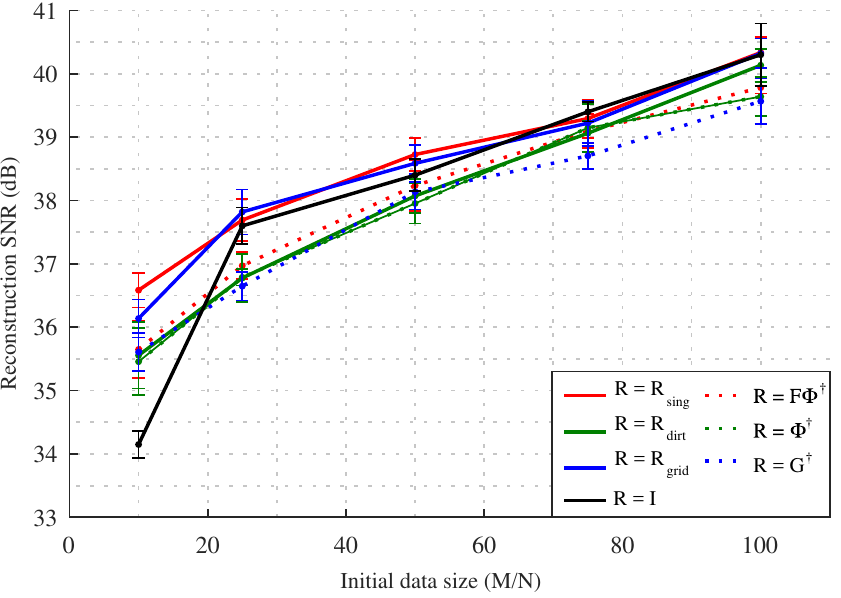}
		\end{subfigure}%
		\begin{subfigure}{.5\linewidth}
  			\centering
  			\includegraphics[trim={0 0 0 0}, clip, width=0.97\columnwidth]{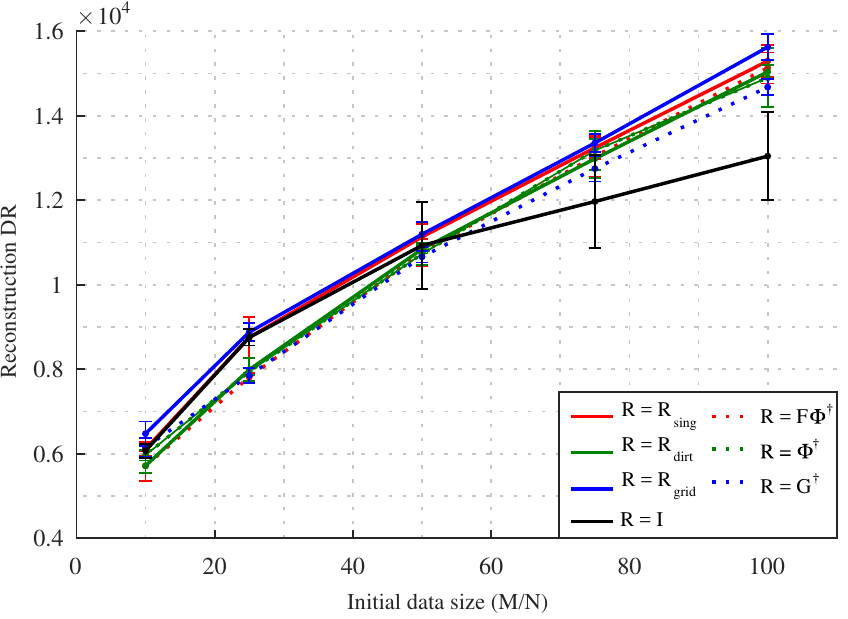}
		\end{subfigure}%
  		\caption{M31}
		\label{fig:m31graphs_uni}
	\end{subfigure}
	\vspace{5pt}
	\begin{subfigure}{1\linewidth}
		\centering
  		\begin{subfigure}{.5\linewidth}
			\centering
  			\includegraphics[trim={0 0 0 0}, clip, width=\columnwidth]{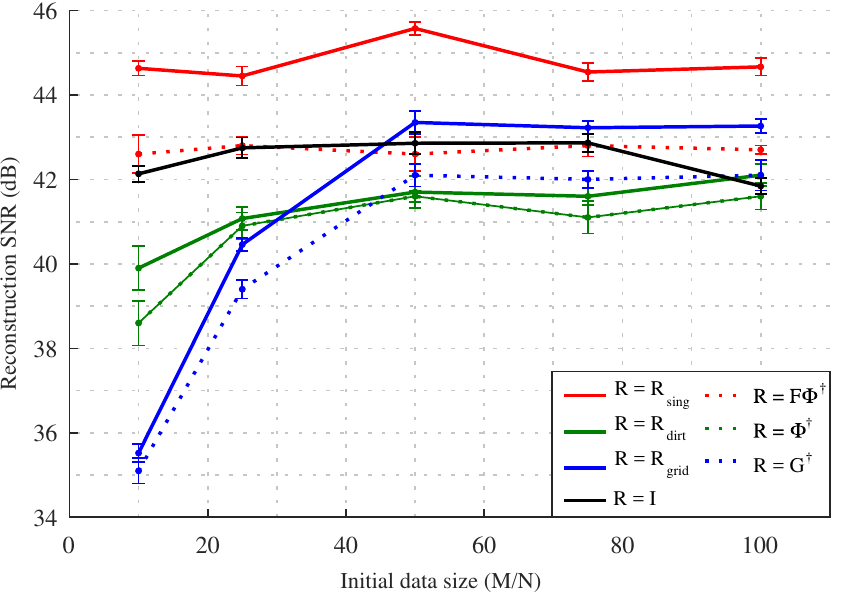}
		\end{subfigure}%
		\begin{subfigure}{.5\linewidth}
  			\centering
  			\includegraphics[trim={0 0 0 0}, clip, width=0.97\columnwidth]{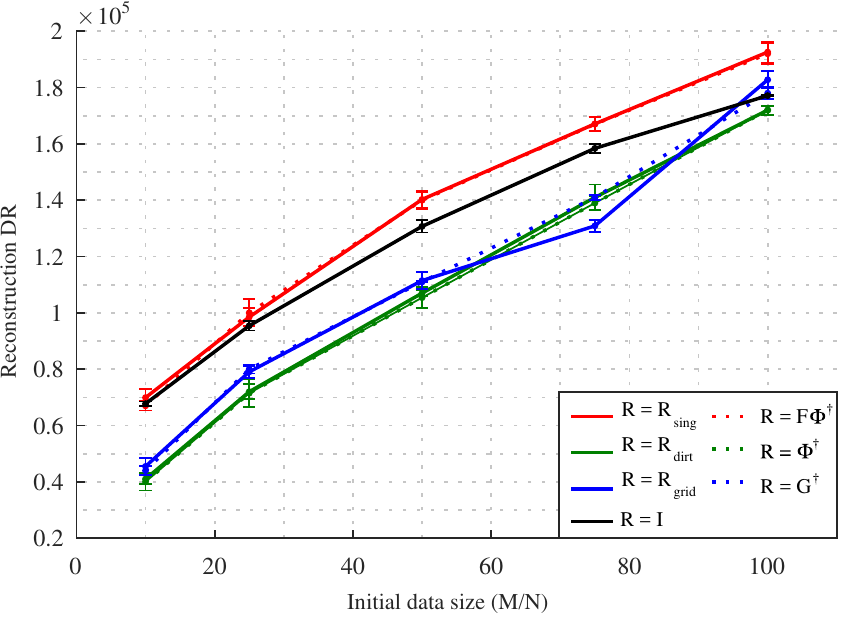}
		\end{subfigure}%
  		\caption{Galaxy cluster}
		\label{fig:galaxyclustergraphs_uni}
	\end{subfigure}
	\vspace{5pt}
	\begin{subfigure}{1\linewidth}
		\centering
  		\begin{subfigure}{.5\linewidth}
			\centering
  			\includegraphics[trim={0 0 0 0}, clip, width=\columnwidth]{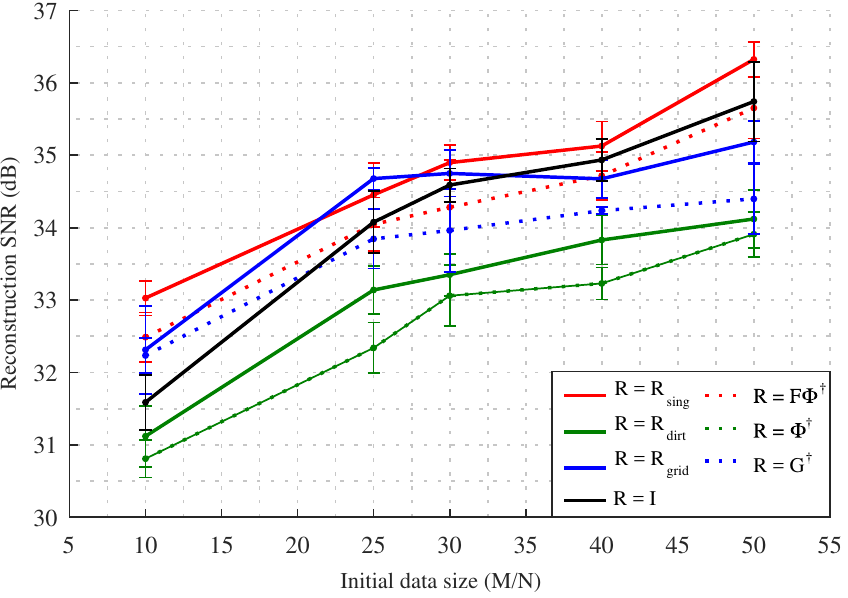}
		\end{subfigure}%
		\begin{subfigure}{.5\linewidth}
  			\centering
  			\includegraphics[trim={0 0 0 0}, clip, width=0.97\columnwidth]{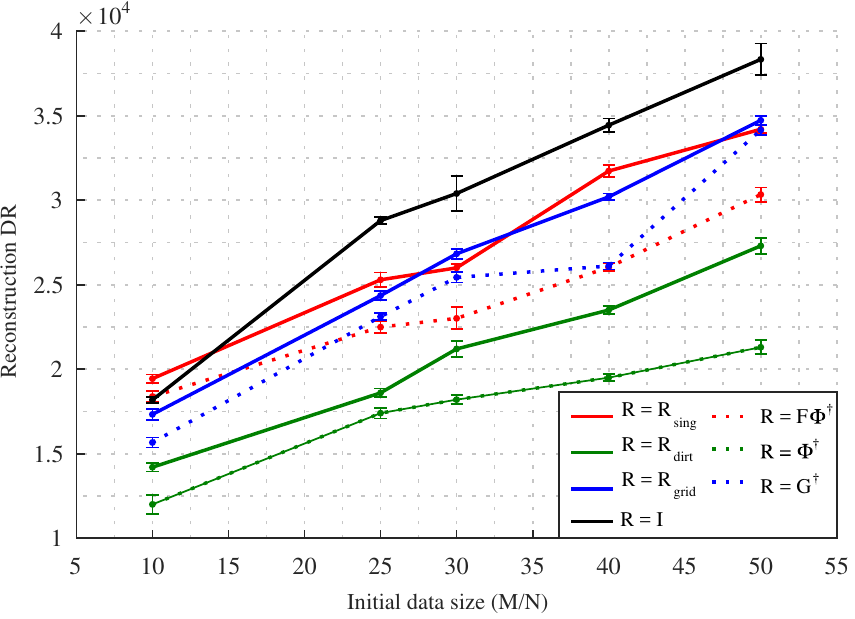}
		\end{subfigure}%
  		\caption{Cygnus A}
		\label{fig:cygagraphs_uni}
	\end{subfigure}
	\caption{Image reconstruction from visibilities using Gaussian random coverage: comparison of different dimensionality reduction methods. Left: SNR; Right: DR, for the recovered image over a range of initial continuous visibilities. Top row: M31 image, initial data size varies from 650,000 to 6.5 million visibilities. Middle row: Galaxy cluster image, initial data size varies from 2.6 million to 26 million visibilities. Bottom row: Cygnus A image, initial data size varies from 4.8 million to 24 million visibilities. Error bar lengths correspond to one standard deviation around the mean over $\sim\!15$ simulations.}
	\label{fig:graphs_uni}
\end{figure*}

\begin{figure*}
	\begin{subfigure}{1\linewidth}
		\centering
  		\begin{subfigure}{.5\linewidth}
			\centering
  			\includegraphics[trim={0 0 0 0}, clip, width=\columnwidth]{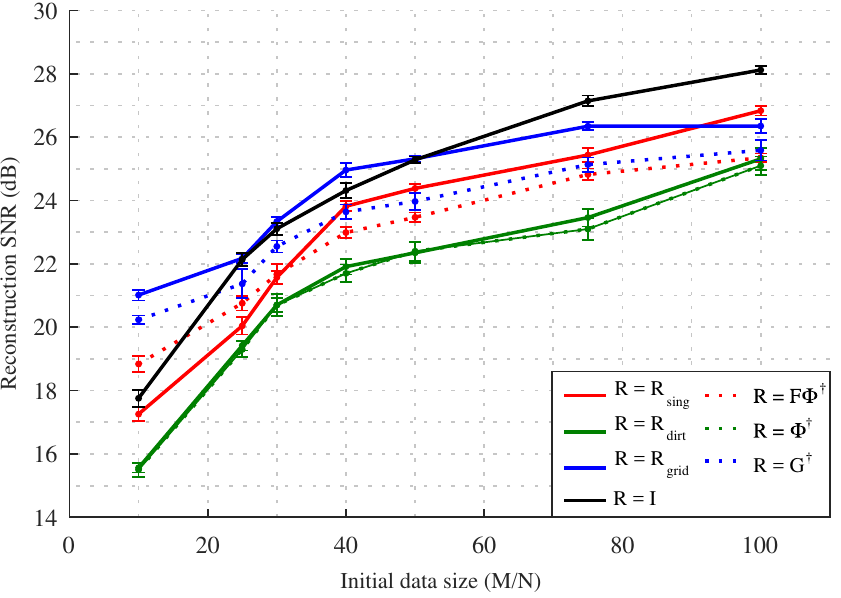}
		\end{subfigure}%
		\begin{subfigure}{.5\linewidth}
  			\centering
  			\includegraphics[trim={0 0 0 0}, clip, width=0.96\columnwidth]{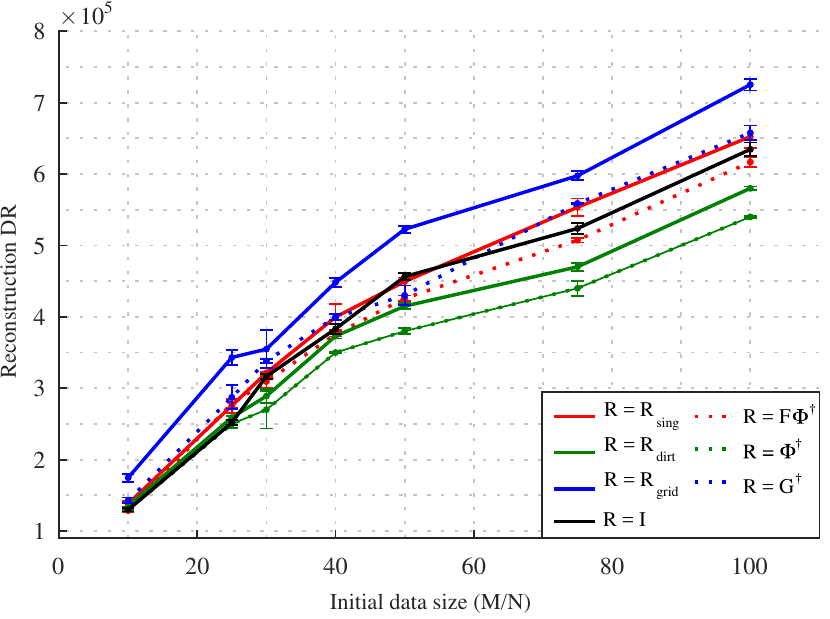}
		\end{subfigure}%
  		\caption{M31}
		\label{fig:m31graphs_ska}
	\end{subfigure}
	\vspace{5pt}
	\begin{subfigure}{1\linewidth}
		\centering
  		\begin{subfigure}{.5\linewidth}
			\centering
  			\includegraphics[trim={0 0 0 0}, clip, width=\columnwidth]{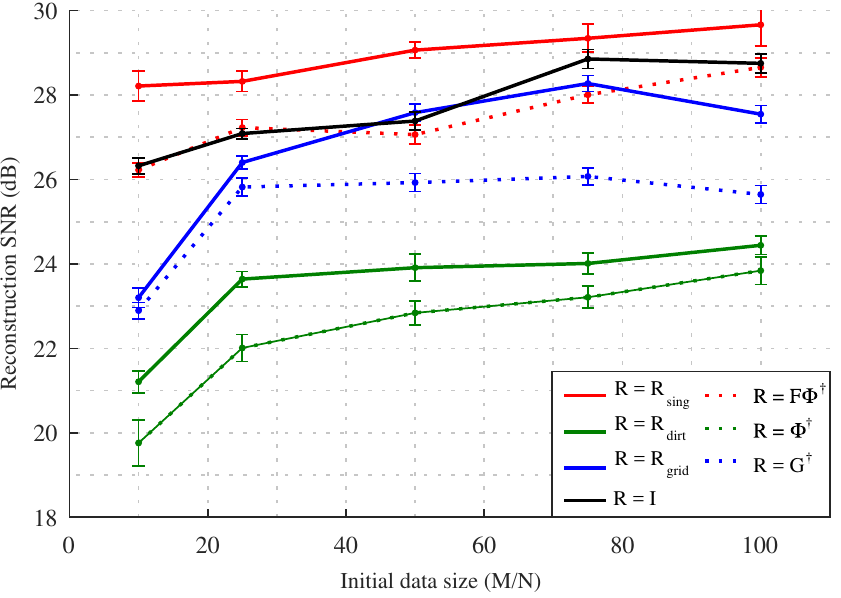}
		\end{subfigure}%
		\begin{subfigure}{.5\linewidth}
  			\centering
  			\includegraphics[trim={0 0 0 0}, clip, width=0.97\columnwidth]{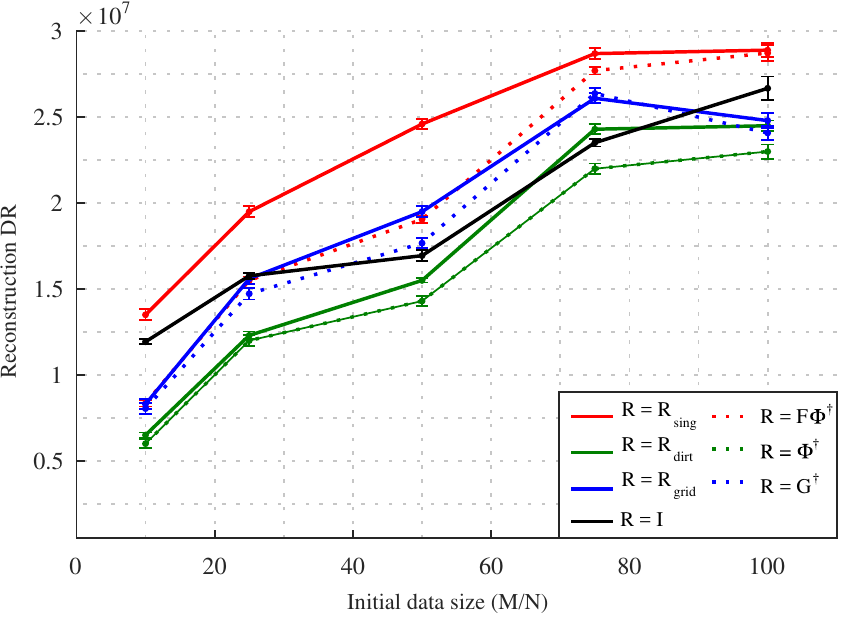}
		\end{subfigure}%
  		\caption{Galaxy cluster}
		\label{fig:galaxyclustergraphs_ska}
	\end{subfigure}
	\vspace{5pt}
	\begin{subfigure}{1\linewidth}
		\centering
  		\begin{subfigure}{.5\linewidth}
			\centering
  			\includegraphics[trim={0 0 0 0}, clip, width=\columnwidth]{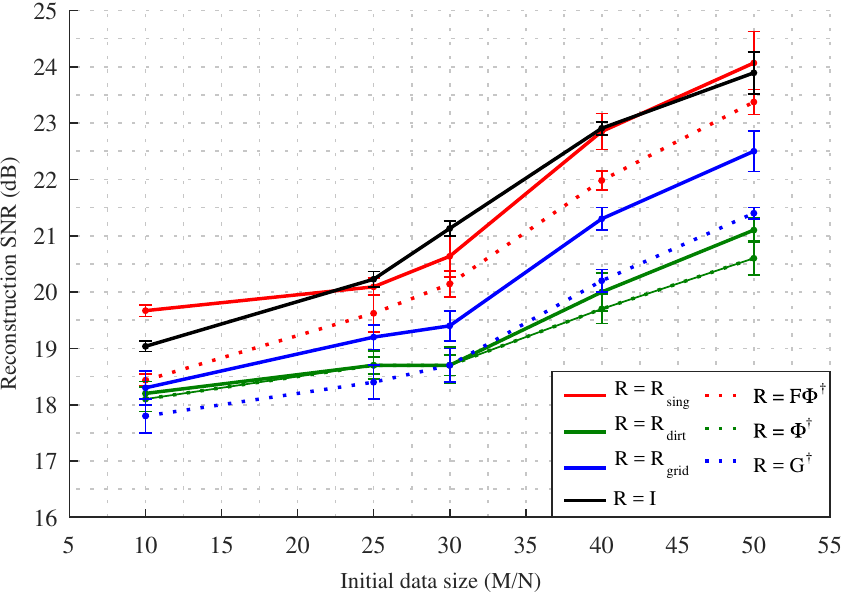}
		\end{subfigure}%
		\begin{subfigure}{.5\linewidth}
  			\centering
  			\includegraphics[trim={0 0 0 0}, clip, width=0.945\columnwidth]{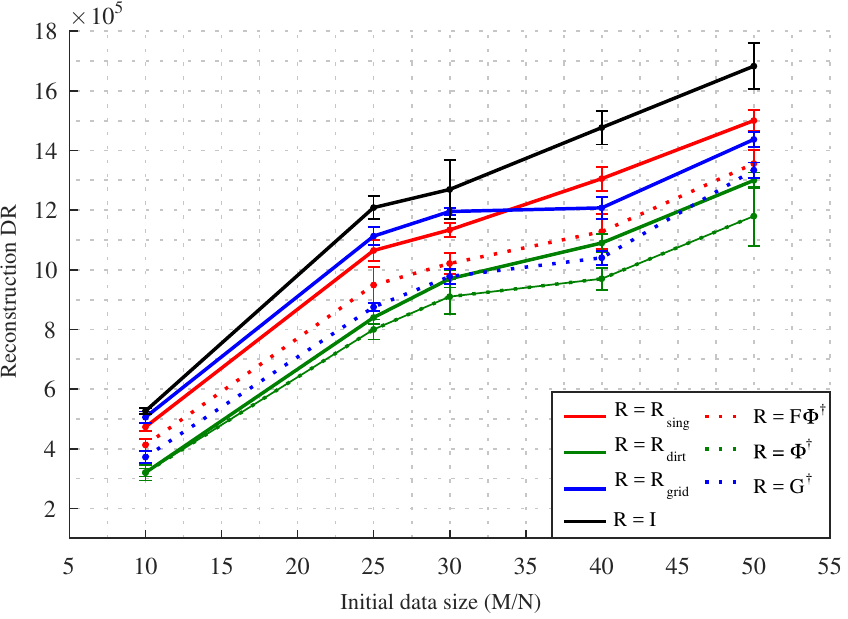}
		\end{subfigure}%
  		\caption{Cygnus A}
		\label{fig:cygagraphs_ska}
	\end{subfigure}
	\caption{Image reconstruction from visibilities using SKA-like coverages: comparison of different dimensionality reduction methods. Left: SNR; Right: DR, for the recovered image over a range of initial continuous visibilities. (a): M31 image, initial data size varies from 650,000 to 6.5 million visibilities. (b): Galaxy cluster image, initial data size varies from 2.6 million to 26 million visibilities. (c): Cygnus A image, initial data size varies from 4.8 million to 24 million visibilities. Error bar lengths correspond to one standard deviation around the mean over $\sim\!15$ simulations.}	\label{fig:graphs_ska}
\end{figure*}

Image reconstruction performance of the different methods over varying simulation settings is discussed here for each test image.
Graphs showing SNR and DR comparisons over the two types of coverage are shown in Figs.~\ref{fig:graphs_uni} and \ref{fig:graphs_ska} respectively. 
For the SKA-like coverages, a visual comparison is also made between the methods $\R=\Ii$,  $\RFPhit$, $\RPhit$, $\RGt$, and \textsc{ms-clean} by showing the reconstructed, error and residual images in $\log_{10}$ scale.
These visual comparisons for the three test images are shown in Figs.~\ref{fig:m31visualcomp}, \ref{fig:galaxyclustervisualcomp} and \ref{fig:cygavisualcomp}.
{\revised 
The reconstructed images shown for \textsc{ms-clean} are obtained by cropping from a 3-4 times larger model image output by WSClean.
For all test images, \textsc{ms-clean} was run with a uniform weighting scheme, set to iterate down to an automatically calculated threshold of two standard deviations of the noise, and with a major loop gain of $0.8$.
The model image in each case was renormalized to have a maximum pixel value {\revisedtwo matching that of the \textsc{ms-clean} model}. 
It may be noted that this does not change the overall visual appearance of the reconstruction and error images as shown in Figs.~\ref{fig:m31visualcomp}, \ref{fig:galaxyclustervisualcomp} and \ref{fig:cygavisualcomp}.
The $\log_{10}$ scale was used to highlight the smallest variations and structures, which inadvertently emphasizes the low-valued artefacts in \textsc{ms-clean} output models; a linear scale would render these artefacts visually indistinguishable from the background.
}
{\revisedtwo Uniformly weighted dirty residual images were generated using WSClean for each output model from the different image recovery methods. Absolute Jy/beam residual values were plotted on a $\log_{10}$ scale, highlighting small variations in structure. The comparatively significant structure seen in the residual images for $\R=\Ii$, $\RFPhit$, $\RPhit$ and $\RGt$ may be due to the absence of negative-valued model components, which are present in the model generated by \textsc{ms-clean} and are compensated for during the computation of the dirty image by WSClean.}

The M31 test image is reconstructed accurately for all the imaging methods. 
Reconstruction with the complete visibilities set reaches 40~dB for Gaussian random coverages and 28~dB for SKA-like coverages at data sizes of $100N$. 
Fig.~\ref{fig:m31graphs_uni} shows that $\RFPhit$ and $\RGt$ perform equally well over Gaussian random coverages, reaching output SNRs of around 40~dB.
Over SKA-like coverages, $\RGt$ results in output SNRs noticeably higher than $\RFPhit$ for most data sizes, e.g., around 25~dB for data sizes of $75-100N$ as shown in Fig.~\ref{fig:m31graphs_ska}.
When comparing DR over Gaussian random coverages, images reconstructed after applying $\RFPhit$ reach $1.6\times 10^4$, $\RGt$ reaches $1.5\times 10^4$ and the complete visibilities set leads to a DR of $1.3\times 10^4$. 
The corresponding values over SKA-like coverages are $7.2\times 10^5$, $6.5\times 10^5$ and $6.3\times 10^5$ respectively.
Fig.~\ref{fig:m31visualcomp} shows a visual comparison of images reconstructed over SKA-like coverages that confirms this trend, where  $\RGt$ results in the lowest error among all methods. 
{\revisedtwo 
}
{\revised The computation time of the ADMM-based algorithm used for image reconstruction shows a clear advantage of using $\RFPhit$, which takes $\approx$1.5 seconds per iteration as opposed to $\approx$18 seconds per iteration without dimensionality reduction, when using all $M=100N$ visibilities.}
{\revised  \textsc{ms-clean} output model images of size $1024\times 1024$ pixels were cropped to $256\times 256$ pixels. 
Image deconvolution took 4 major iterations, and the output model shown In Fig.~\ref{fig:m31visualcomp} contains 8458 components. 
{\revisedtwo 
}

\begin{figure*}
	\centering
	\includegraphics[trim={0px 0px 0px 0px}, clip, height=0.24\linewidth]{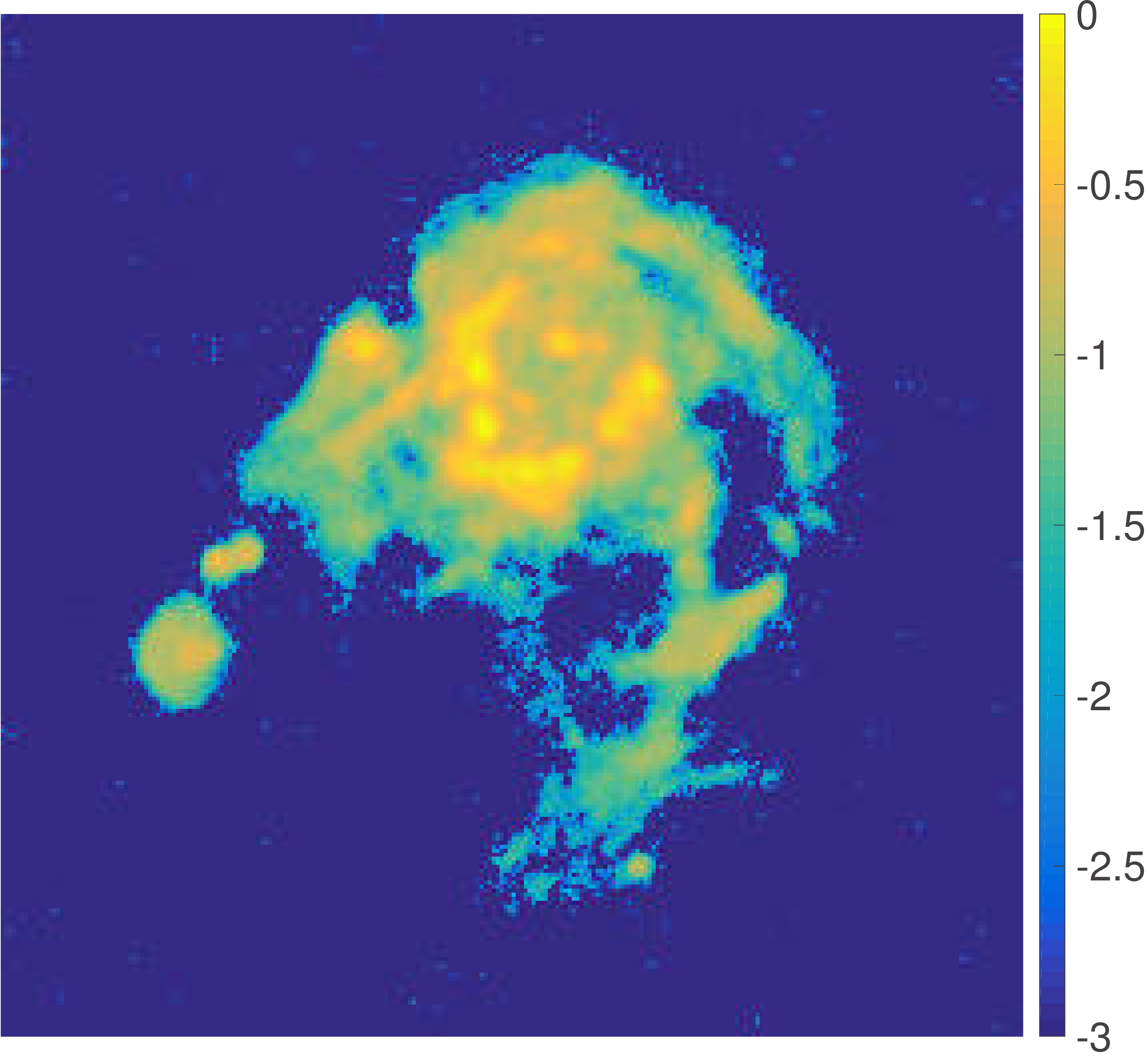}\hspace{2pt}
	\includegraphics[trim={0px 0px 0px 0px}, clip, height=0.24\linewidth]{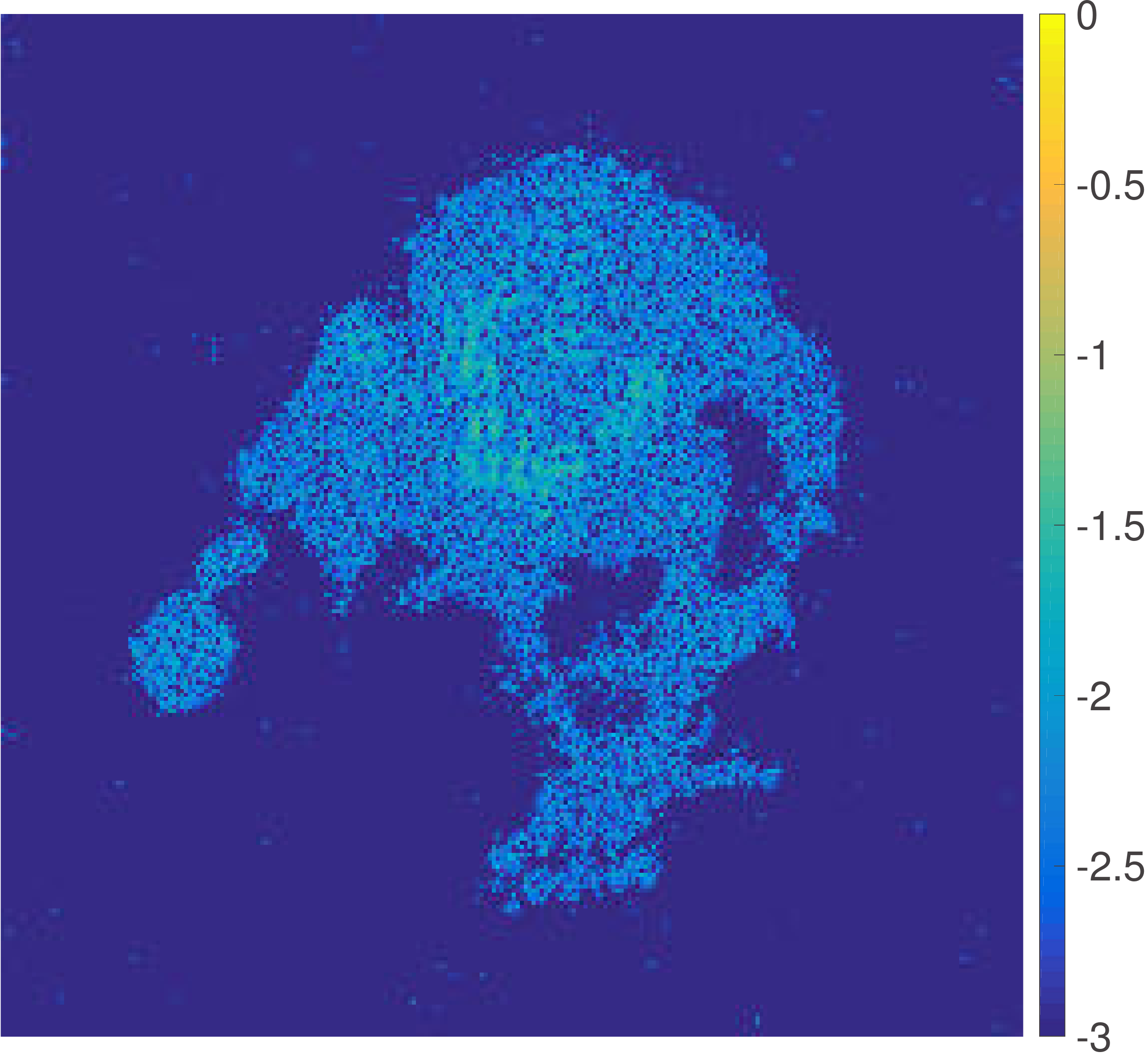}\hspace{2pt}
	\includegraphics[trim={0px 0px 0px 0px}, clip, height=0.24\linewidth]{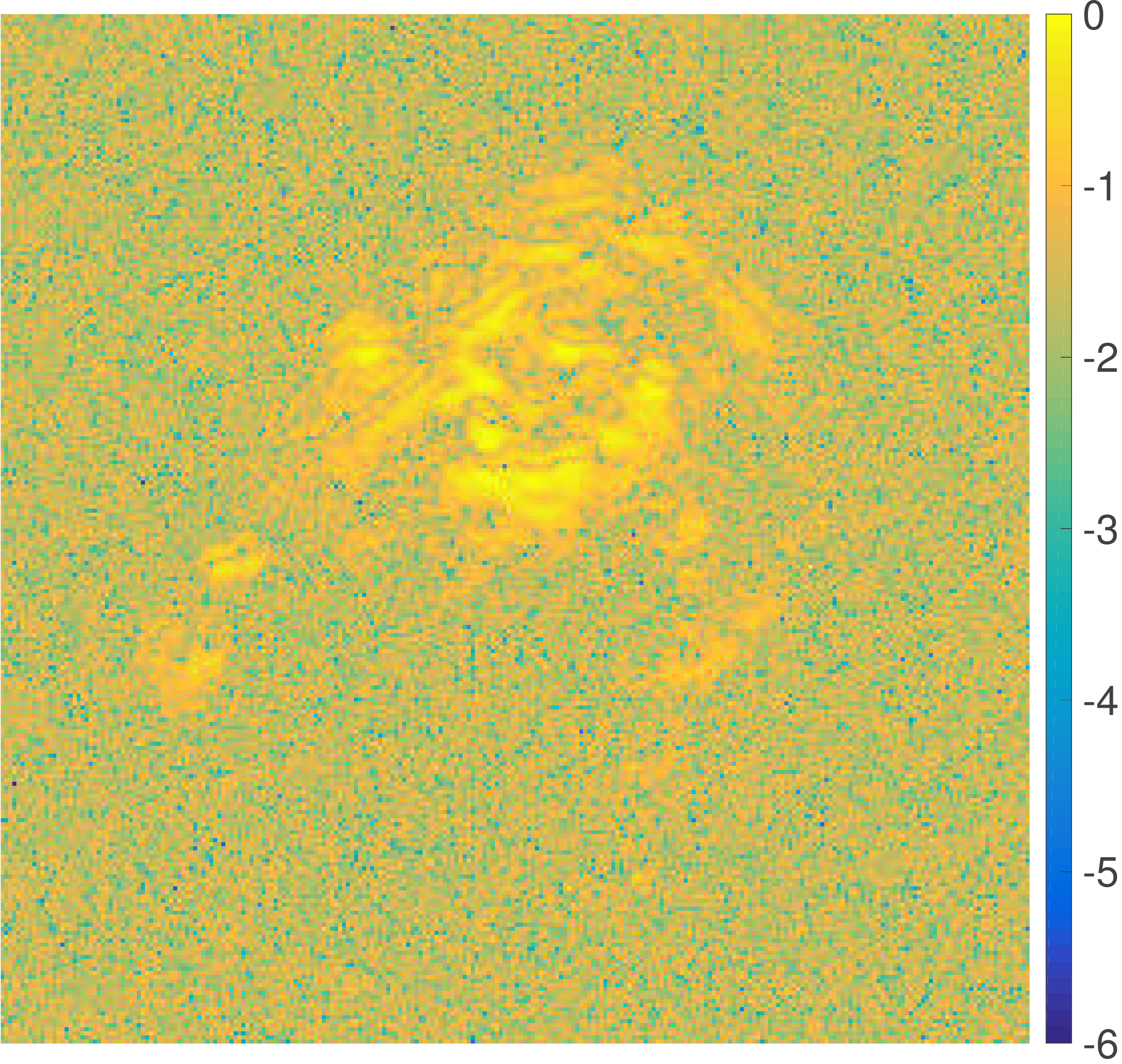}

	\vspace{3pt}
		
	\includegraphics[trim={0px 0px 0px 0px}, clip, height=0.24\linewidth]{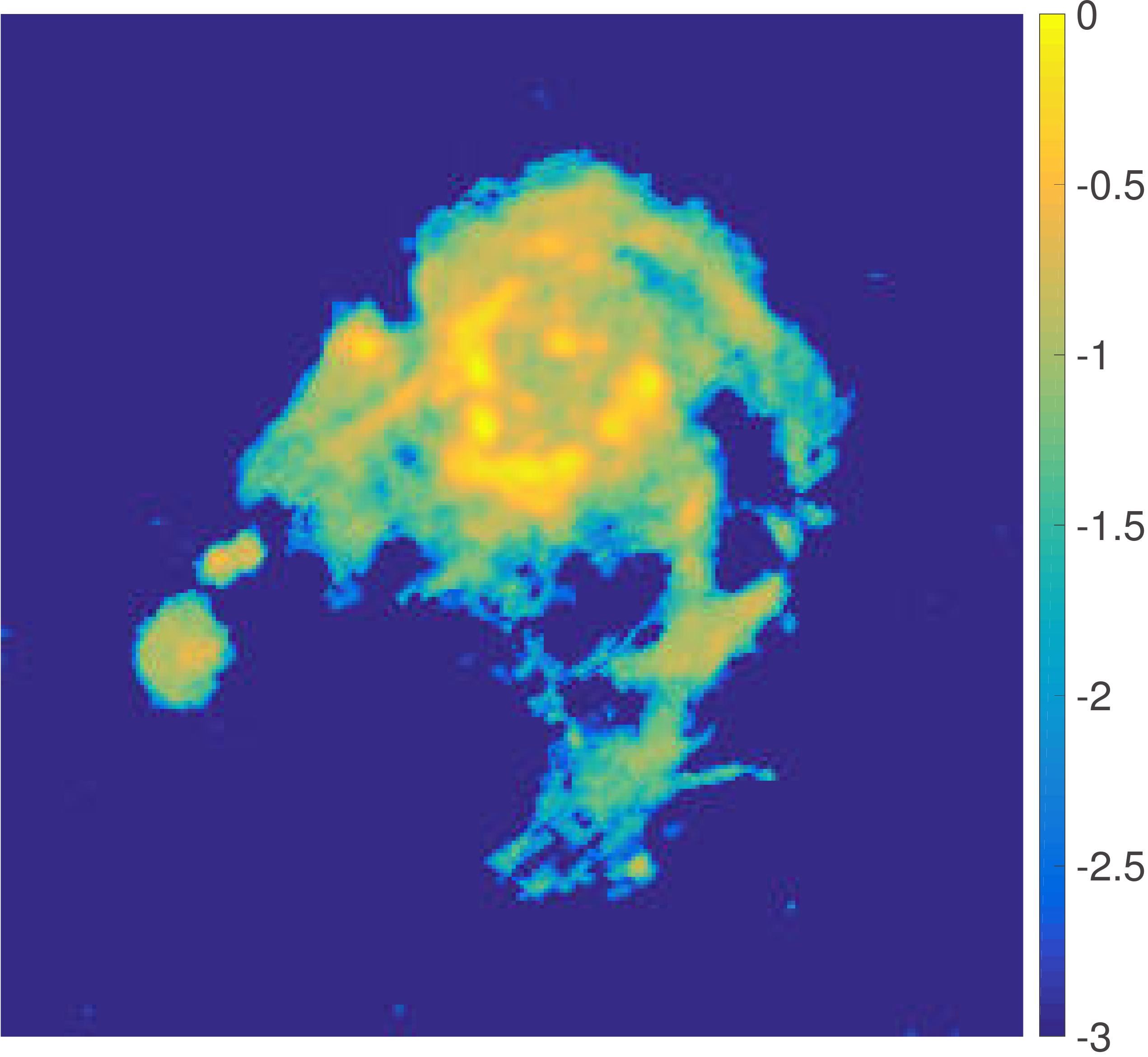}\hspace{2pt}
	\includegraphics[trim={0px 0px 0px 0px}, clip, height=0.24\linewidth]{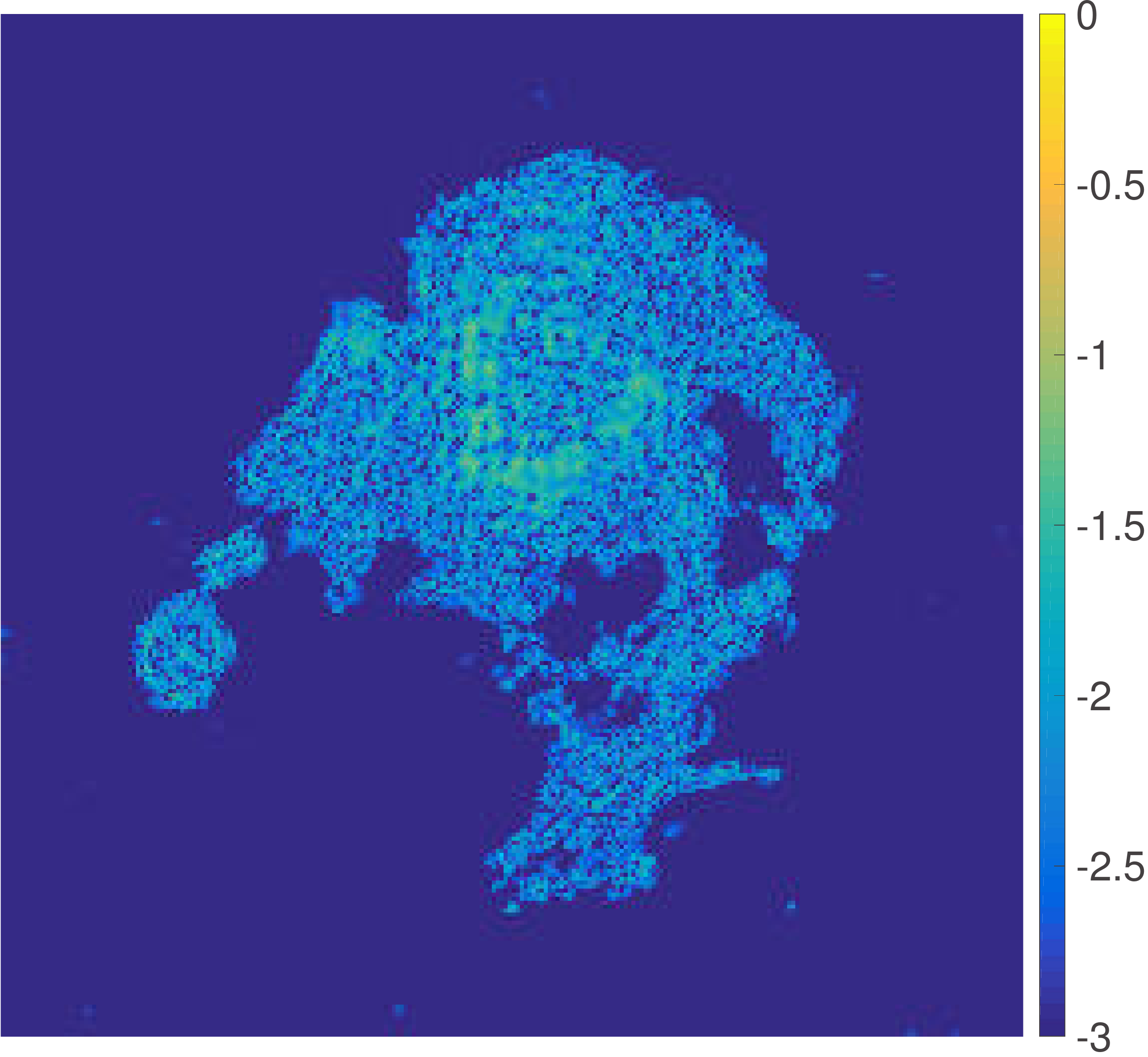}\hspace{2pt}
	\includegraphics[trim={0px 0px 0px 0px}, clip, height=0.24\linewidth]{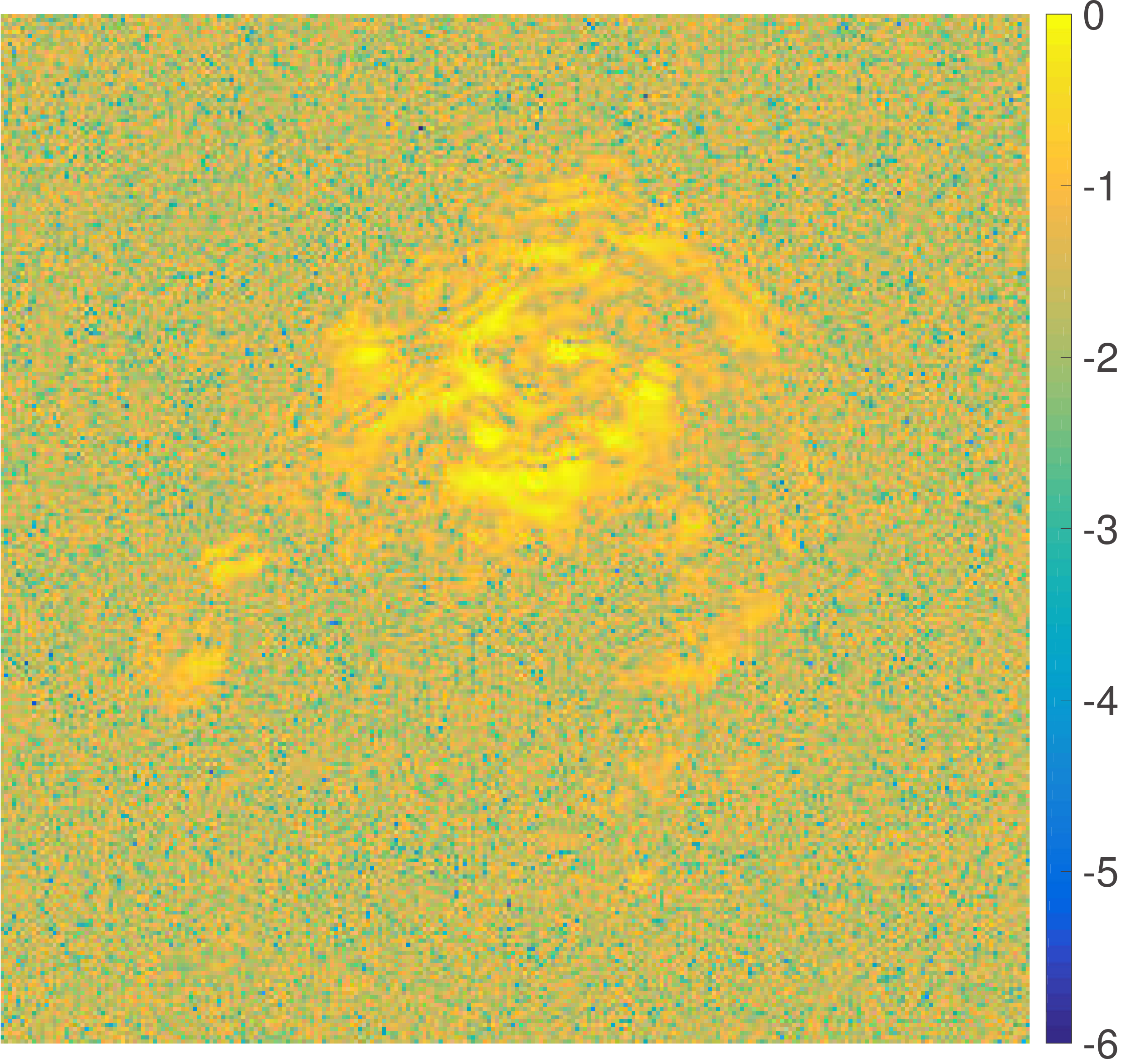}

	\vspace{3pt}
		
	\includegraphics[trim={0px 0px 0px 0px}, clip, height=0.24\linewidth]{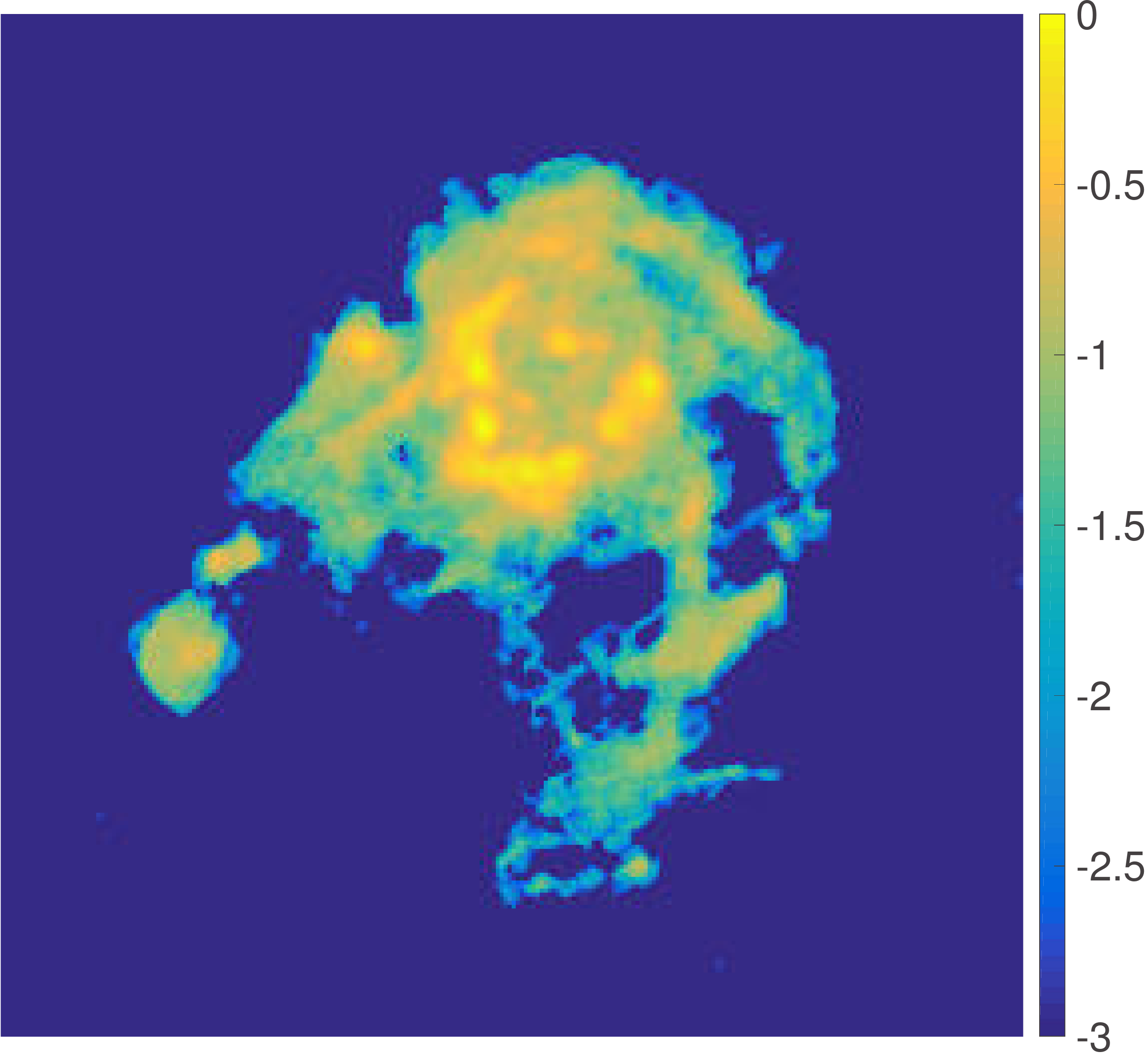}\hspace{2pt}
	\includegraphics[trim={0px 0px 0px 0px}, clip, height=0.24\linewidth]{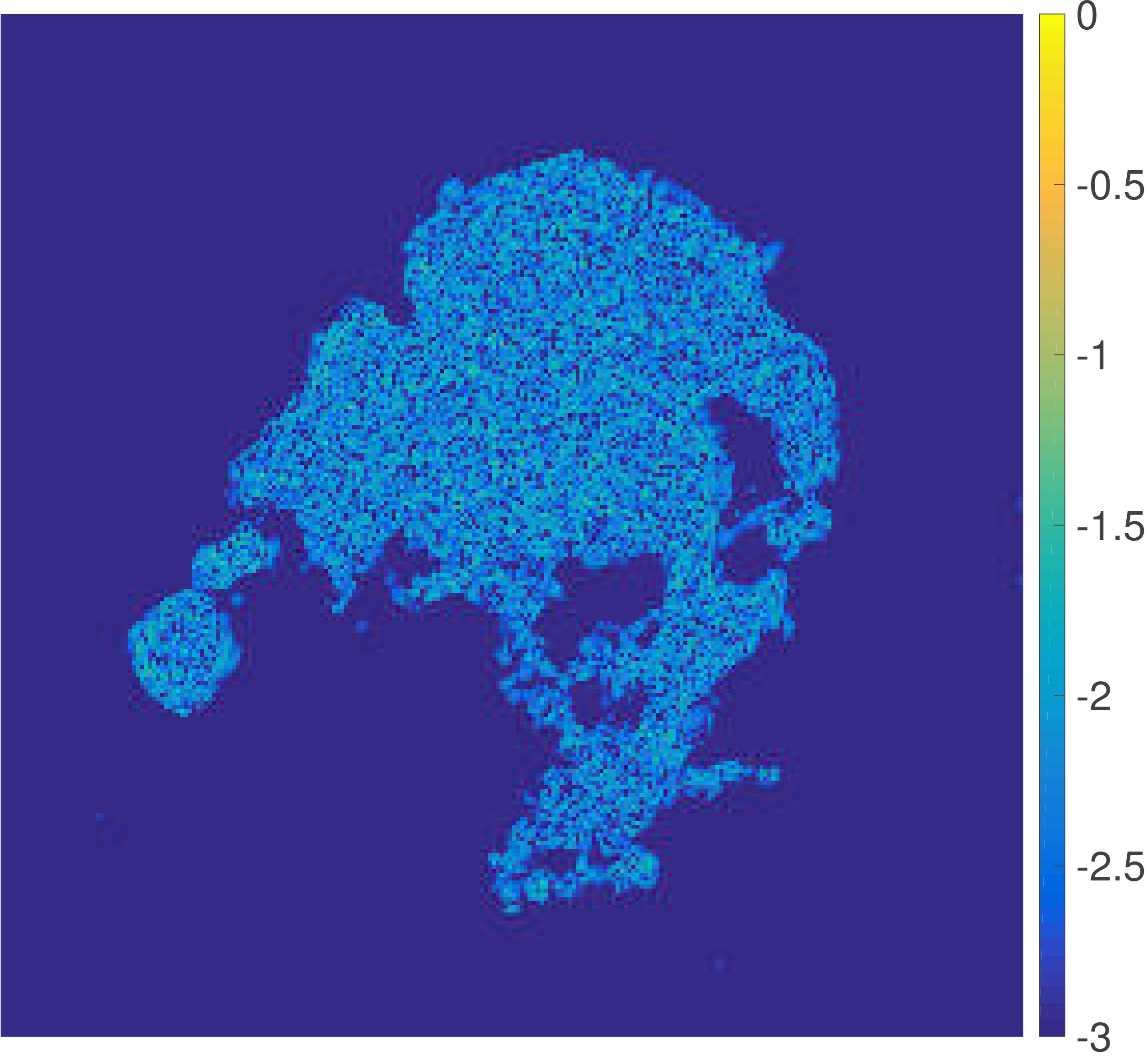}\hspace{2pt}
	\includegraphics[trim={0px 0px 0px 0px}, clip, height=0.24\linewidth]{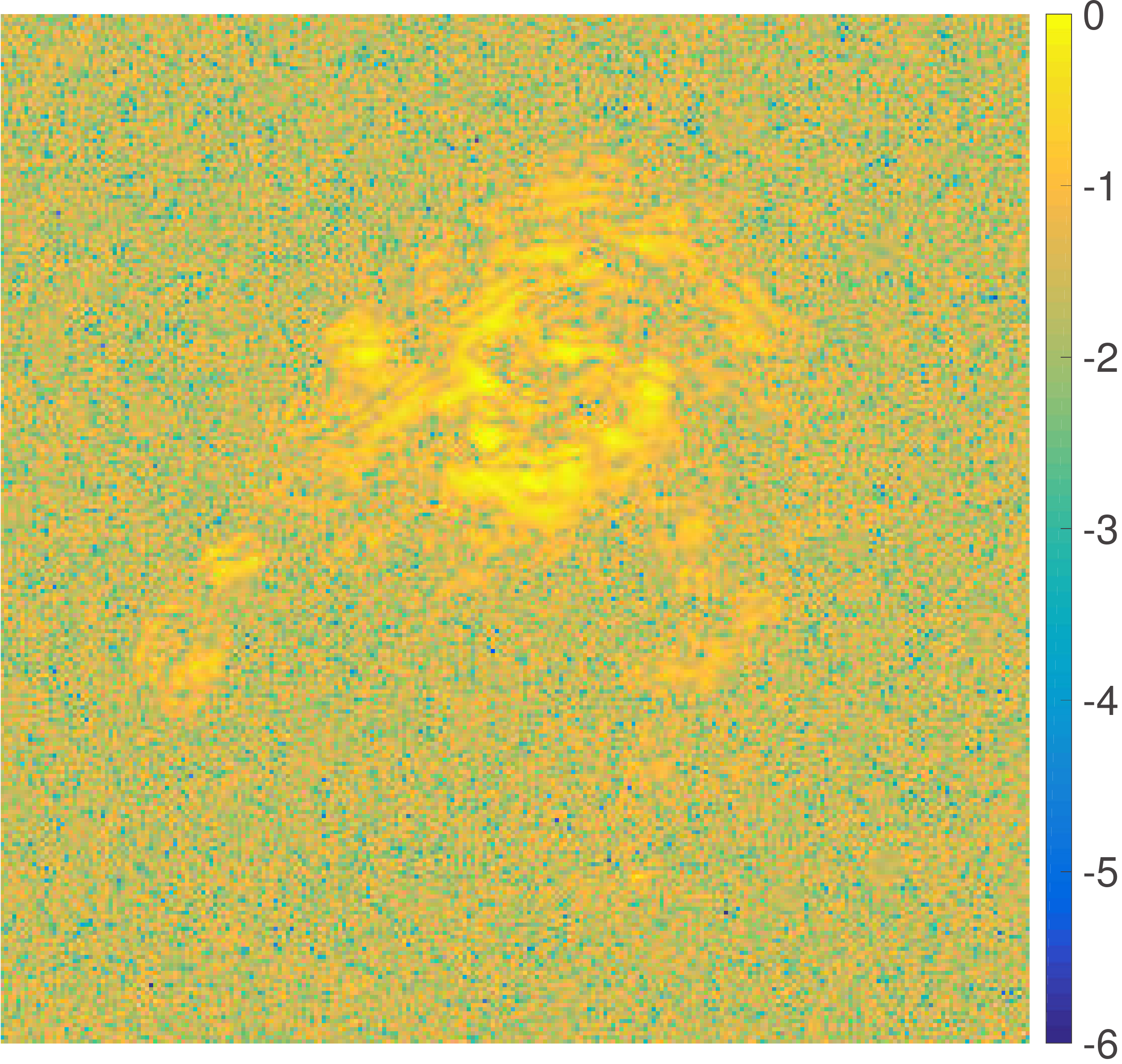}

	\vspace{3pt}
		
	\includegraphics[trim={0px 0px 0px 0px}, clip, height=0.24\linewidth]{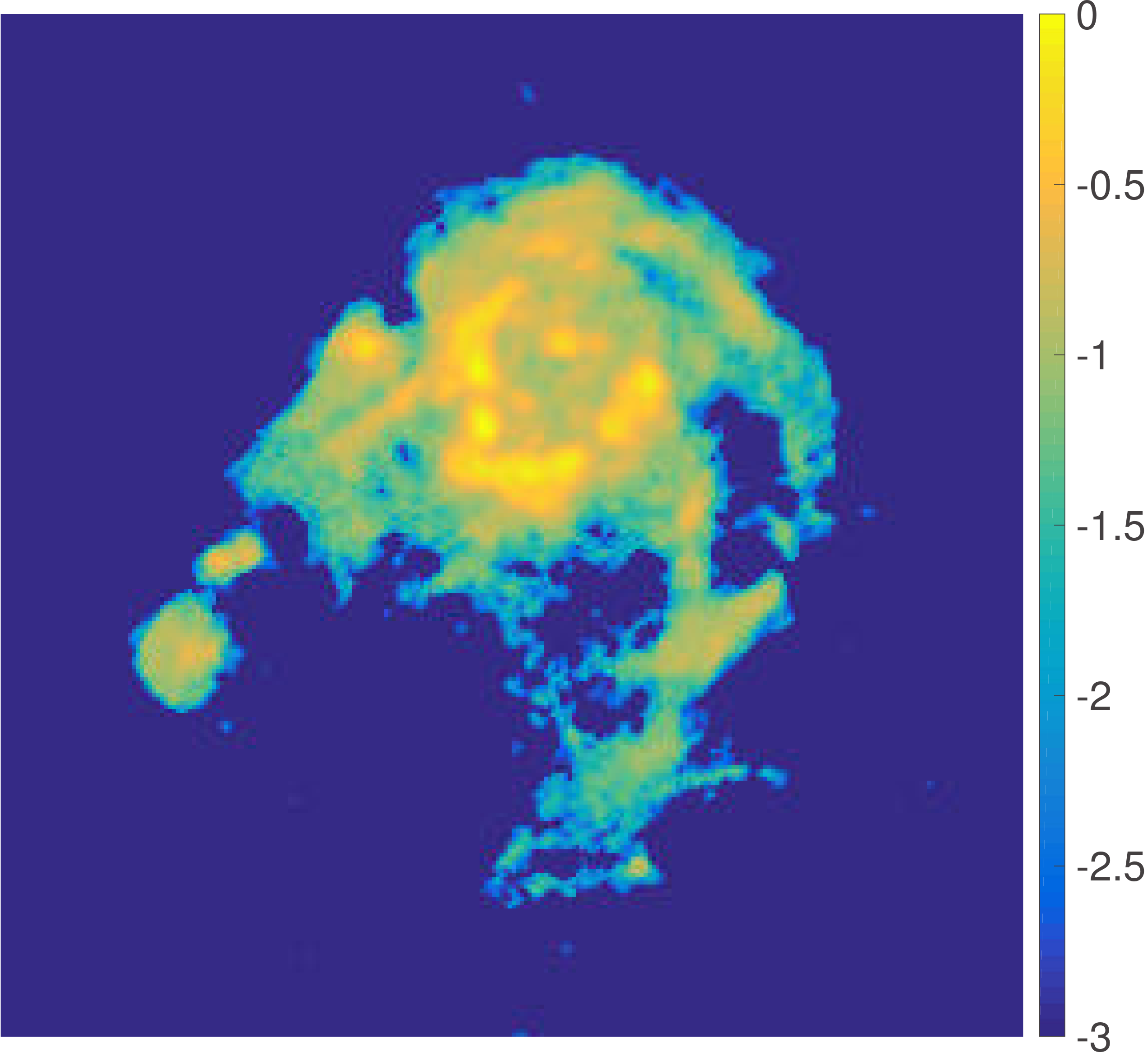}\hspace{2pt}
	\includegraphics[trim={0px 0px 0px 0px}, clip, height=0.24\linewidth]{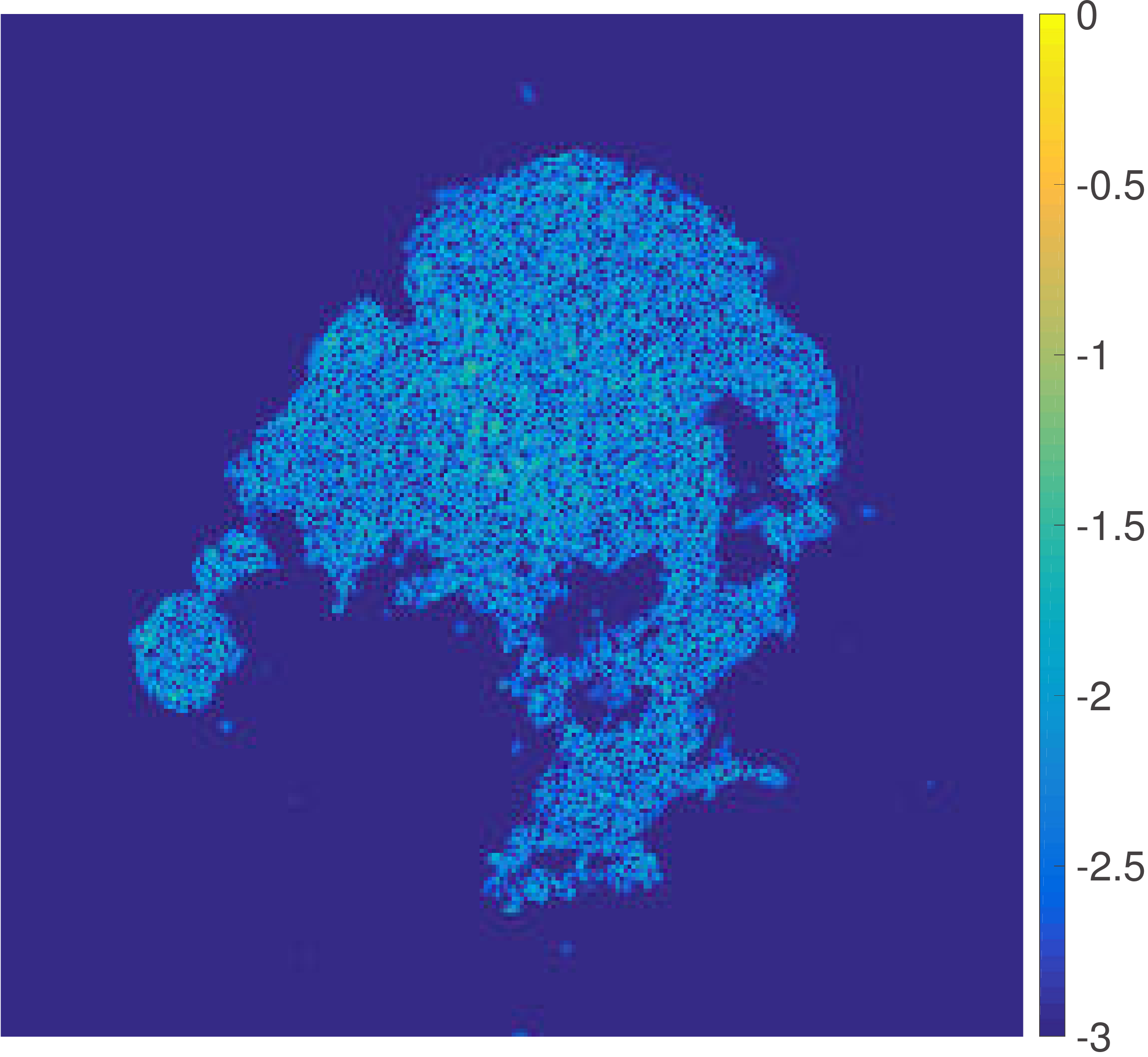}\hspace{2pt}
	\includegraphics[trim={0px 0px 0px 0px}, clip, height=0.24\linewidth]{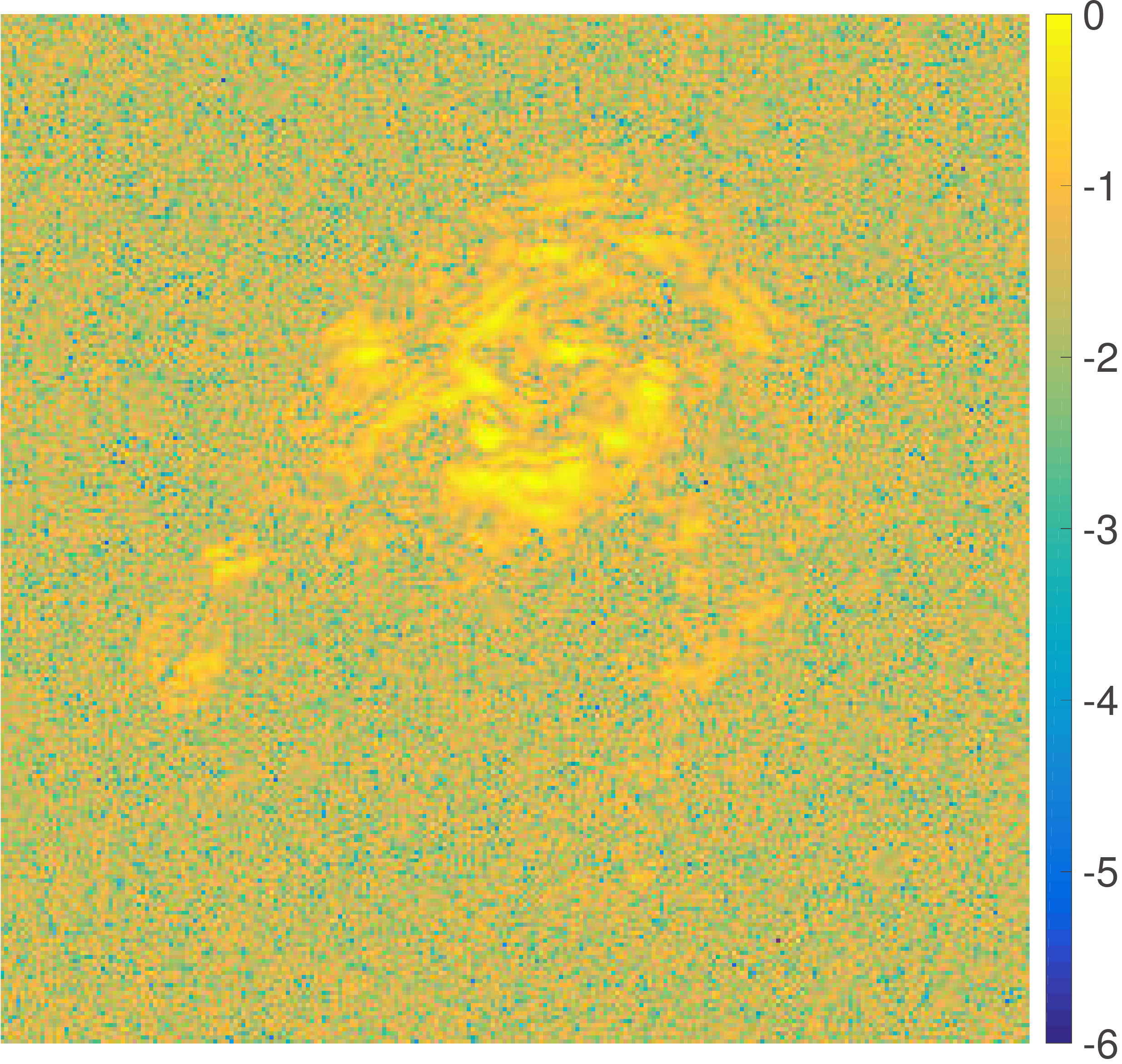}
	
	\vspace{3pt}
		
	\includegraphics[trim={0px 0px 0px 0px}, clip, height=0.24\linewidth]{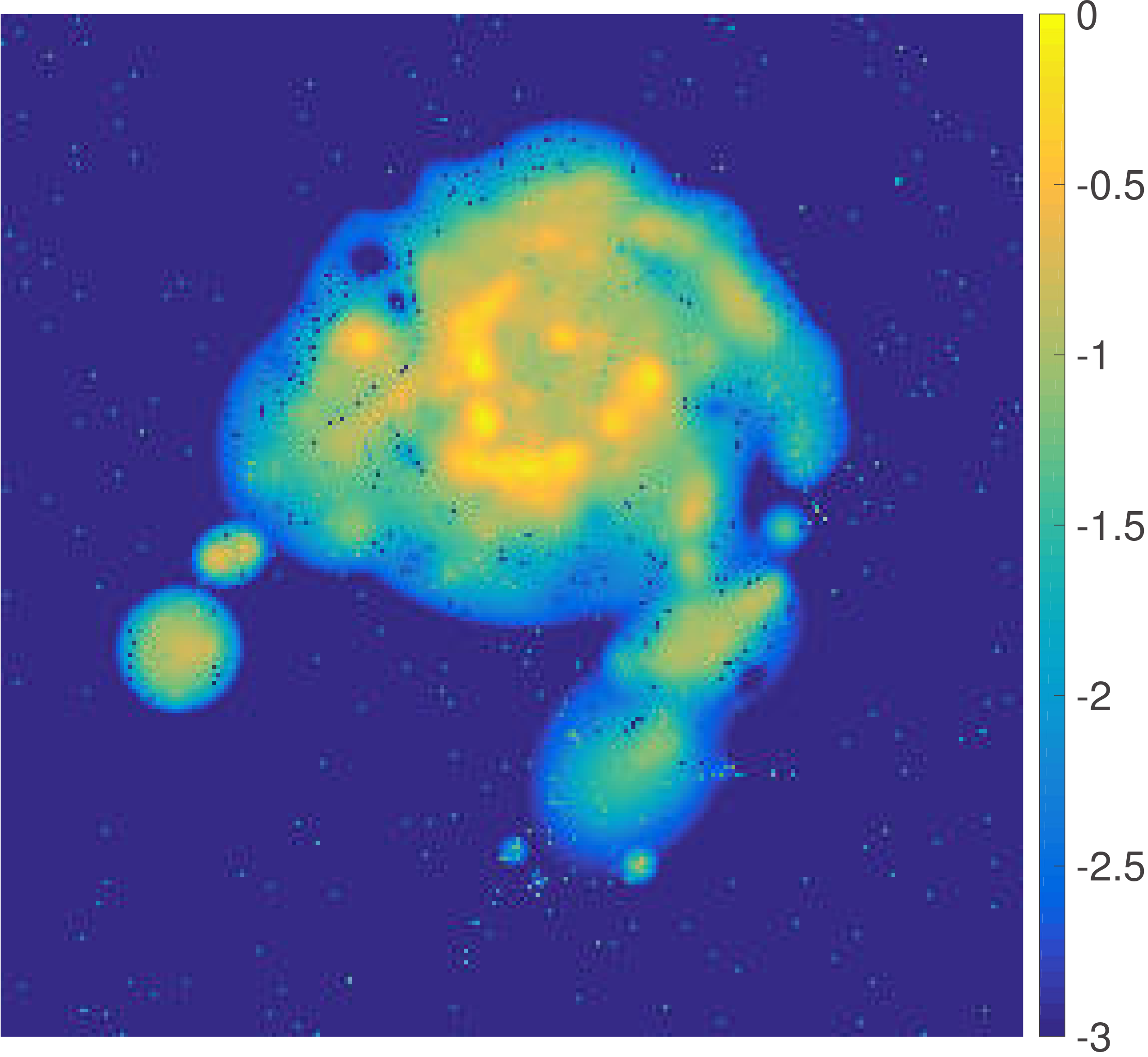}\hspace{2pt}
	\includegraphics[trim={0px 0px 0px 0px}, clip, height=0.24\linewidth]{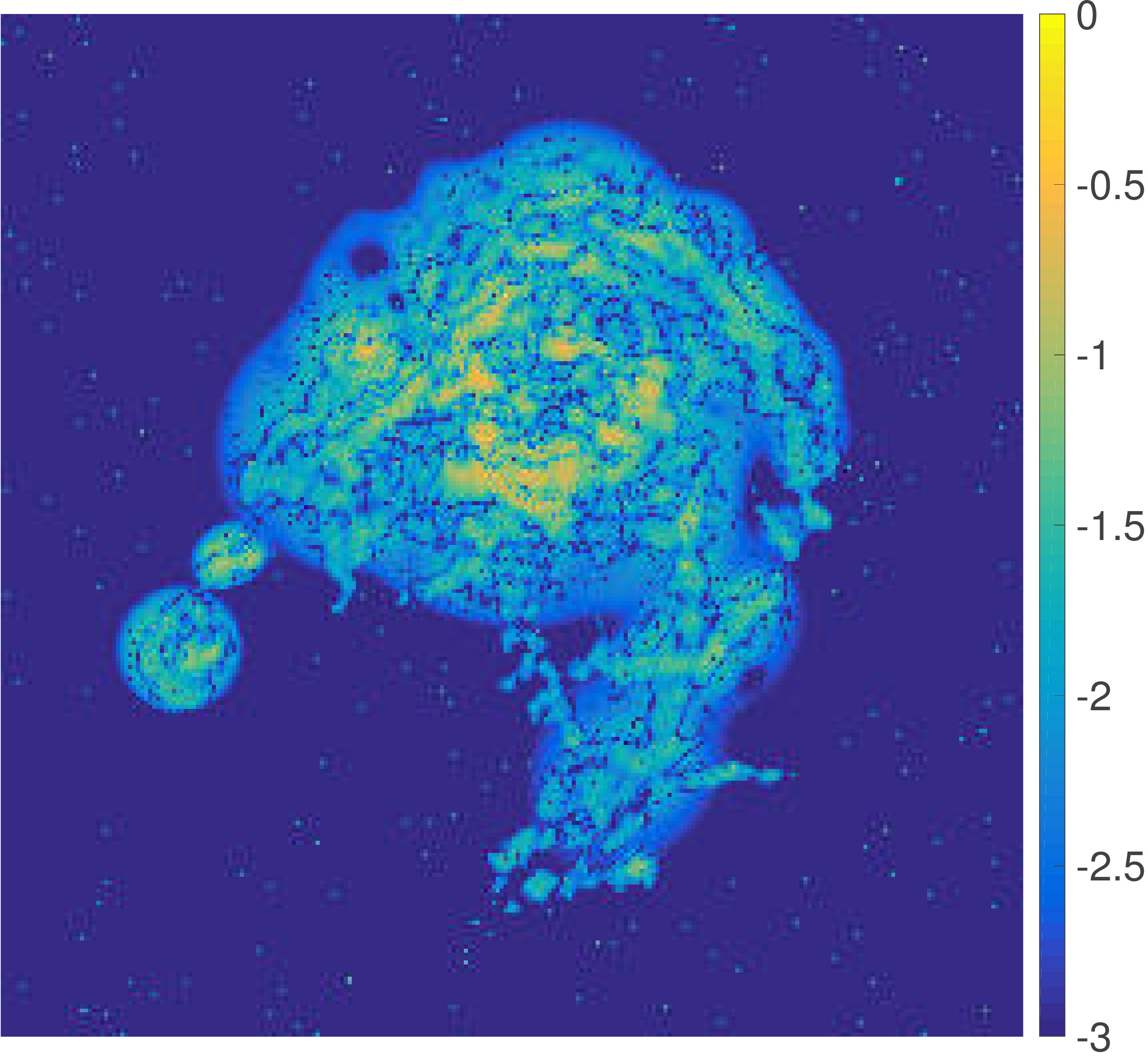}\hspace{2pt}
	\includegraphics[trim={0px 0px 0px 0px}, clip, height=0.24\linewidth]{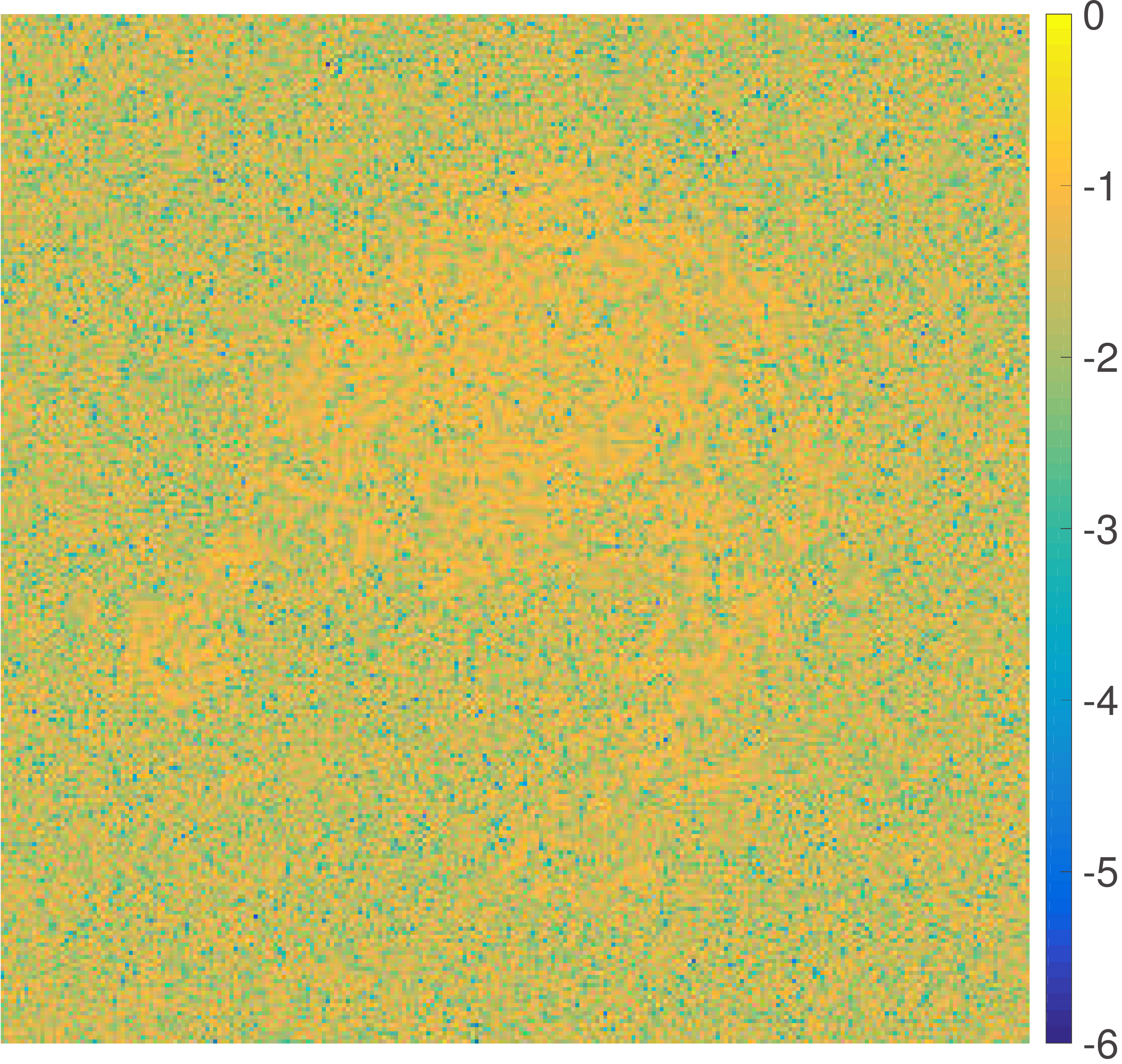}
	
	\caption{Visual comparison of reconstruction results for the M31 test image with $M=100N$. From left to right: reconstructed, error and residual images in $\log_{10}$ scale. First four rows from top to bottom: reconstruction performed with all visibilities, `reduced' visibilities after performing dimensionality reduction with $\RFPhit$, $\RPhit$, and $\RGt$ respectively. Last row: reconstruction using \textsc{ms-clean} with a uniform weighting scheme.}
	\label{fig:m31visualcomp}
\end{figure*}

The galaxy cluster test image was chosen for its high dynamic range, and simulations show that $\RFPhit$ results in much better reconstruction than all other methods, both in terms of DR and SNR.
As seen in Fig.~\ref{fig:galaxyclustergraphs_uni}, the SNR from $\RFPhit$ on Gaussian random coverages is more than 2~dB higher than the complete visibilities set, on average, reaching consistently up to 45~dB and outperforming all other dimensionality reduction methods.
Over SKA-like coverages, the image reconstruction trend as shown in Fig.~\ref{fig:galaxyclustergraphs_ska} is similar, with $\RFPhit$  providing output SNRs reaching almost 30~dB while $\RGt$ usually has SNRs 2~dB below this value.
The DR curves confirm this trend, and high DR values of $2.8\times 10^7$ are reached with $\RFPhit$ over SKA-like coverages.
In comparison, $\RGt$ and the complete visibilities set reach $2.5\times 10^7$ and $2.6\times 10^7$ respectively.
The DR values are comparatively low for Gaussian random coverages, but the relative improvement is maintained. The respective values are $1.9\times 10^5$, $1.8\times 10^5$ and $1.7\times 10^5$.
A visual comparison in Fig.~\ref{fig:galaxyclustervisualcomp} highlights the sharper structure of the bright sources and detail in the reconstructed image corresponding to $\RFPhit$ when compared to the other methods.
Fig.~\ref{fig:galaxyclustervisualcomp} also shows very low values over the residual images in general across the methods, but an extremely sensitive scale is set to forcibly show the structure present in the residual.
{\revisedtwo 
}
The actual residual image values are close to the numerical precision for these computations.
{\revised Reconstruction using $M=50N$ visibilities took $\approx$12 seconds per iteration without dimensionality reduction, whereas it only took $\approx$1.8 seconds per iteration using $\RFPhit$.}
{\revised Image deconvolution with \textsc{ms-clean} took 6 major iterations, and the model shown in Fig.~\ref{fig:galaxyclustervisualcomp} contains 30649 components.
Output model images of size $2048\times 2048$ pixels were cropped to $512\times 512$ pixels. 
The reconstruction is closer to the other methods, and the residual image contains only minor structure. 
Some high-flux regions are visible in the error image that were not appropriately modelled in the reconstruction.}

\begin{figure*}
	\centering
	\includegraphics[trim={0px 0px 0px 0px}, clip, height=0.24\linewidth]{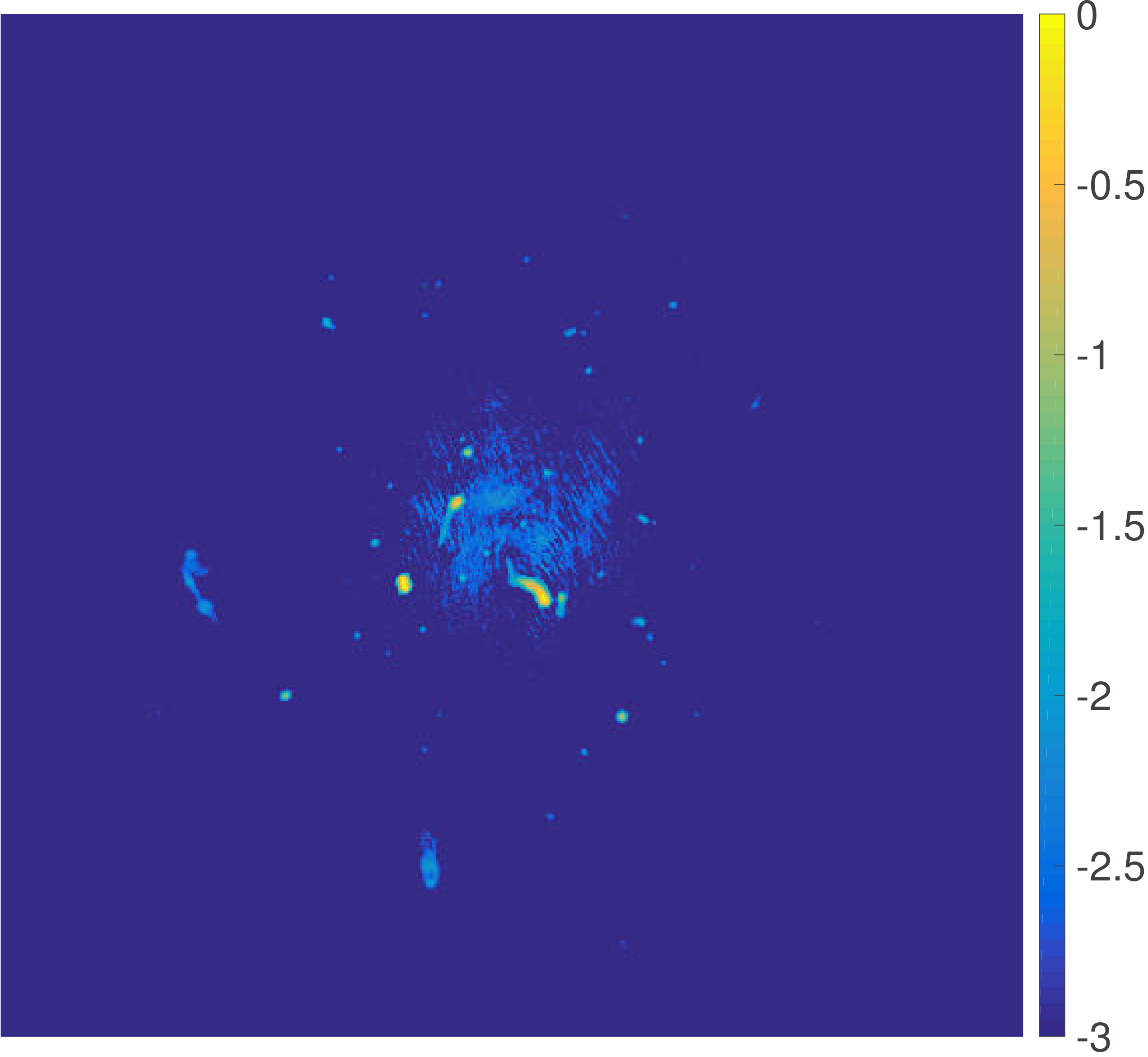}\hspace{2pt}
	\includegraphics[trim={0px 0px 0px 0px}, clip, height=0.24\linewidth]{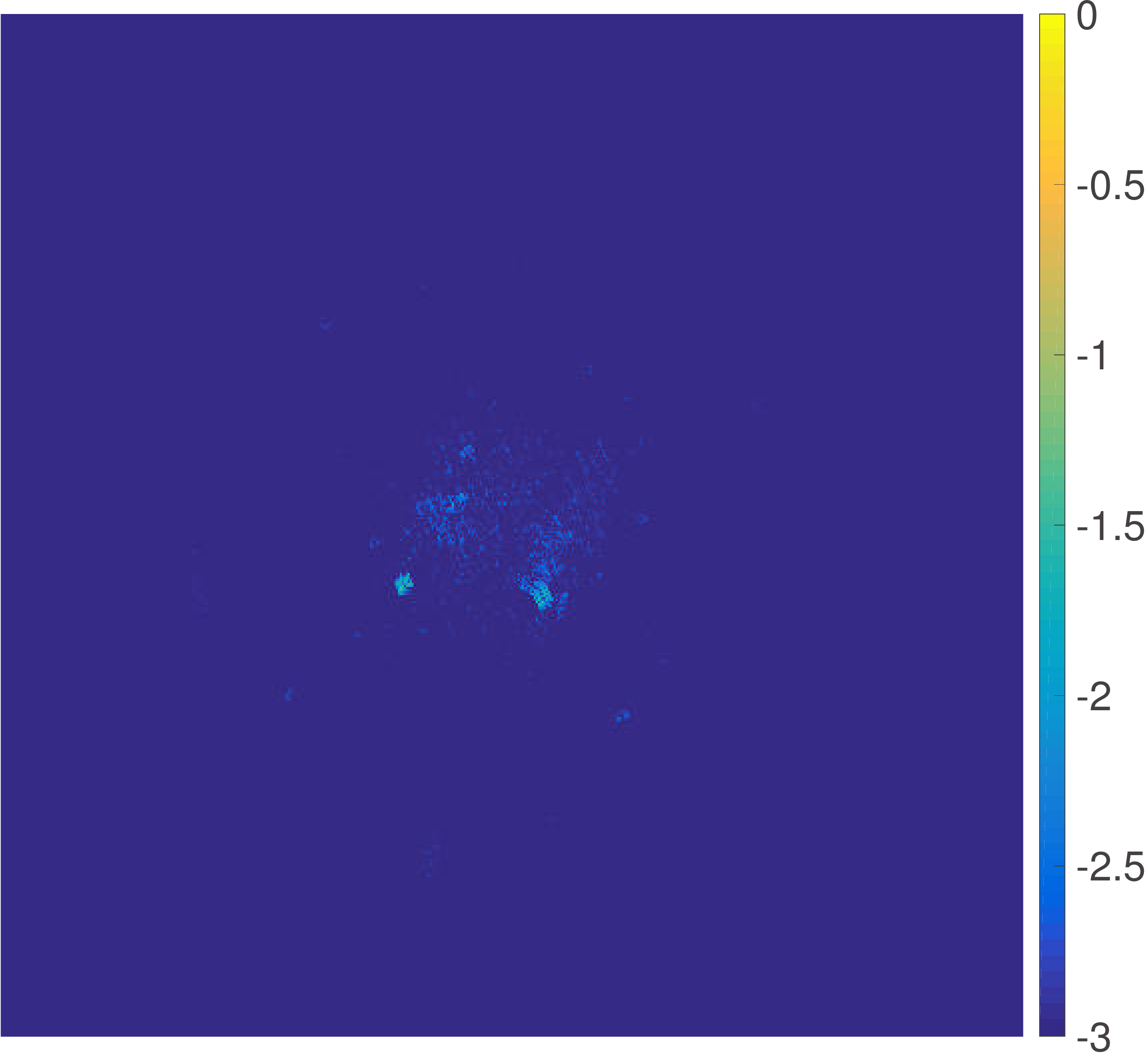}\hspace{2pt}
	\includegraphics[trim={0px 0px 0px 0px}, clip, height=0.24\linewidth]{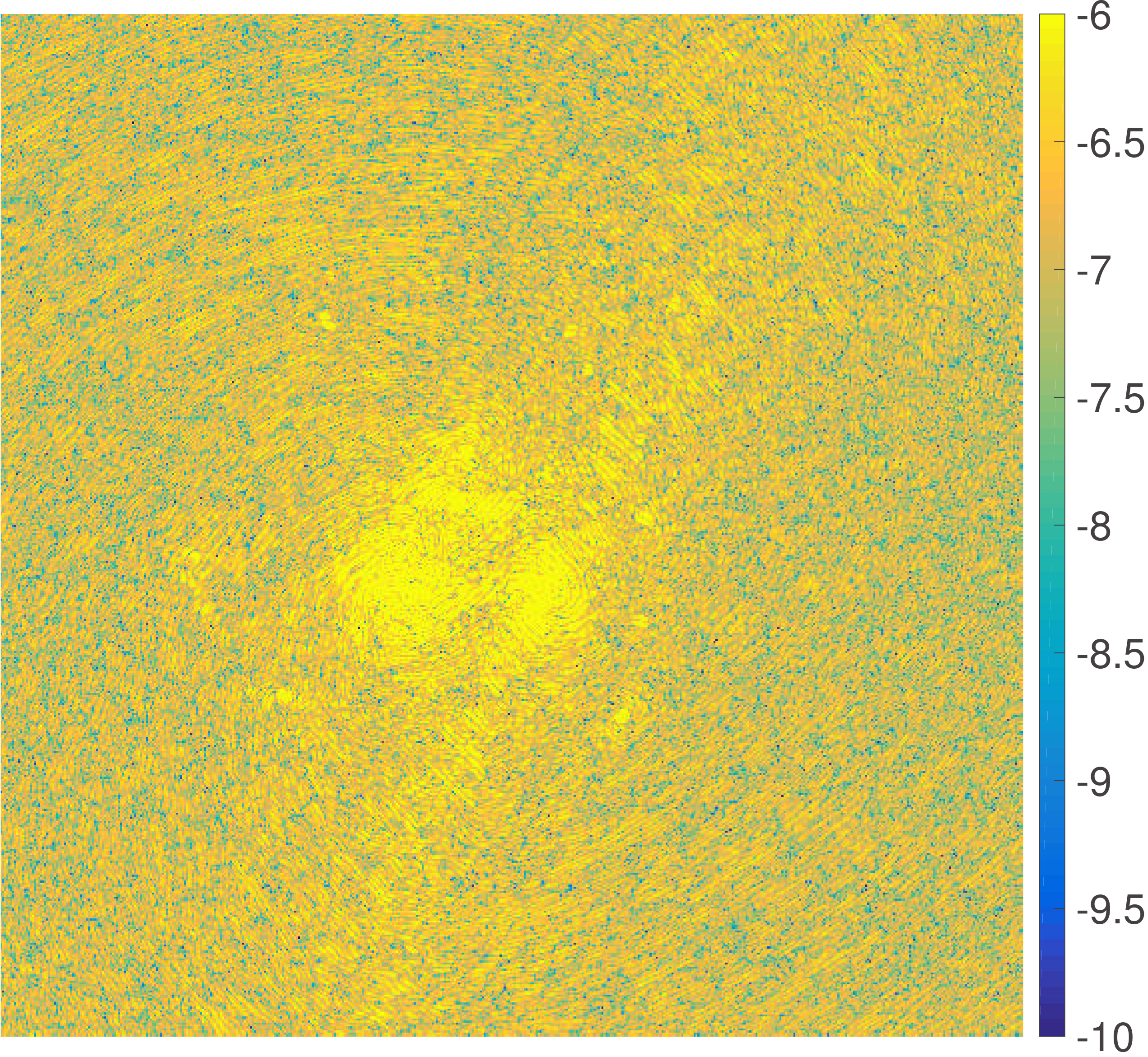}

	\vspace{3pt}
		
	\includegraphics[trim={0px 0px 0px 0px}, clip, height=0.24\linewidth]{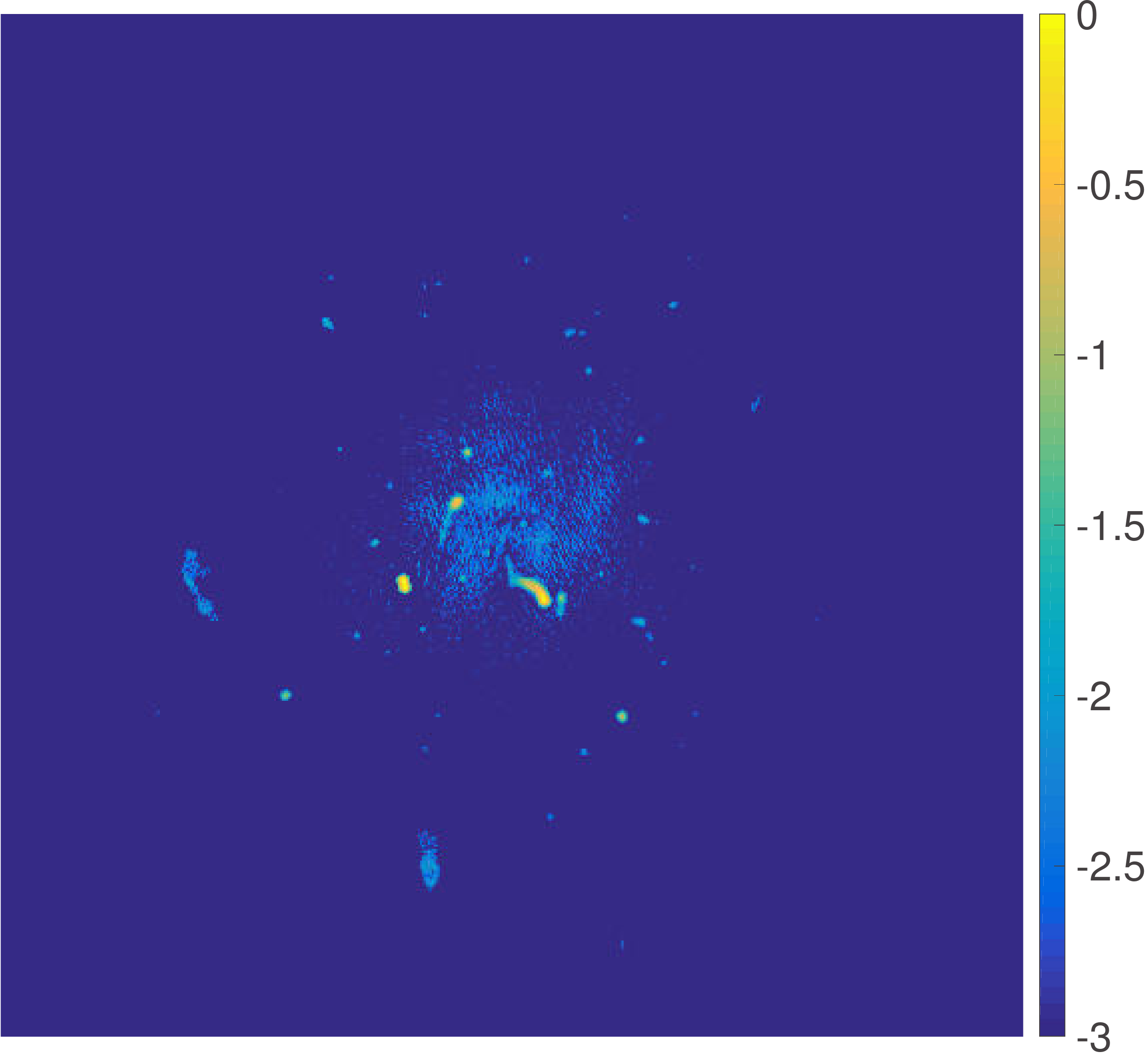}\hspace{2pt}
	\includegraphics[trim={0px 0px 0px 0px}, clip, height=0.24\linewidth]{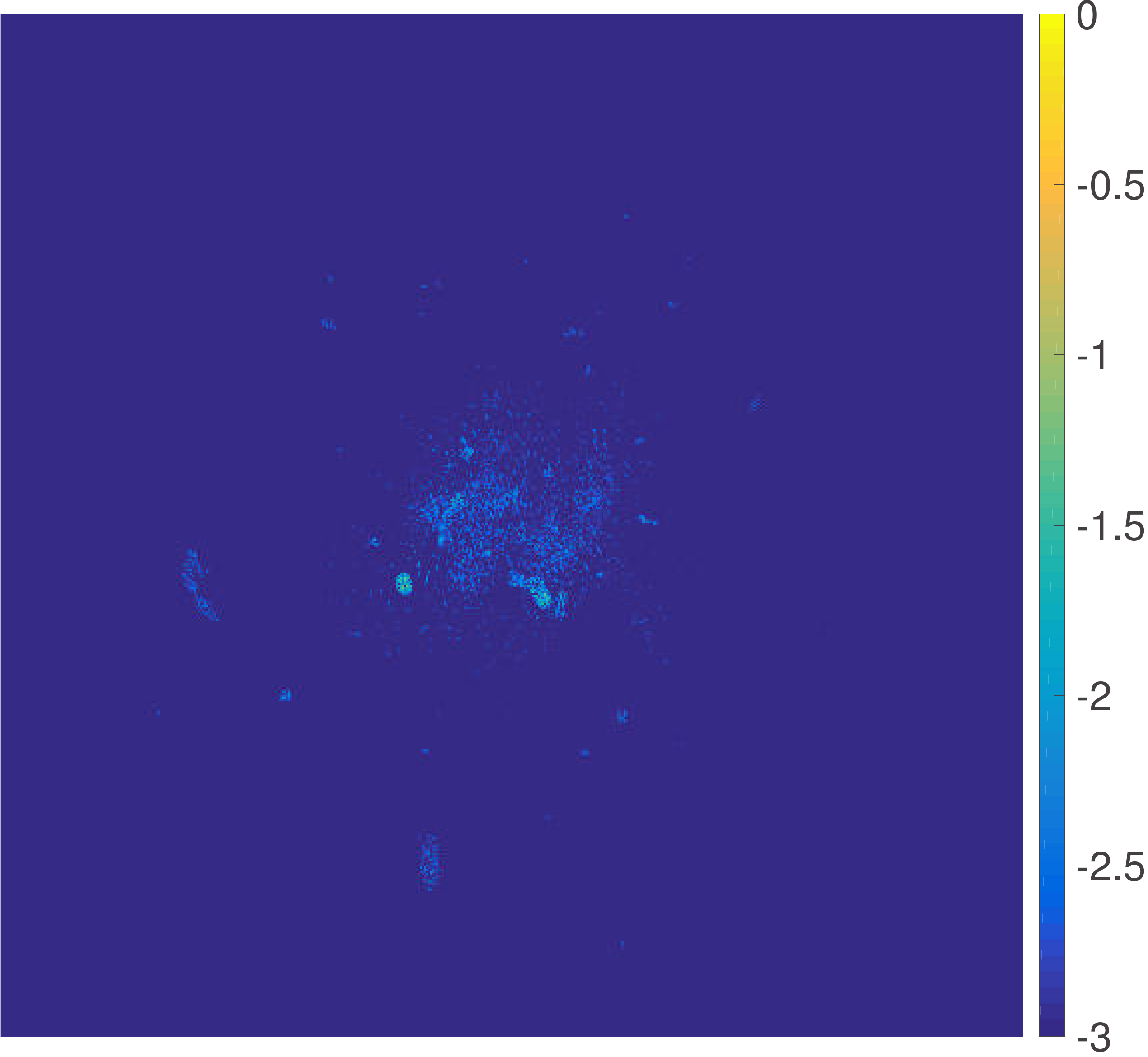}\hspace{2pt}
	\includegraphics[trim={0px 0px 0px 0px}, clip, height=0.24\linewidth]{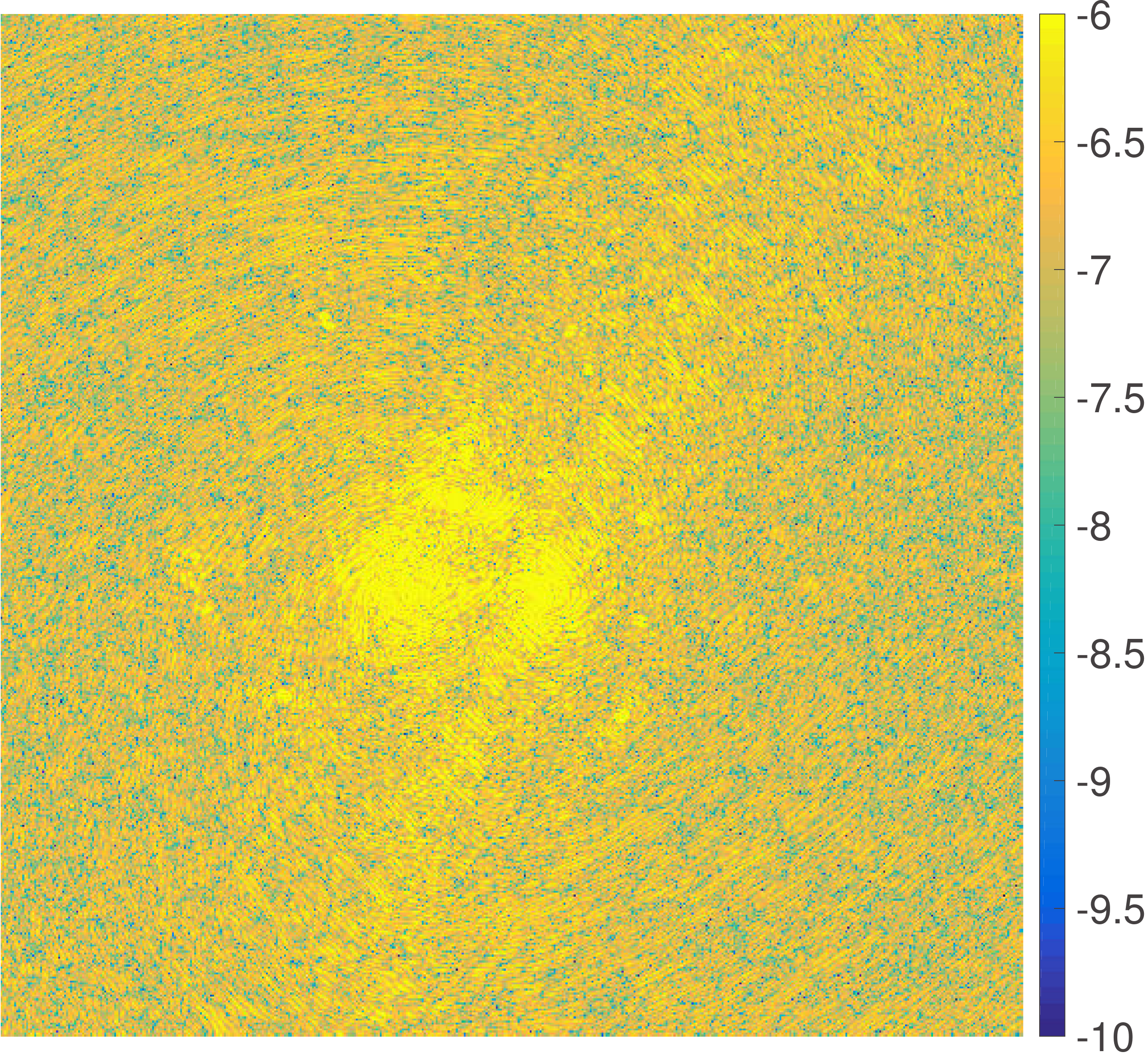}

	\vspace{3pt}
	
	\includegraphics[trim={0px 0px 0px 0px}, clip, height=0.24\linewidth]{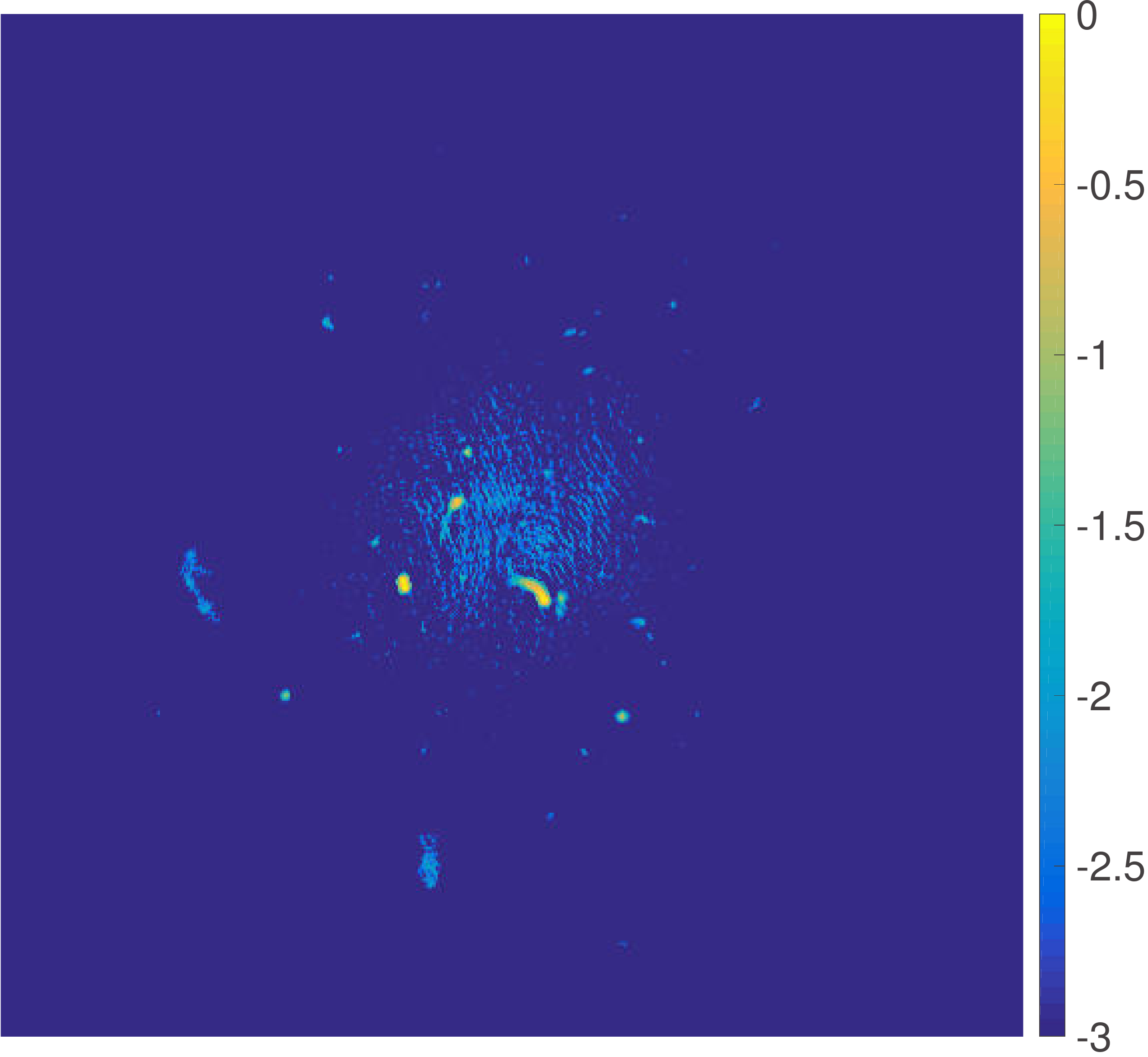}\hspace{2pt}
	\includegraphics[trim={0px 0px 0px 0px}, clip, height=0.24\linewidth]{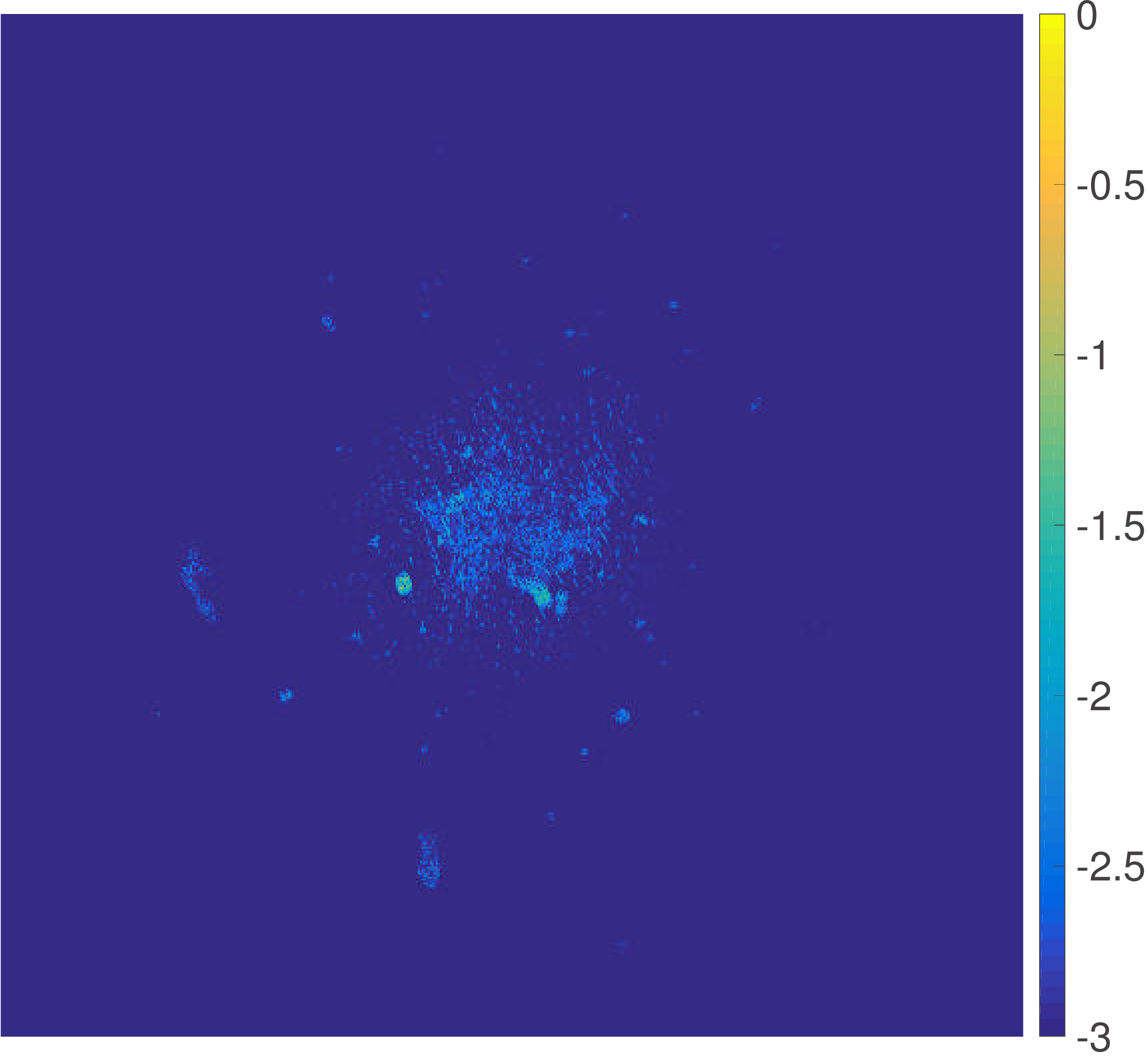}\hspace{2pt}
	\includegraphics[trim={0px 0px 0px 0px}, clip, height=0.24\linewidth]{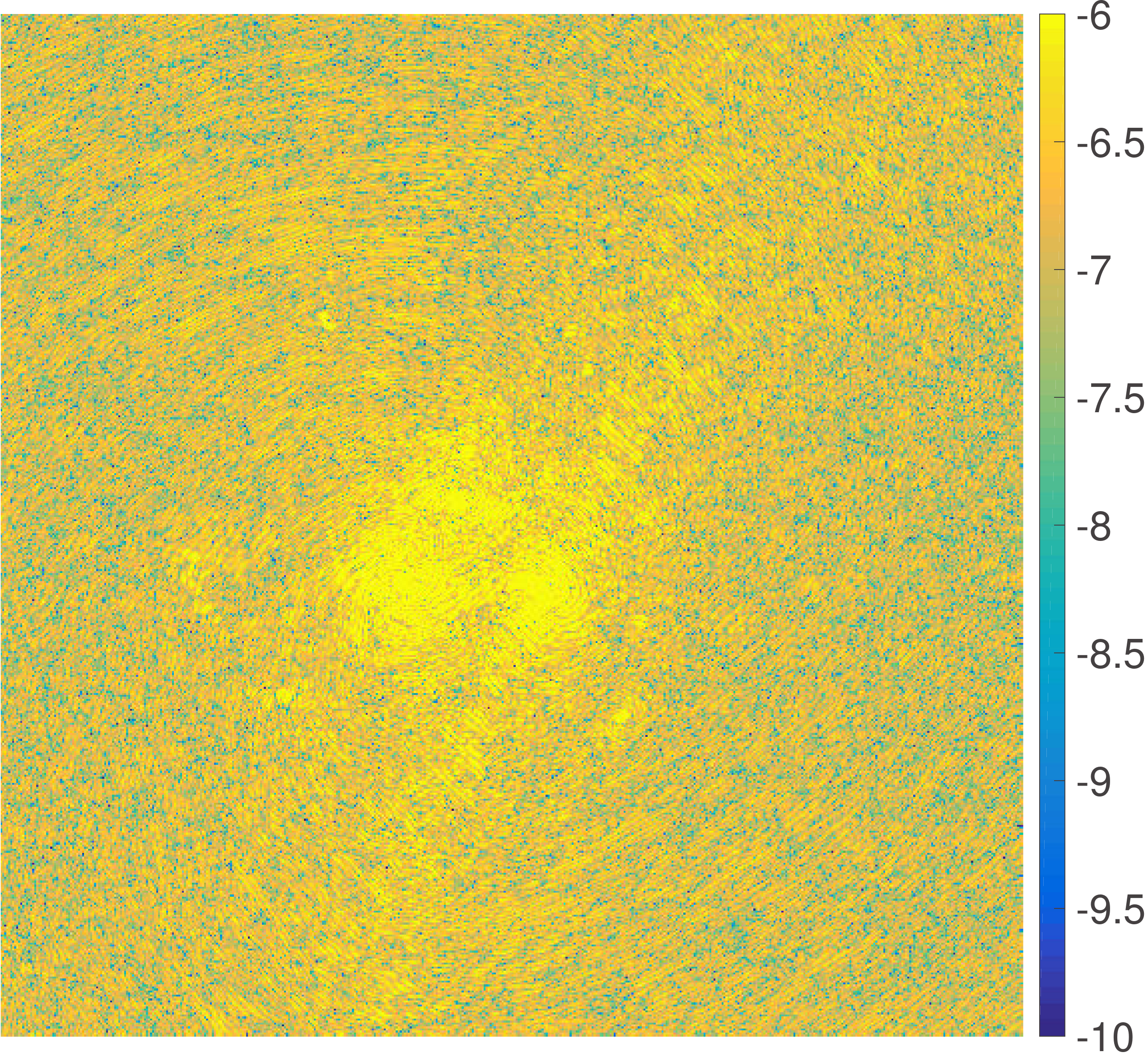}

	\vspace{3pt}
		
	\includegraphics[trim={0px 0px 0px 0px}, clip, height=0.24\linewidth]{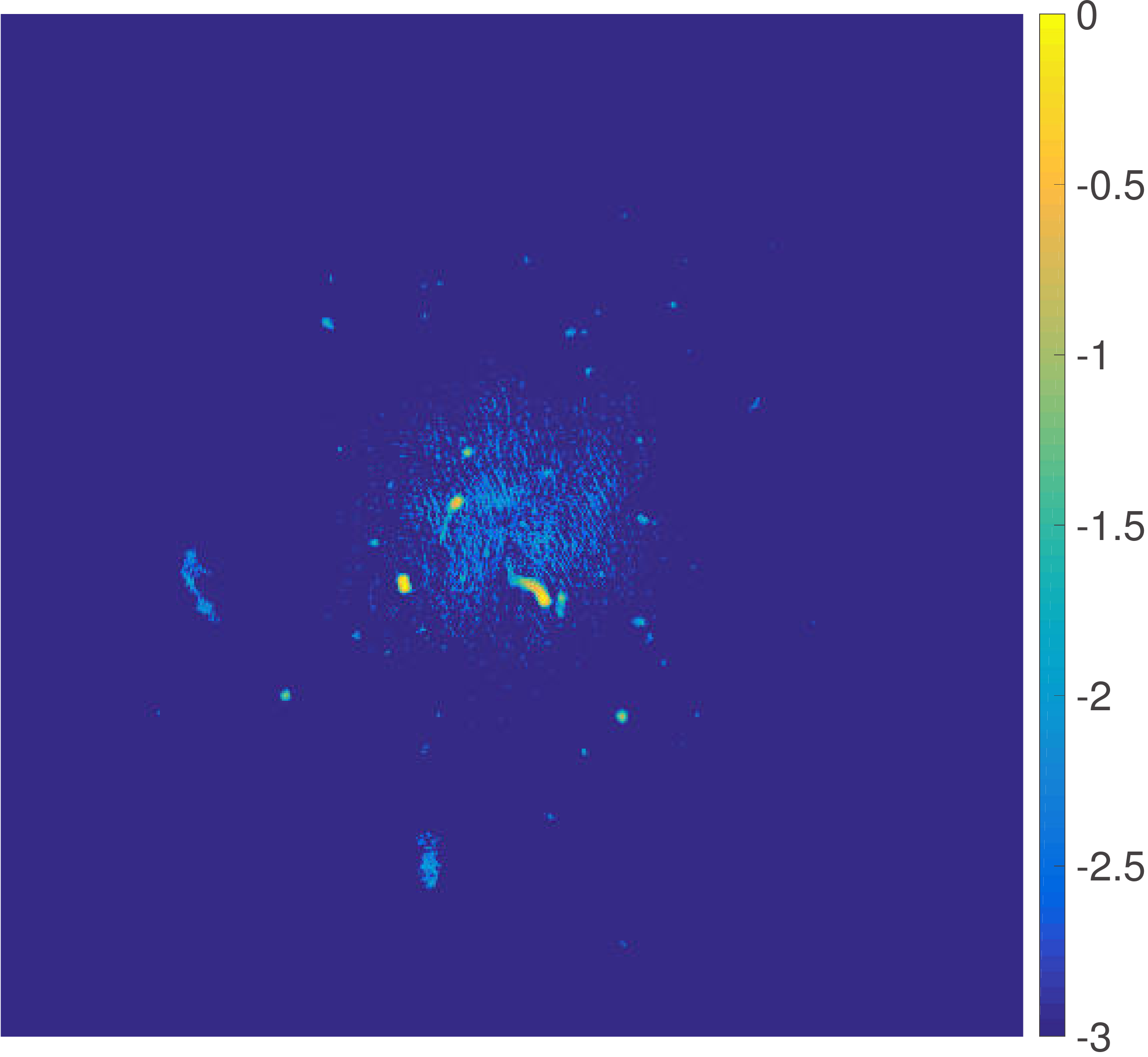}\hspace{2pt}
	\includegraphics[trim={0px 0px 0px 0px}, clip, height=0.24\linewidth]{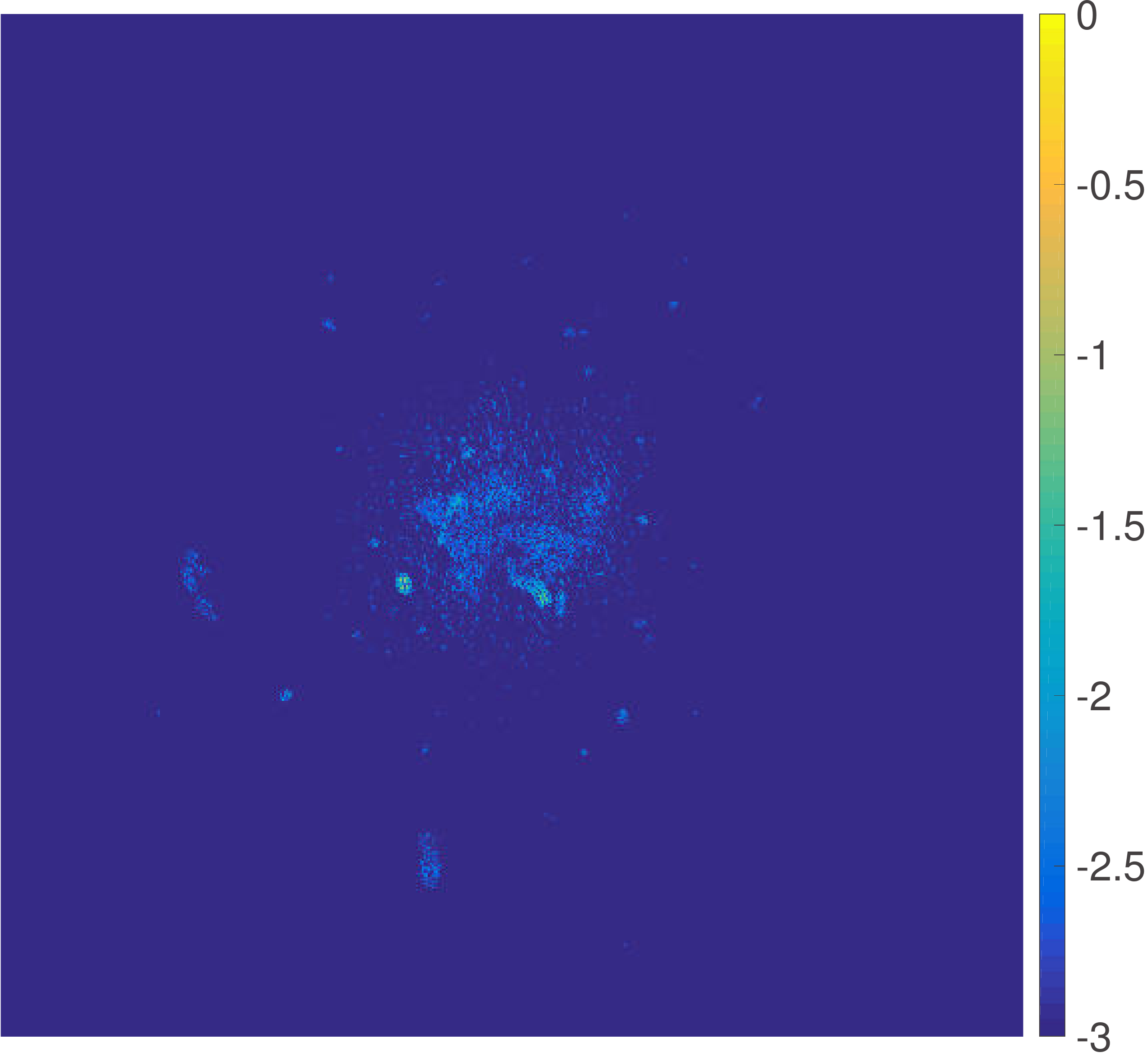}\hspace{2pt}
	\includegraphics[trim={0px 0px 0px 0px}, clip, height=0.24\linewidth]{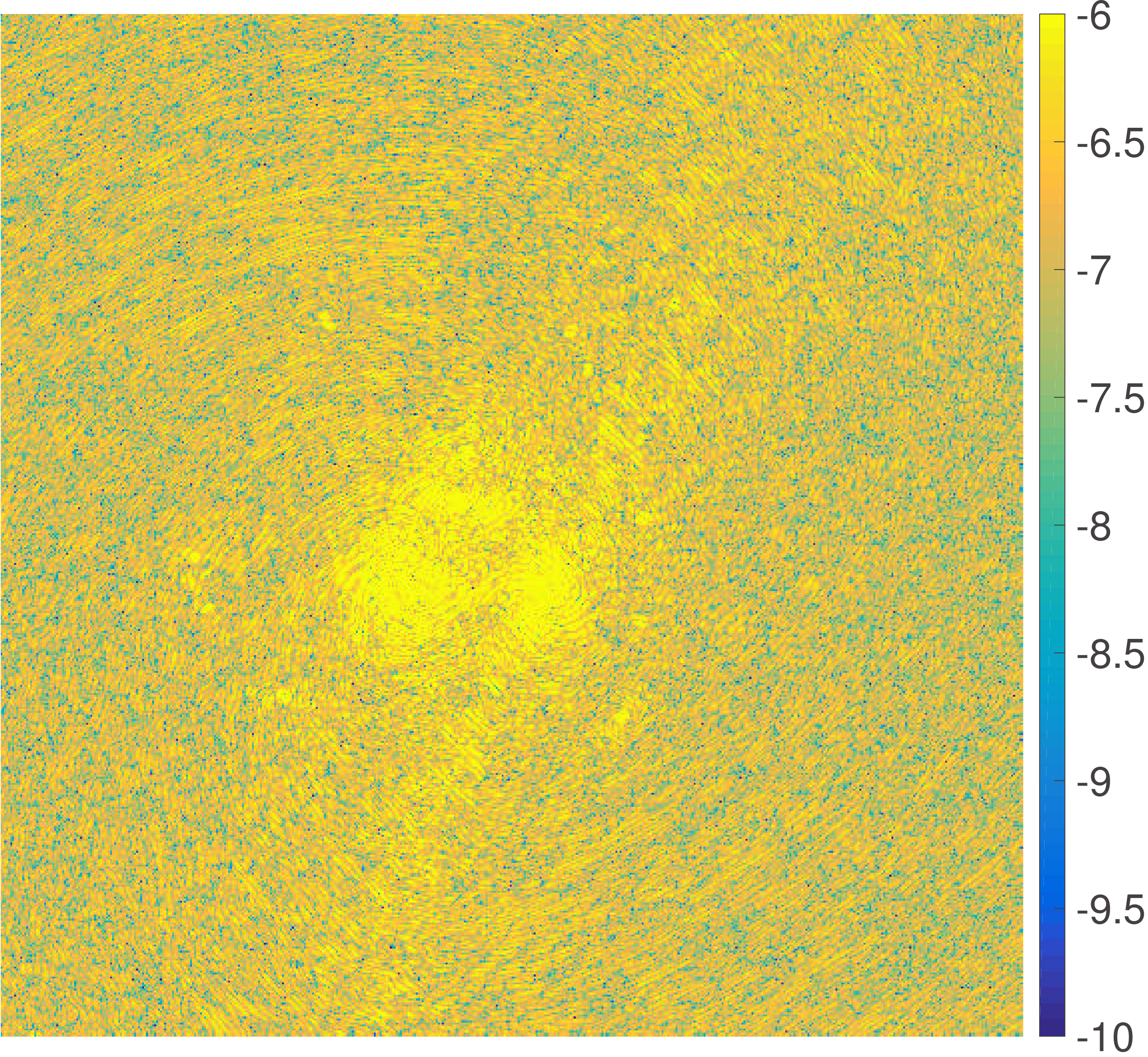}
	
	\vspace{3pt}
		
	\includegraphics[trim={0px 0px 0px 0px}, clip, height=0.24\linewidth]{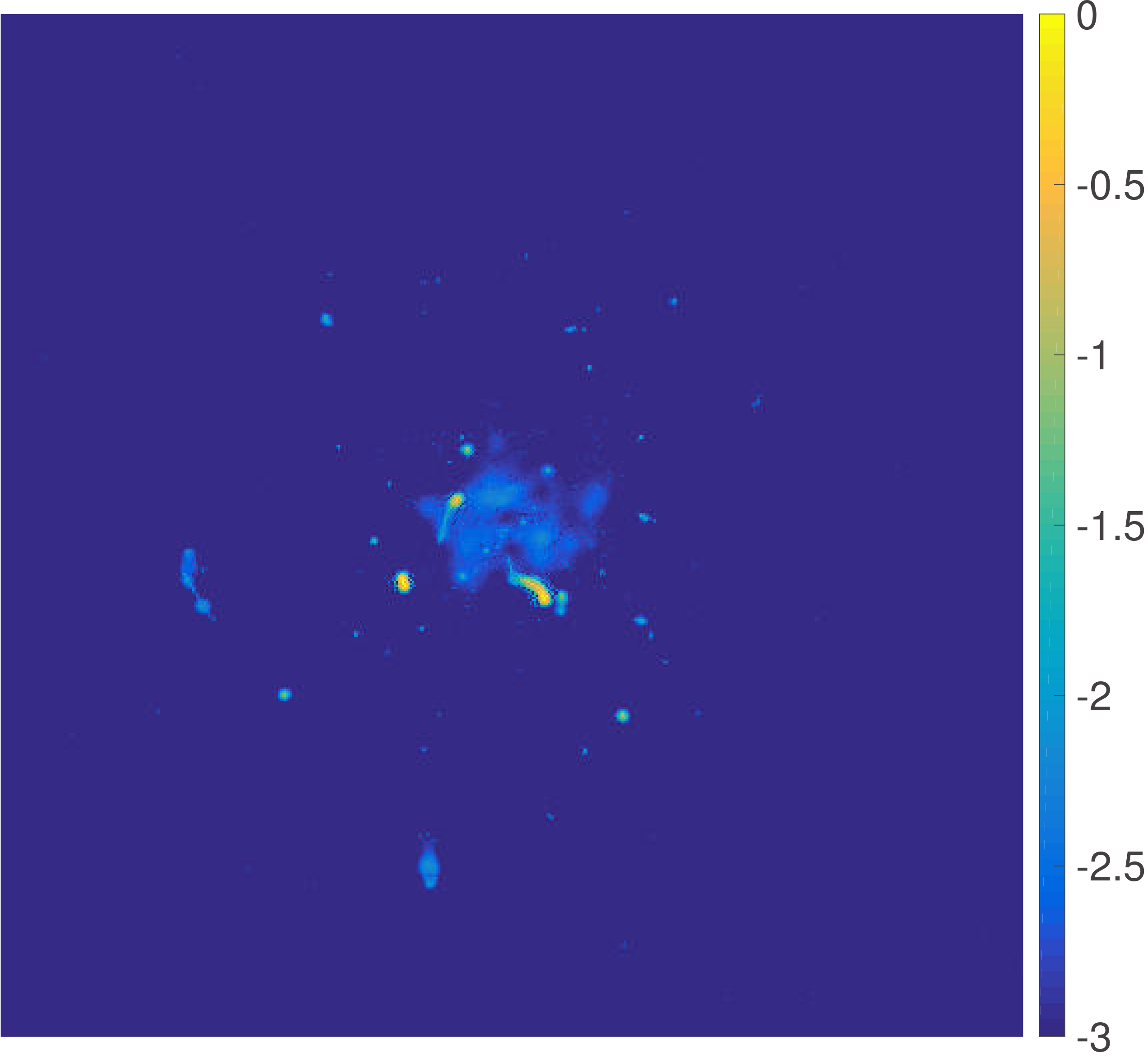}\hspace{2pt}
	\includegraphics[trim={0px 0px 0px 0px}, clip, height=0.24\linewidth]{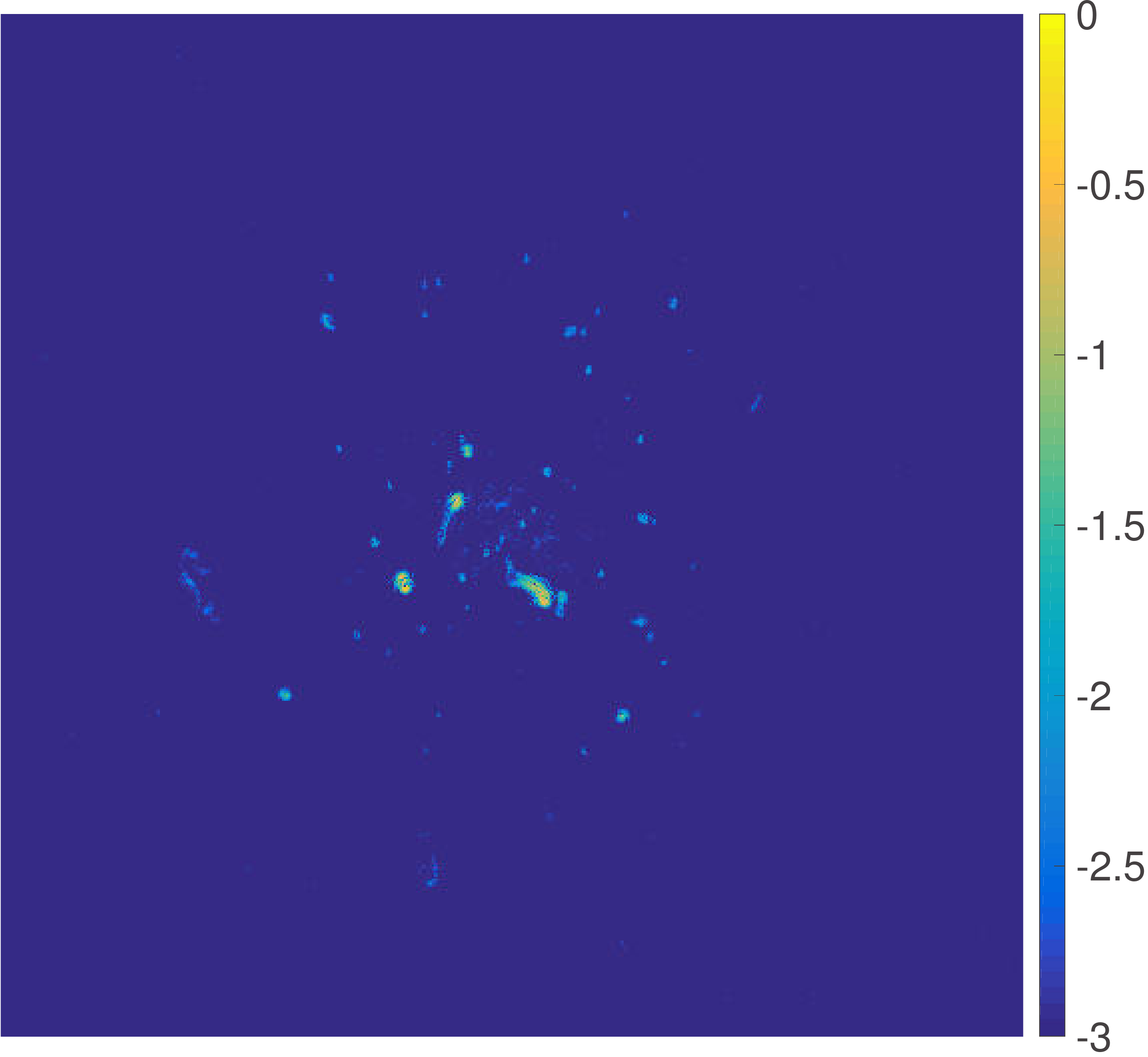}\hspace{2pt}
	\includegraphics[trim={0px 0px 0px 0px}, clip, height=0.24\linewidth]{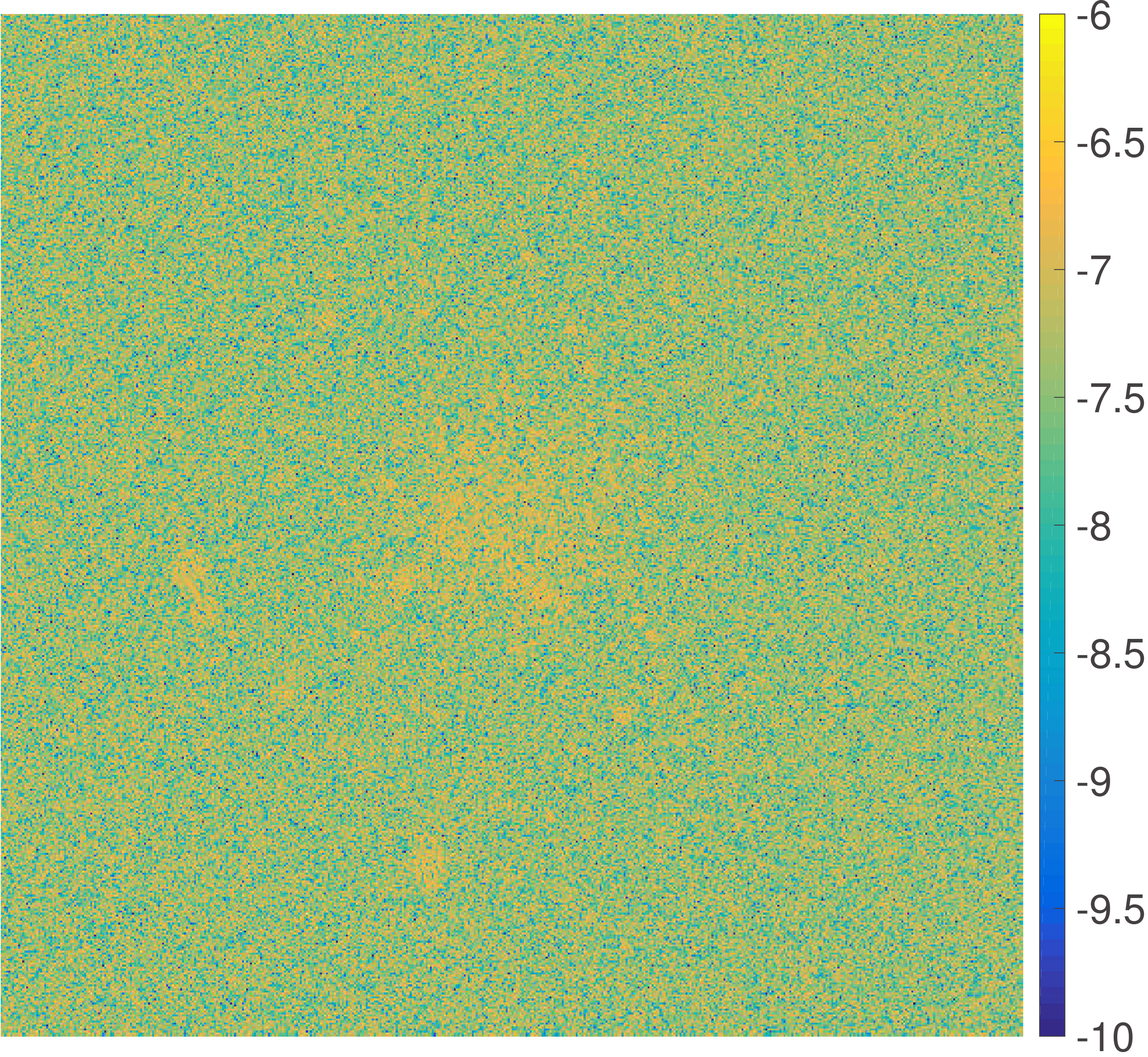}
	
	\caption{Visual comparison of reconstruction results for the `Galaxy cluster' test image with $M=50N$. From left to right: reconstructed, error and residual images in $\log_{10}$ scale. From top to bottom: reconstruction performed with all visibilities using ADMM, and `reduced' visibilities after performing dimensionality reduction with $\RFPhit$, $\RPhit$, and $\RGt$ respectively. Last row: reconstruction using \textsc{ms-clean} with a uniform weighting scheme.}
	\label{fig:galaxyclustervisualcomp}
\end{figure*}

The Cygnus A image was chosen for the varied structure present in the different parts of the image. 
Fig.~\ref{fig:cygagraphs_uni} shows output SNR for reconstruction with $\RFPhit$ marginally higher than that with the complete visibilities set over Gaussian random coverages, at 36.3~dB and 35.7~dB respectively. 
$\RGt$ leads to 35.2~dB output SNR.
Over SKA-like coverages, $\RFPhit$ performs better than the complete visibilities set over low data sizes, but the difference is made up for larger data sizes and the output SNRs reach comparable values, at 24~dB and 23.8~dB respectively, as seen in Fig.~\ref{fig:cygagraphs_ska}. 
DR values are clearly highest when using the complete visibilities set. 
Over Gaussian random coverages, it reaches up to $3.8\times 10^4$ whereas both $\RFPhit$ and $\RGt$ show similar trends across data sizes, reaching up to $3.4\times 10^4$. 
Over SKA-like coverages, the DR values are $1.7\times 10^6$, $1.5\times 10^6$ and $1.4\times 10^6$ respectively. 
{\revisedtwo 
}
The error images reflect the trend seen in Figs.~\ref{fig:cygagraphs_uni} and \ref{fig:cygagraphs_ska}. $\RGt$ has higher errors, and the complete visibilities set also fails to recover all the diffuse structure in the two lobes. 
The recovered images show that $\RFPhit$ is able to faithfully recover the diffuse structure as well as the bright point-like sources present in the image.
{\revised The computation time per iteration for reconstruction using $M=10N$ visibilities was $\approx$20 seconds when imaging without prior dimensionality reduction, which decreased to $\approx$7.5 seconds per iteration using $\RFPhit$.}
{\revised  \textsc{ms-clean} output model images of size $1431\times 3075$ pixels were cropped to $477\times 1025$ pixels.
The model image shown in Fig.~\ref{fig:cygavisualcomp} contains 21507 components and took 5 major iterations.
The reconstruction shows smooth regions instead of the details of the diffuse structure present in the test image.
The error image illustrates missing features with details of the test image that were not captured by the model.
}

{\revised The higher SNR achieved in some cases after dimensionality reduction with $\RFPhit$ or $\RGt$, as compared to $\R=\Ii$, is a secondary effect -- of the approximated projection to the singular vectors of the measurement operator $\PPhi$ in the case of $\RFPhit$ and of an effective `averaging' over neighbouring $uv$ points in the case of $\RGt$. It may be attributed to the retention of signal information through the null space of $\PPhi$ while effectively reducing noise content. It must be noted here, however, that this is a by-product of the design of $\RFPhit$ and $\RGt$. The extent of this apparent denoising depends on the type of image and the actual coverage under consideration, as seen in Figs.~\ref{fig:graphs_uni} and \ref{fig:graphs_ska}. In all cases, the reconstruction from $\RFPhit$ is seen to be at least as accurate as that from the complete visibilities set.}

An overarching trend across images, coverages and data sizes is that using an identity matrix to approximate the initial noise covariance matrix results in consistently poorer reconstruction, qualified both in values of output SNR and reconstruction DR. 
This trend supports our understanding that appropriate handling and a justified approximation of the noise covariance are essential for accurate performance of the image reconstruction algorithms used in this work.

\begin{figure*}
	\centering
	\includegraphics[trim={0px 0px 0px 0px}, clip, height=0.14\linewidth]{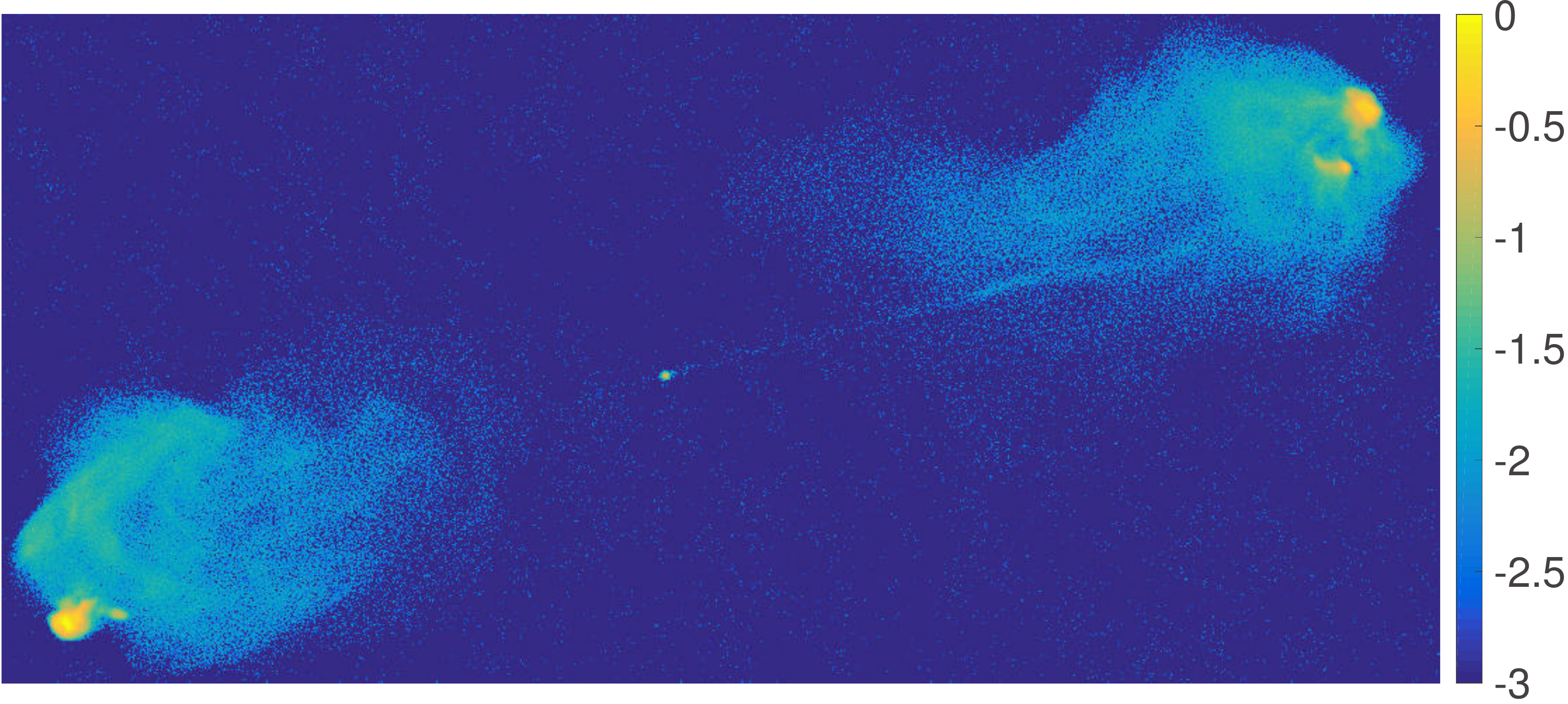}\hspace{2pt}
	\includegraphics[trim={0px 0px 0px 0px}, clip, height=0.14\linewidth]{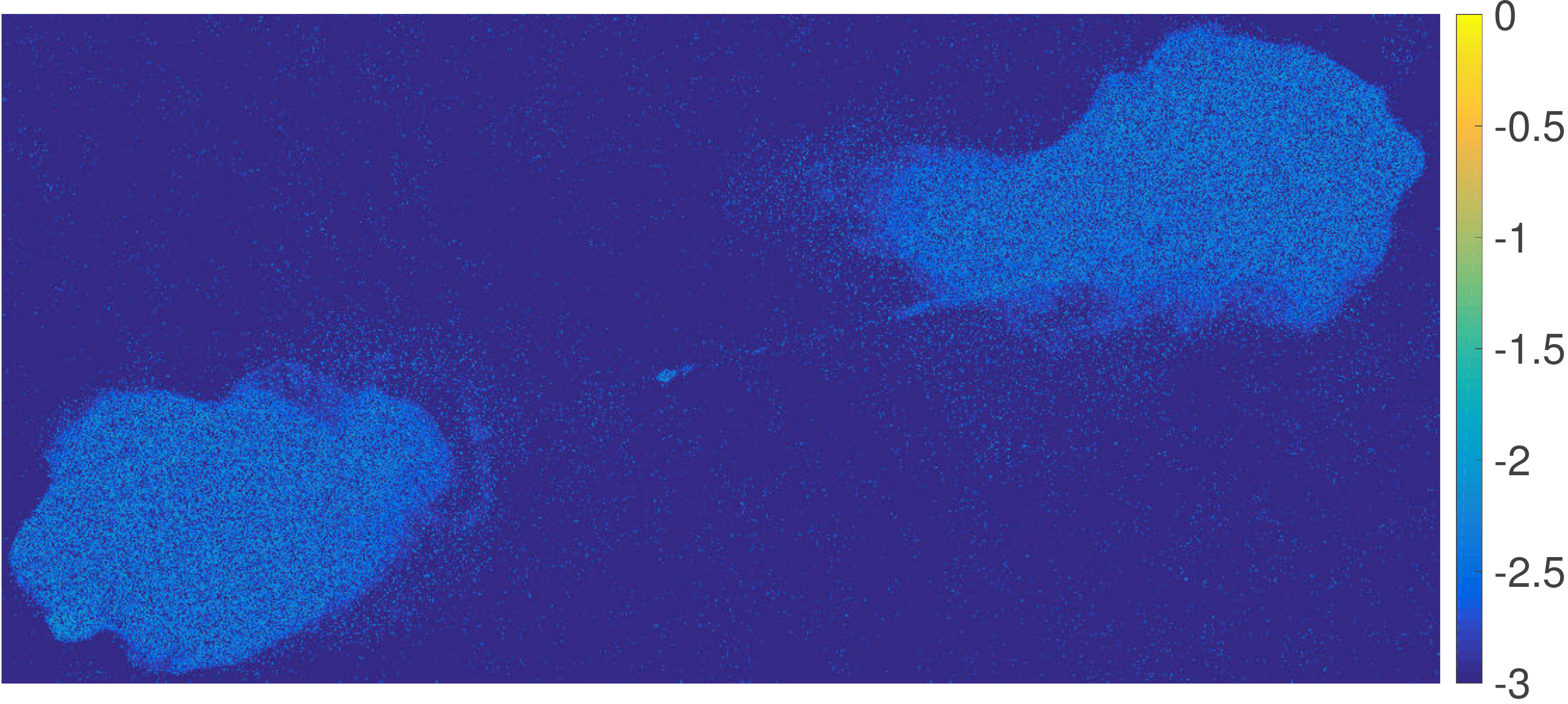}\hspace{2pt}
	\includegraphics[trim={0px 0px 0px 0px}, clip, height=0.14\linewidth]{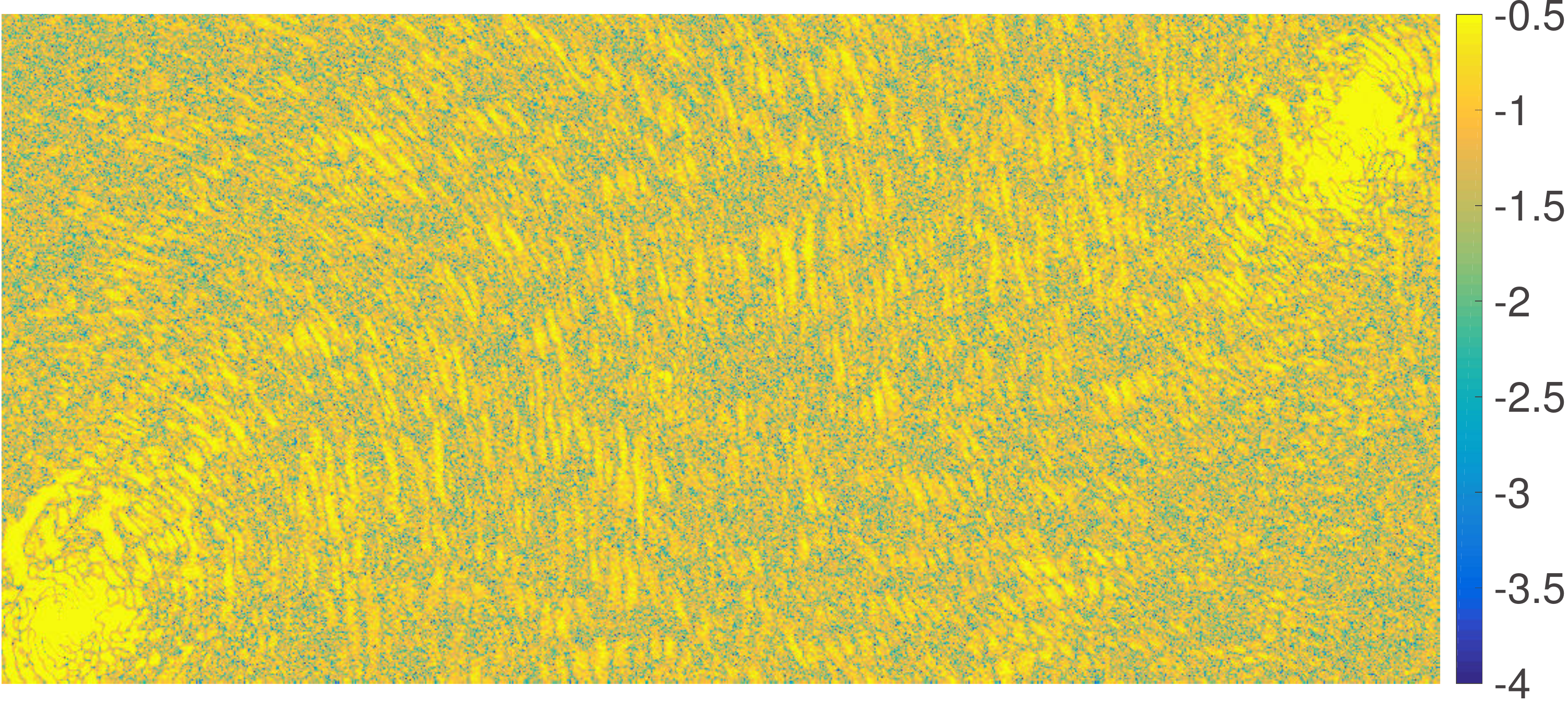}

	\vspace{3pt}
		
	\includegraphics[trim={0px 0px 0px 0px}, clip, height=0.14\linewidth]{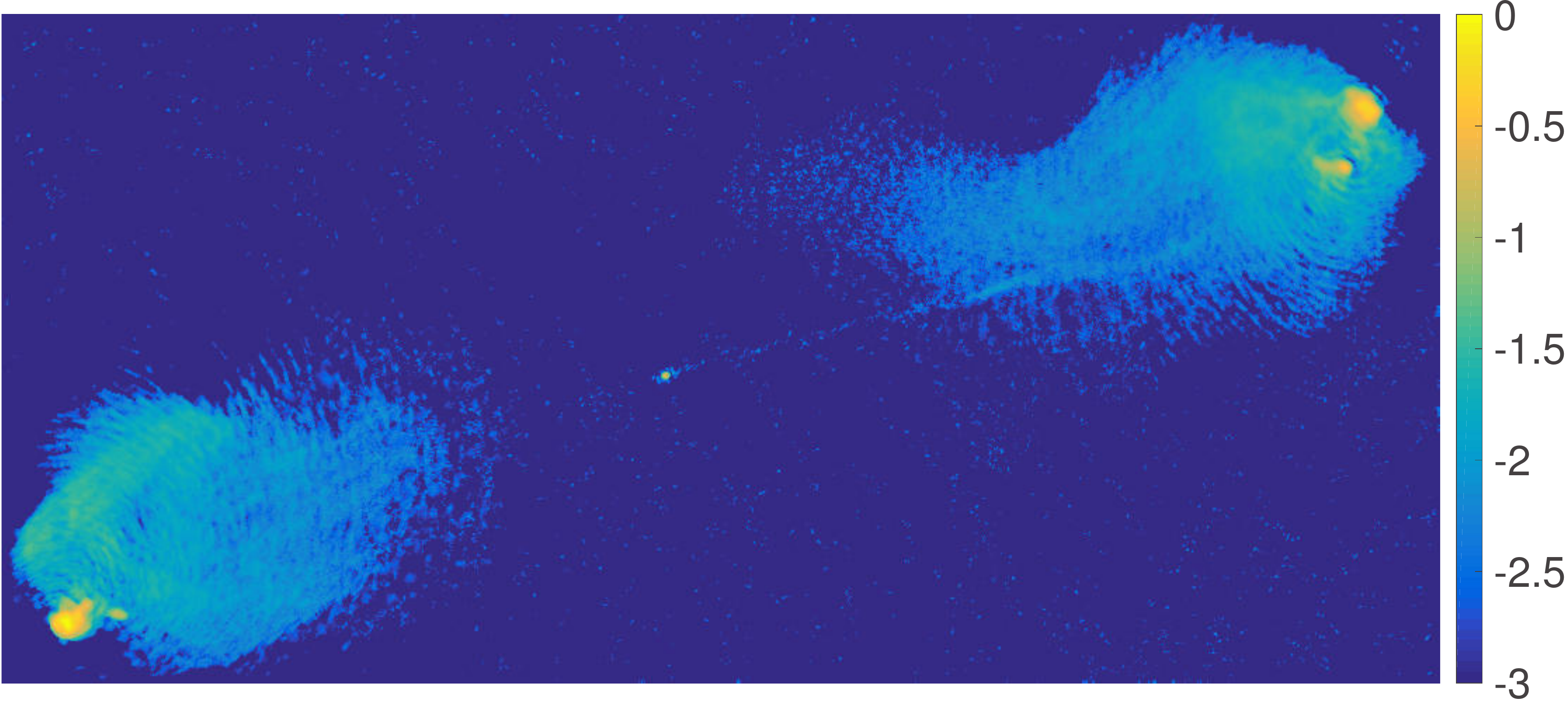}\hspace{2pt}
	\includegraphics[trim={0px 0px 0px 0px}, clip, height=0.14\linewidth]{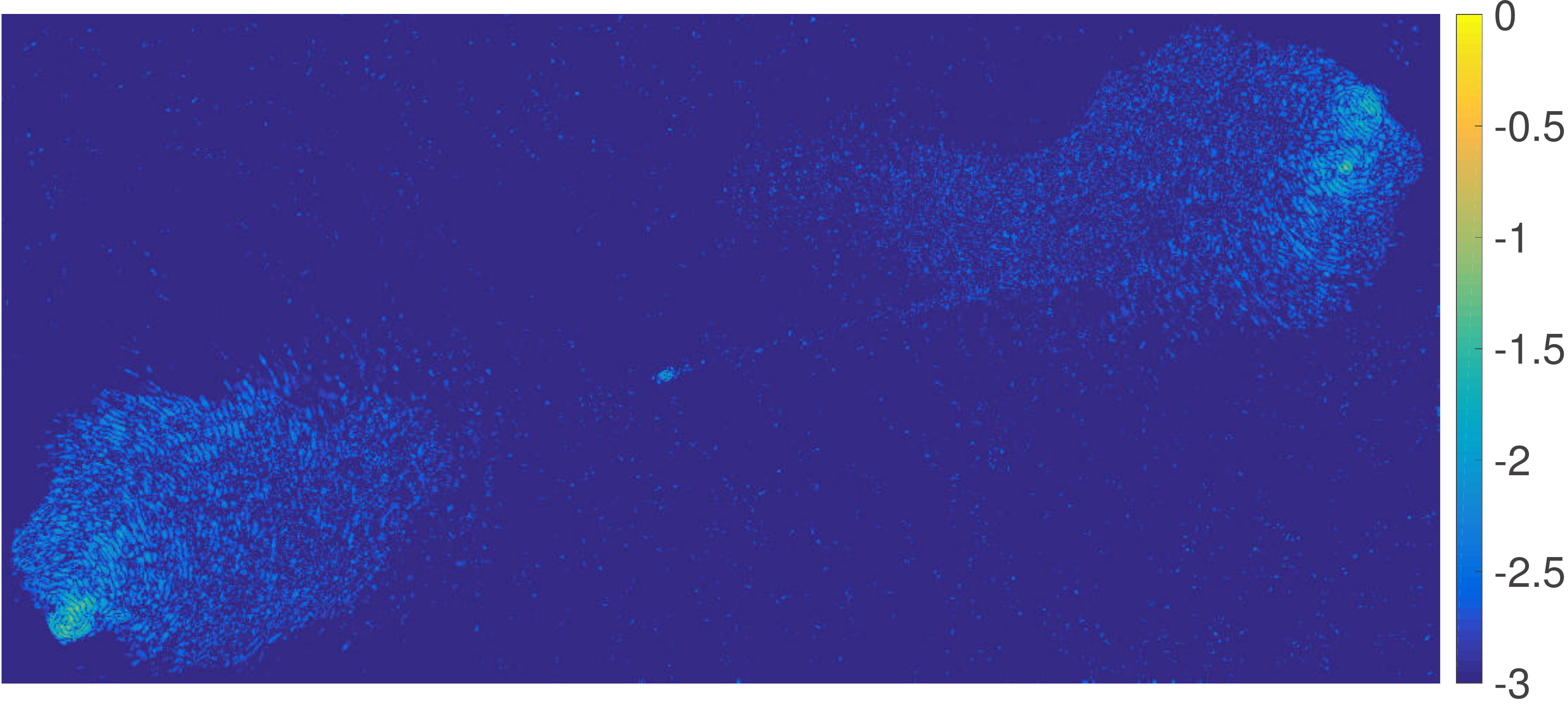}\hspace{2pt}
	\includegraphics[trim={0px 0px 0px 0px}, clip, height=0.14\linewidth]{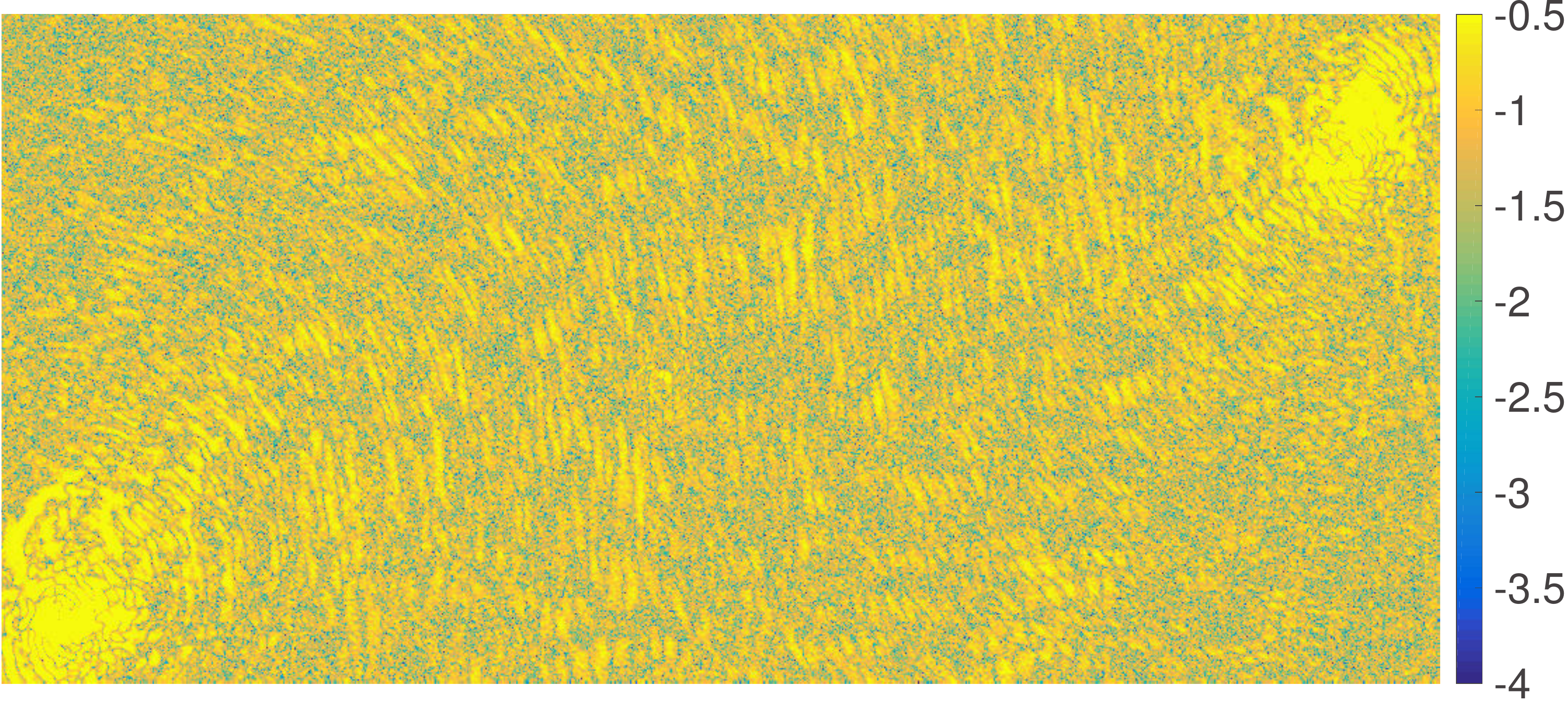}

	\vspace{3pt}
		
	\includegraphics[trim={0px 0px 0px 0px}, clip, height=0.14\linewidth]{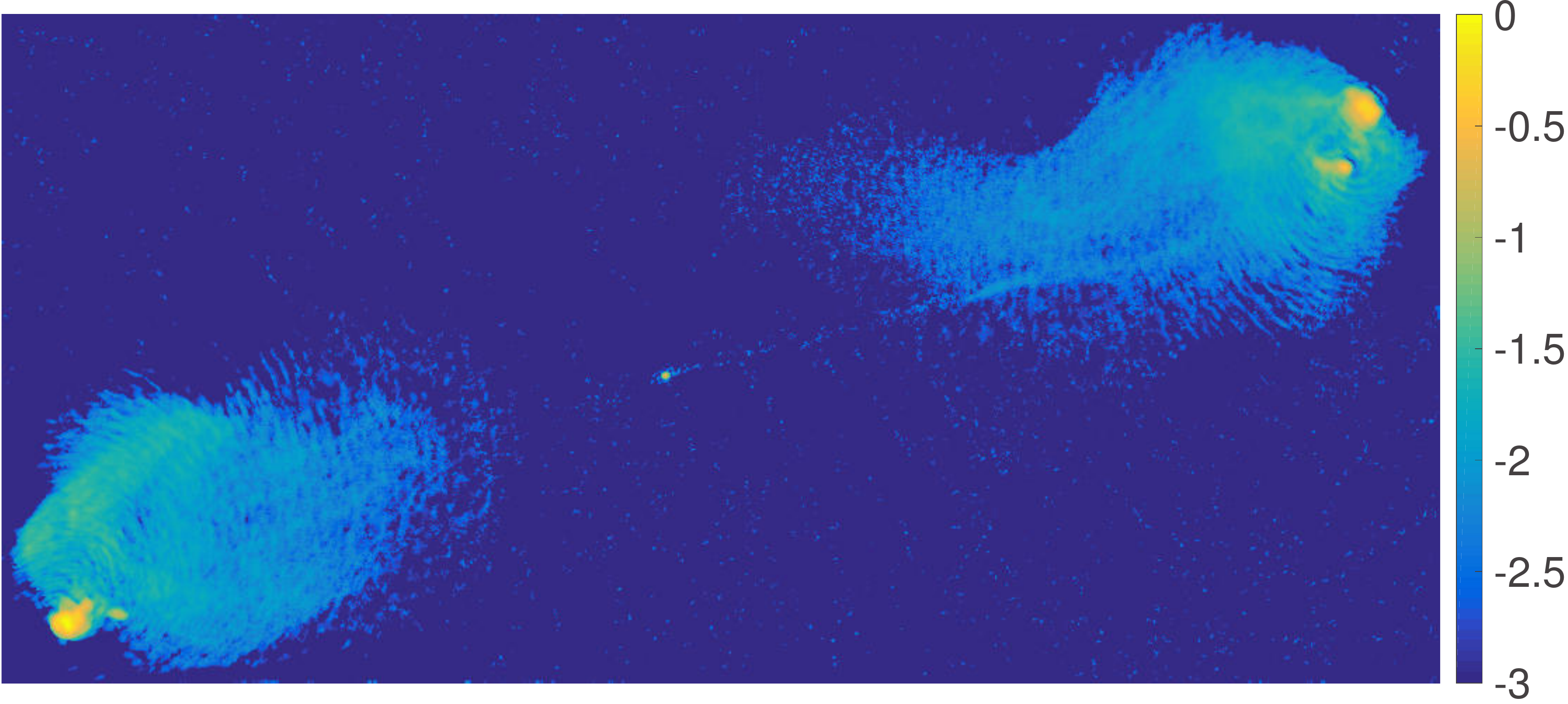}\hspace{2pt}
	\includegraphics[trim={0px 0px 0px 0px}, clip, height=0.14\linewidth]{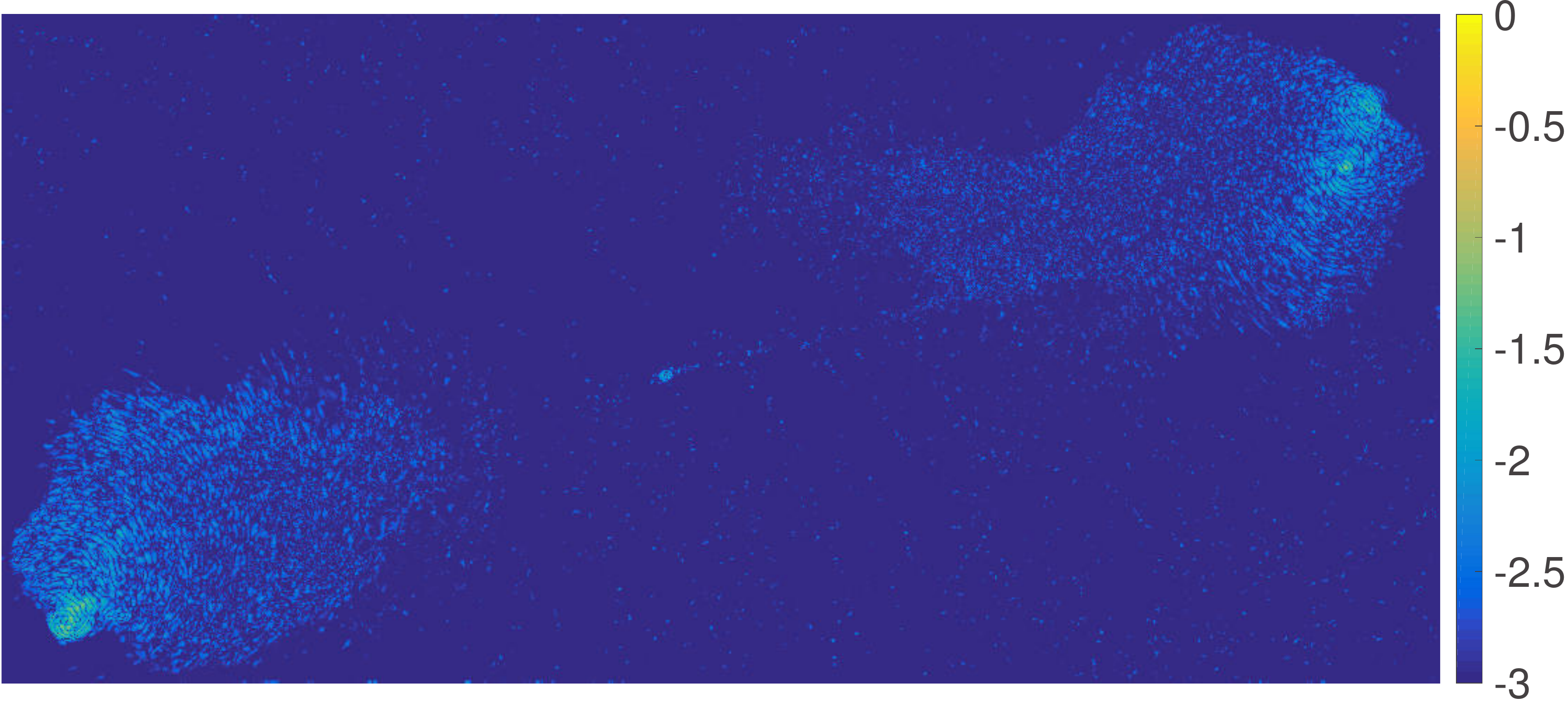}\hspace{2pt}
	\includegraphics[trim={0px 0px 0px 0px}, clip, height=0.14\linewidth]{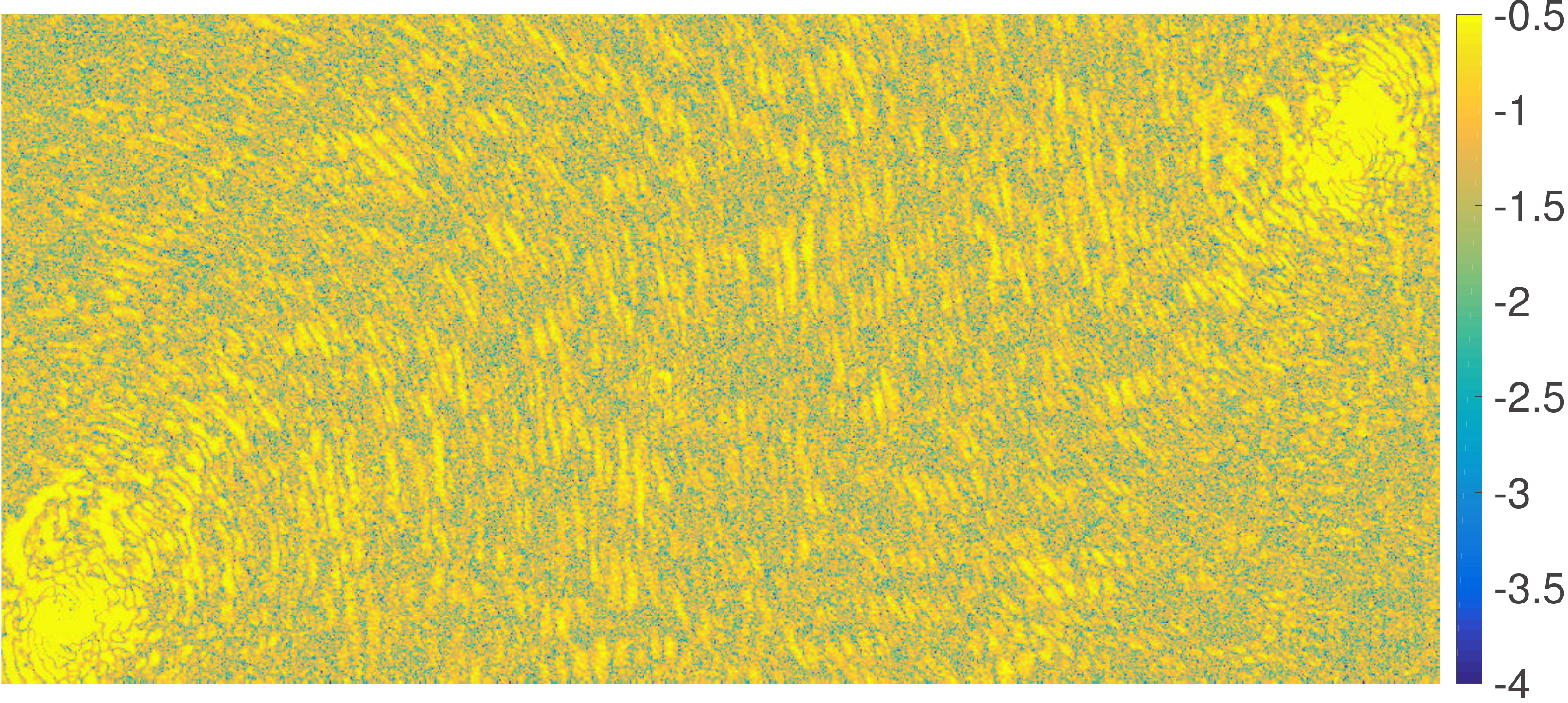}

	\vspace{3pt}
		
	\includegraphics[trim={0px 0px 0px 0px}, clip, height=0.14\linewidth]{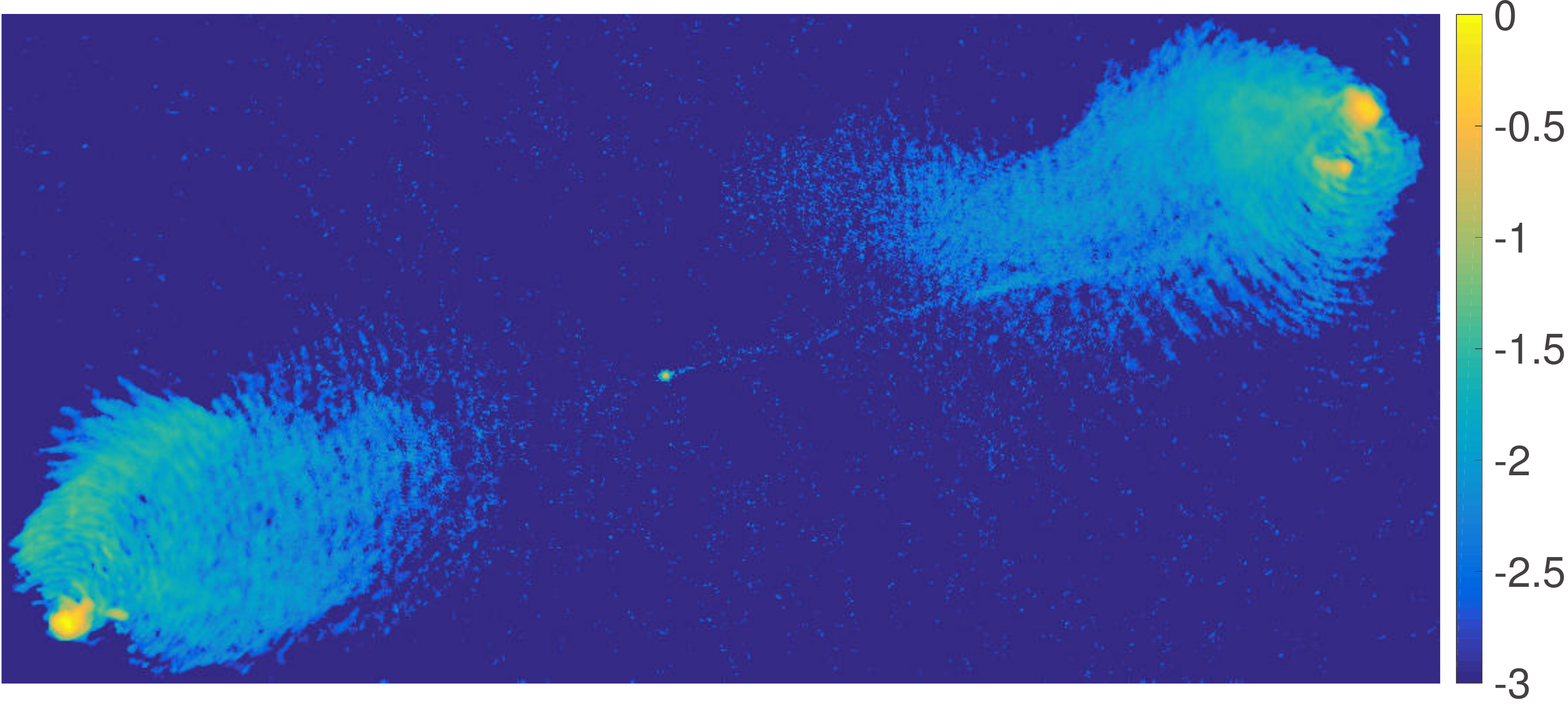}\hspace{2pt}
	\includegraphics[trim={0px 0px 0px 0px}, clip, height=0.14\linewidth]{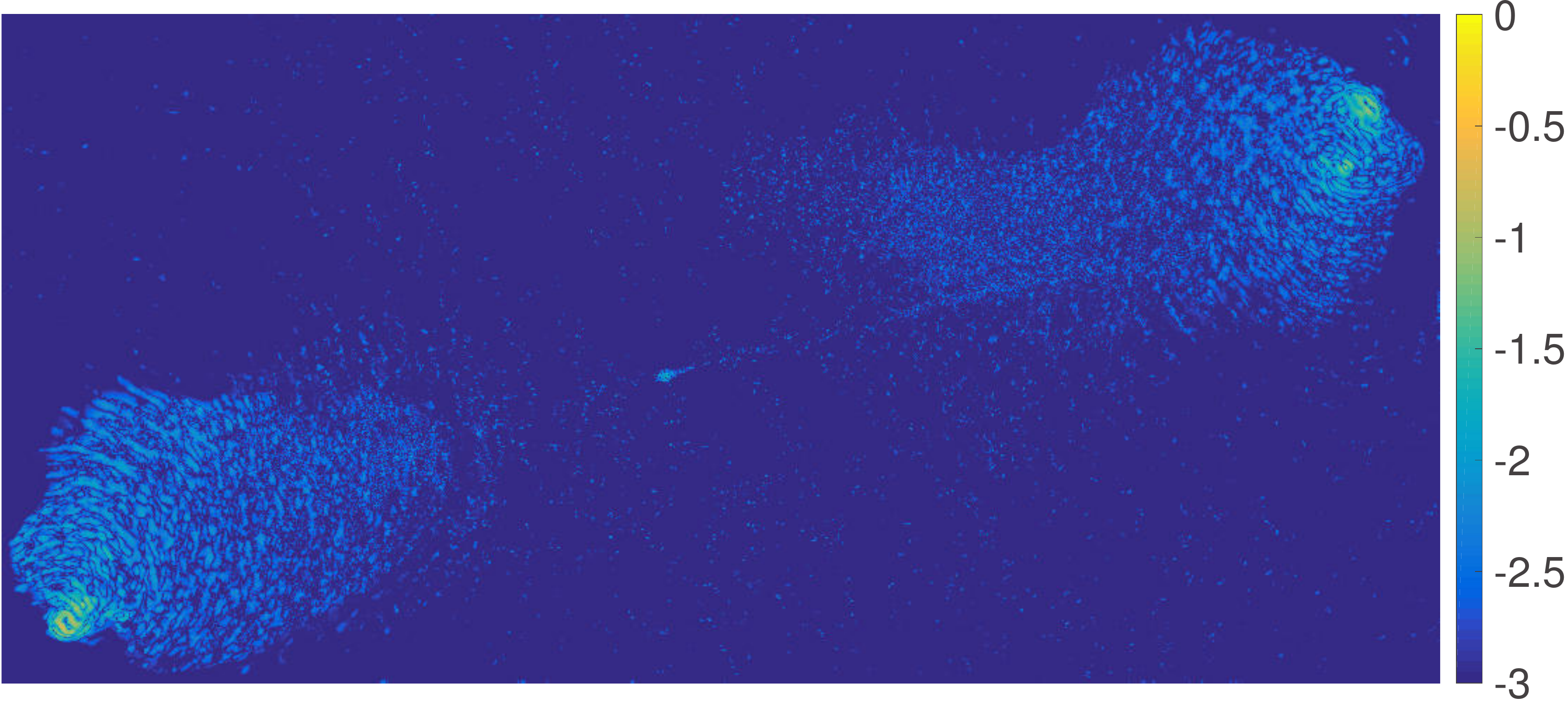}\hspace{2pt}
	\includegraphics[trim={0px 0px 0px 0px}, clip, height=0.14\linewidth]{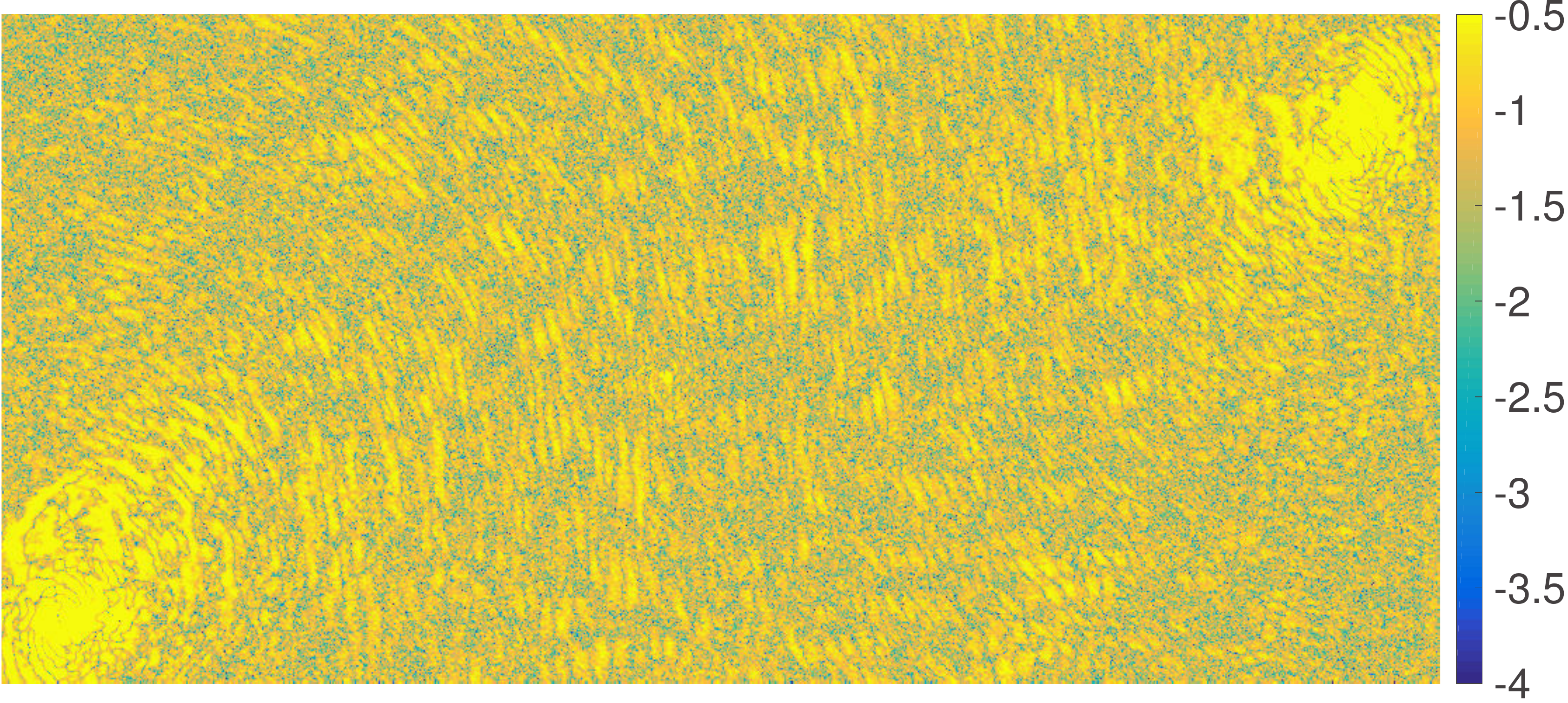}
	
	\vspace{3pt}

	\includegraphics[trim={0px 0px 0px 0px}, clip, height=0.14\linewidth]{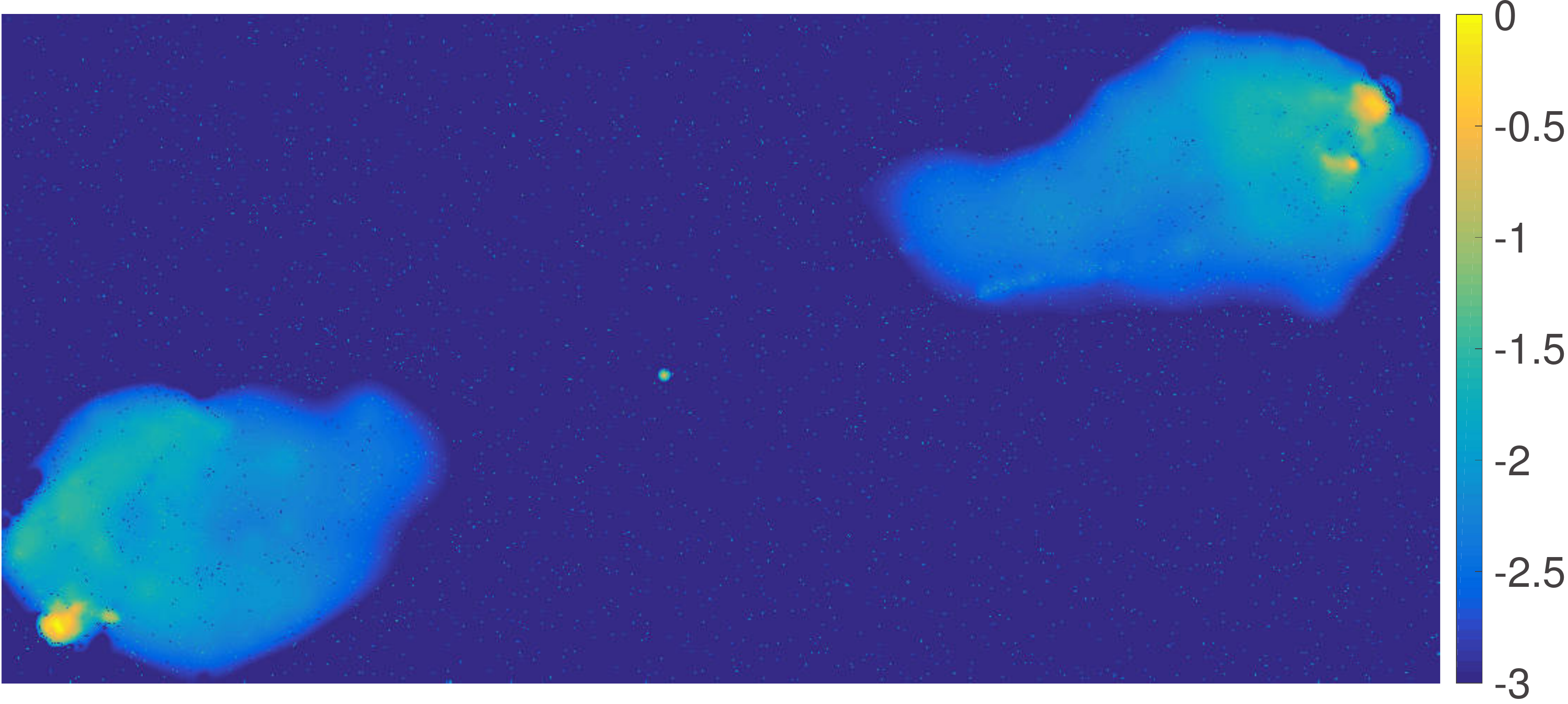}\hspace{2pt}
	\includegraphics[trim={0px 0px 0px 0px}, clip, height=0.14\linewidth]{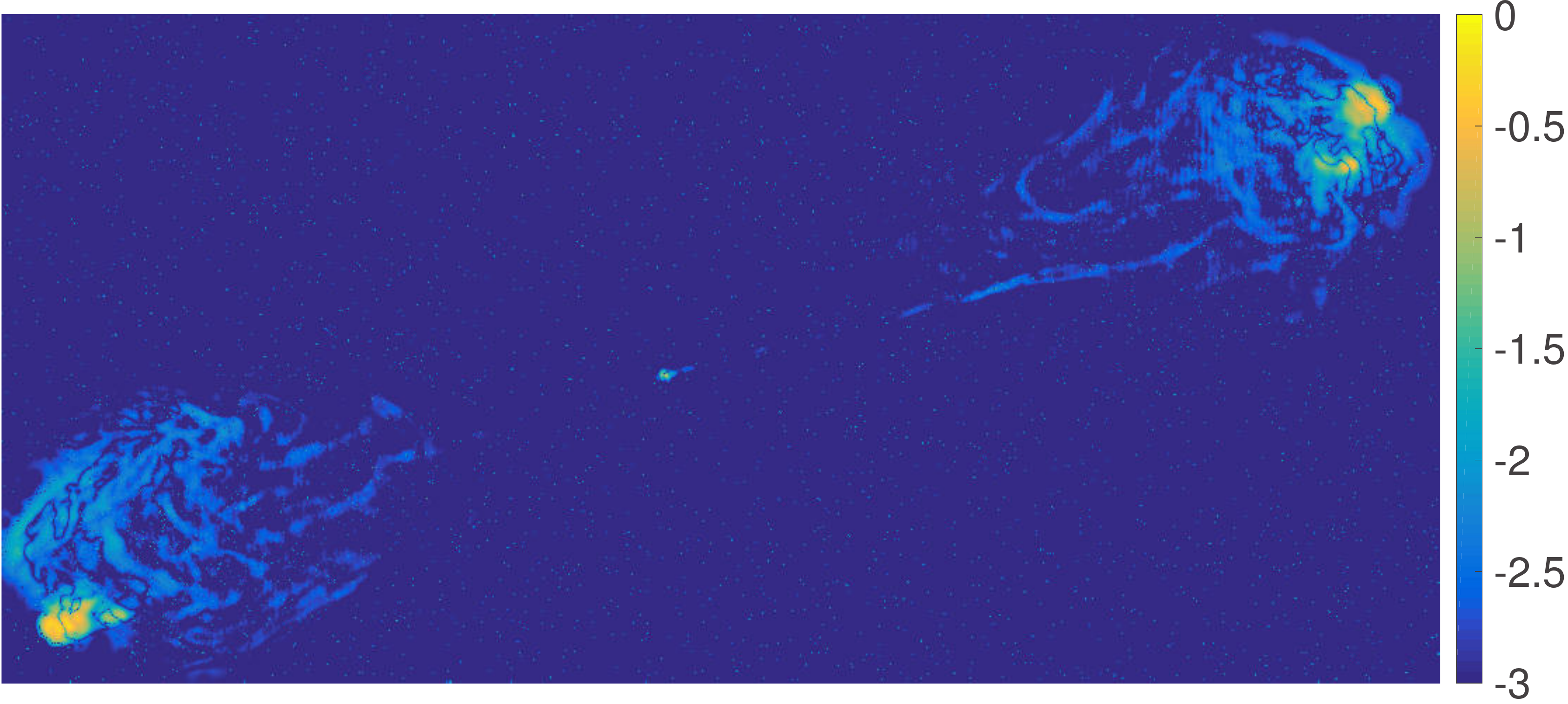}\hspace{2pt}
	\includegraphics[trim={0px 0px 0px 0px}, clip, height=0.14\linewidth]{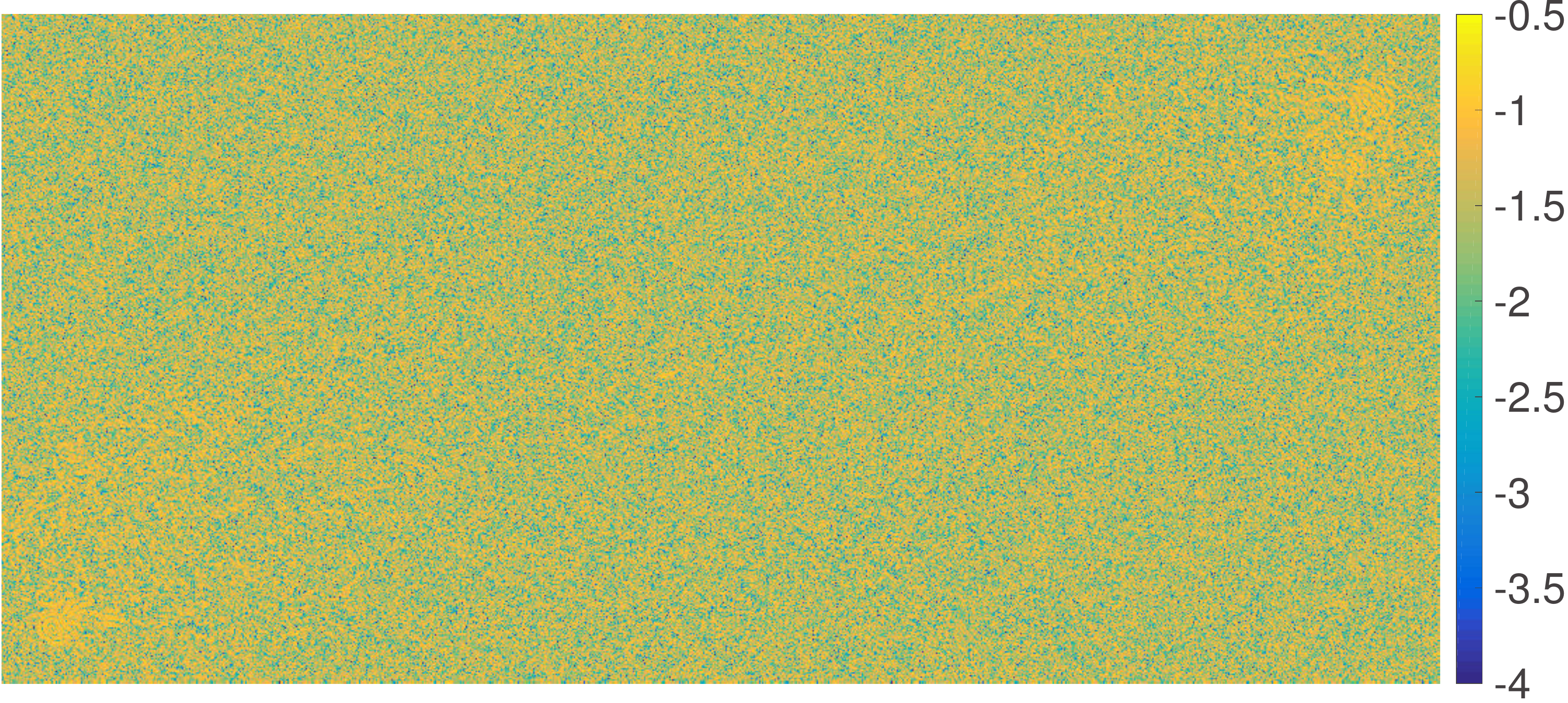}
	
	\caption{Visual comparison of reconstruction results for the `Cygnus A' test image with $M=10N$. From left to right: reconstructed, error and residual images in $\log_{10}$ scale. From top to bottom: reconstruction performed with all visibilities using ADMM, and `reduced' visibilities after performing dimensionality reduction with $\RFPhit$, $\RPhit$, and $\RGt$ respectively. Last row: reconstruction using \textsc{ms-clean} with a uniform weighting scheme.}
	\label{fig:cygavisualcomp}
\end{figure*}

\subsubsection{Including noise-dependent thresholding}
\label{subsec:discardeigenvaluessimulations}
We investigated a further dimensionality reduction to very small sizes $\Nz \ll N$ and $\nn \ll 4N$ for $\RFPhit$ and $\RGt$ respectively.
We performed image reconstruction using reduced data obtained through $\RFPhit$ and $\RGt$, and observed the effects of reducing dimensionality to particularly low values. 
Fig.~\ref{fig:snrs_discardeigenvalues} shows the SNR of reconstructed images from data reduced to sizes ranging from $4N$ all the way down to $0.05N$, which translates to a final low-dimensional data vector of approximately 4,000, 13,000 and 25,000 `reduced' visibilities for the M31, galaxy cluster and Cygnus A images respectively.
Reconstruction was performed using initial continuous visibilities of size $10N$, $25N$ and $50N$, simulated over the same SKA-like coverages that were used for obtaining the results shown in Figs.~\ref{fig:graphs_ska}, \ref{fig:m31visualcomp}, \ref{fig:galaxyclustervisualcomp} and \ref{fig:cygavisualcomp}.
The point of `diminishing returns' with respect to reduced data dimension and corresponding SNR can be seen as the inflection points in Fig.~\ref{fig:snrs_discardeigenvalues} where the SNR no longer remains unaffected by discarding further content.

\begin{figure}
	\begin{subfigure}{0.94\columnwidth}
		\centering
		\includegraphics[trim={0 0 0 0}, clip, width=\columnwidth]{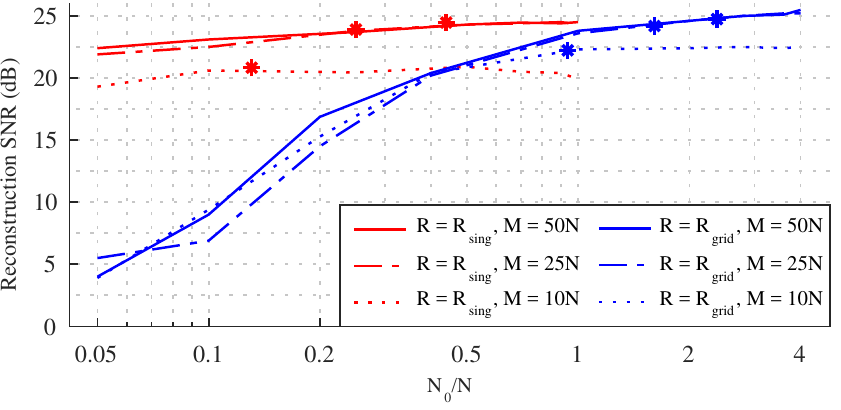}
		\caption{M31}
		\label{fig:m31discard_ska}
	\end{subfigure}
	\vspace{5pt}
	\begin{subfigure}{0.94\columnwidth}
		\centering
		\includegraphics[trim={0 0 0 0}, clip, width=\columnwidth]{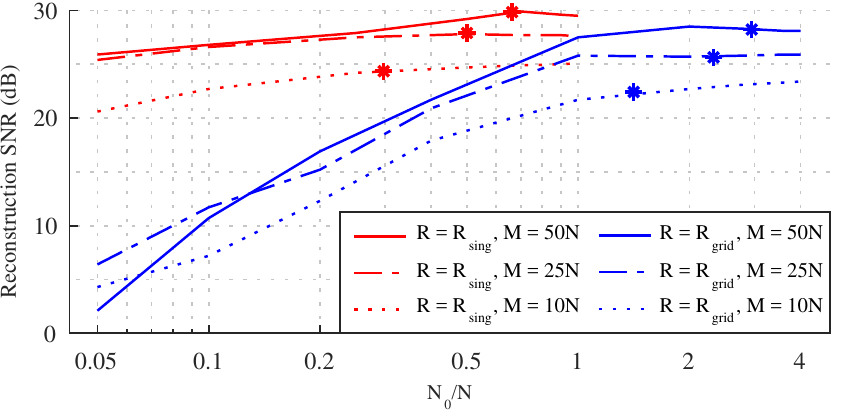}
		\caption{Galaxy cluster}
		\label{fig:galaxyclusterdiscard_ska}
	\end{subfigure}
	\vspace{5pt}
	\begin{subfigure}{0.94\columnwidth}
		\centering
		\includegraphics[trim={0 0 0 0}, clip, width=\columnwidth]{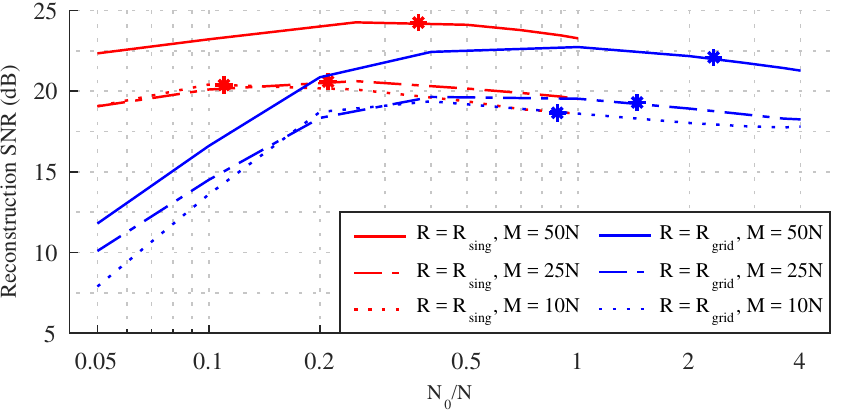}
		\caption{Cygnus A}
		\label{fig:cygadiscard_ska}
	\end{subfigure}
	\caption{Evolution of reconstruction quality from data reduced to below image size. (a) M31; (b) Galaxy cluster; (c) Cygnus A test images. Initial data size ranges from 650,000 to 12 million visibilities, simulated on SKA-like coverages. Continuous visibilities contain 30~dB additive noise. The analytically computed noise-based threshold marked as `*' shows the values of $\Nz$ and $\nn$ for $\gamma=1$, which corresponds to the minimum value of $\Nz$ or $\nn$ that ensures that no discarded data is more significant than noise fluctuation as given by Eq.~(\ref{eq:diagthreshold}) and Eq.~(\ref{eq:diagthresholdgt}).}
	\label{fig:snrs_discardeigenvalues}
\end{figure}

We find that the dimensionality reduction method $\RFPhit$ is much more robust to reducing data size below image size, and that we are able to reduce data from an initial visibilities dimension of $50N$ to a final data size of $0.05N$ while decreasing the SNR by less than 5~dB. 
The method $\RGt$, however, is seen to be affected adversely from significant dimensionality reduction, and the reconstruction quality dips strongly with decreasing data size to values much below image size.
A visual comparison of the artefacts introduced in the reconstruction due to an extremely low data size can be seen in Fig.~\ref{fig:discardeigenvalues_comp} on M31, the galaxy cluster and a zoomed-in portion of the Cygnus A image, highlighting the robustness of image reconstruction after reducing dimensionality with $\RFPhit$ as compared with $\RGt$.

\begin{figure*}
	\begin{subfigure}{1\linewidth}
		\centering		
		\includegraphics[trim={0px 0px 0px 0px}, clip, height=0.21\linewidth]{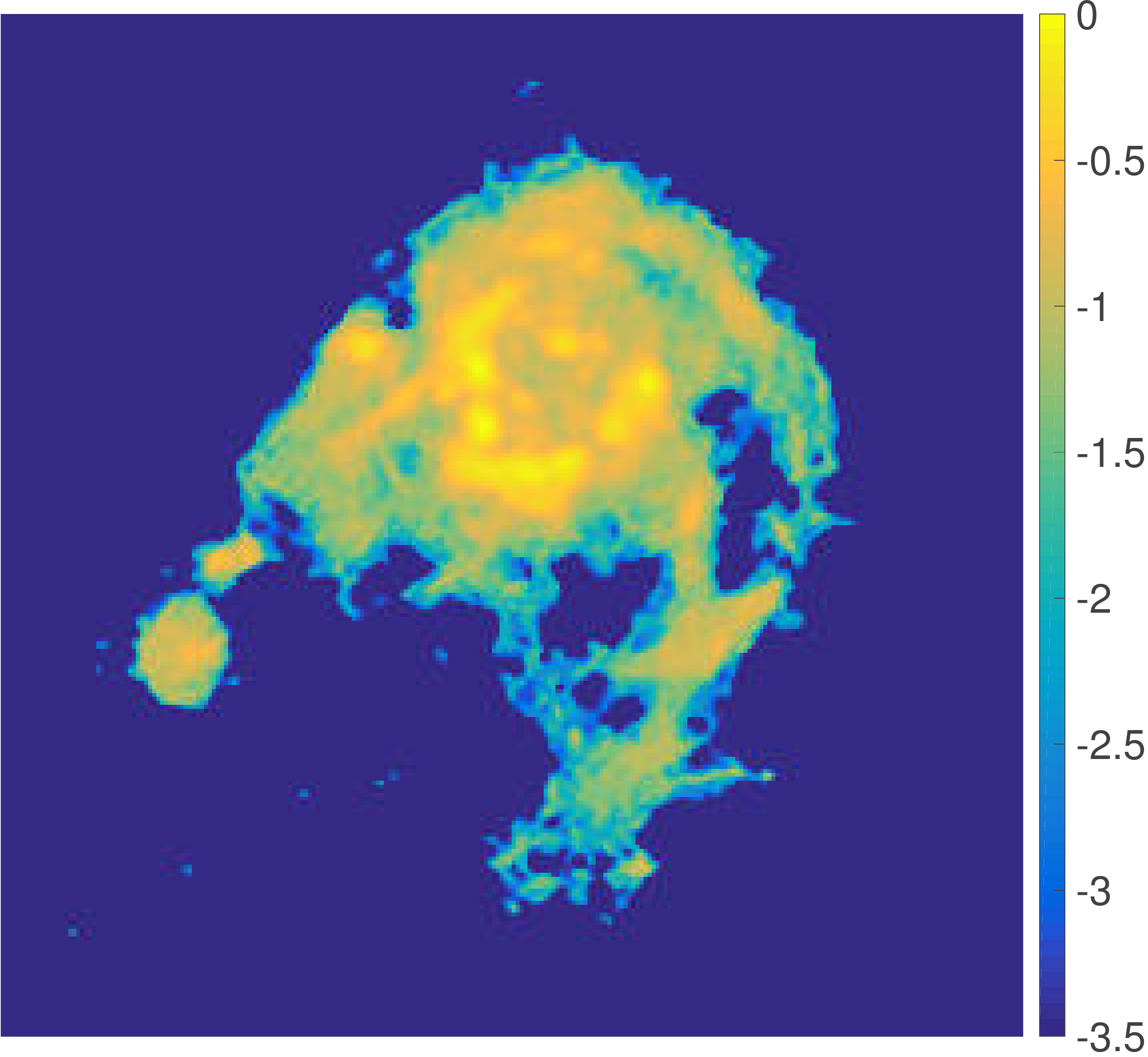}\hspace{2pt}
		\includegraphics[trim={0px 0px 0px 0px}, clip, height=0.21\linewidth]{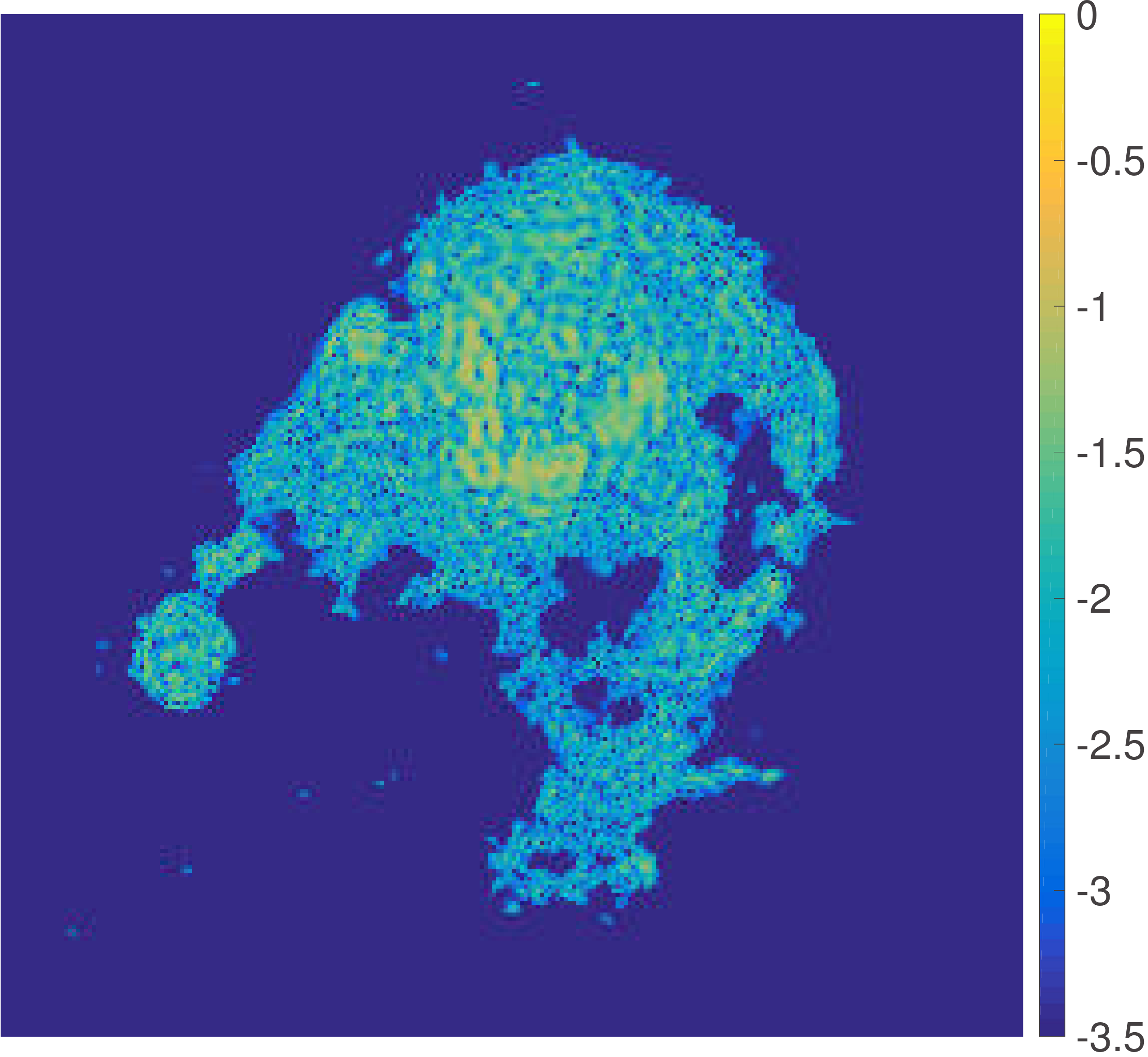}\hspace{2pt}
		\includegraphics[trim={0px 0px 0px 0px}, clip, height=0.21\linewidth]{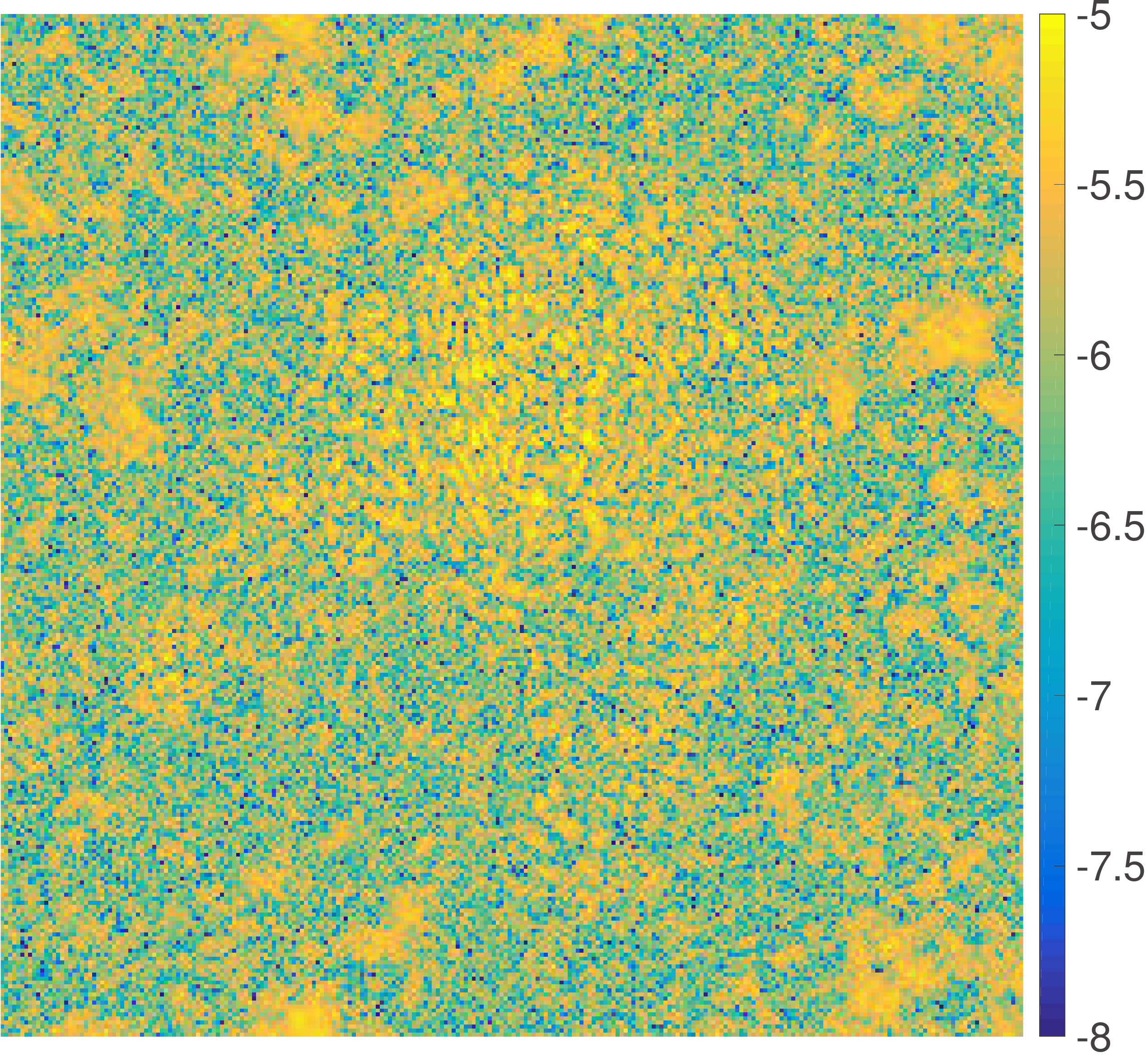} \\
		\vspace{3pt}		
		\includegraphics[trim={0px 0px 0px 0px}, clip, height=0.21\linewidth]{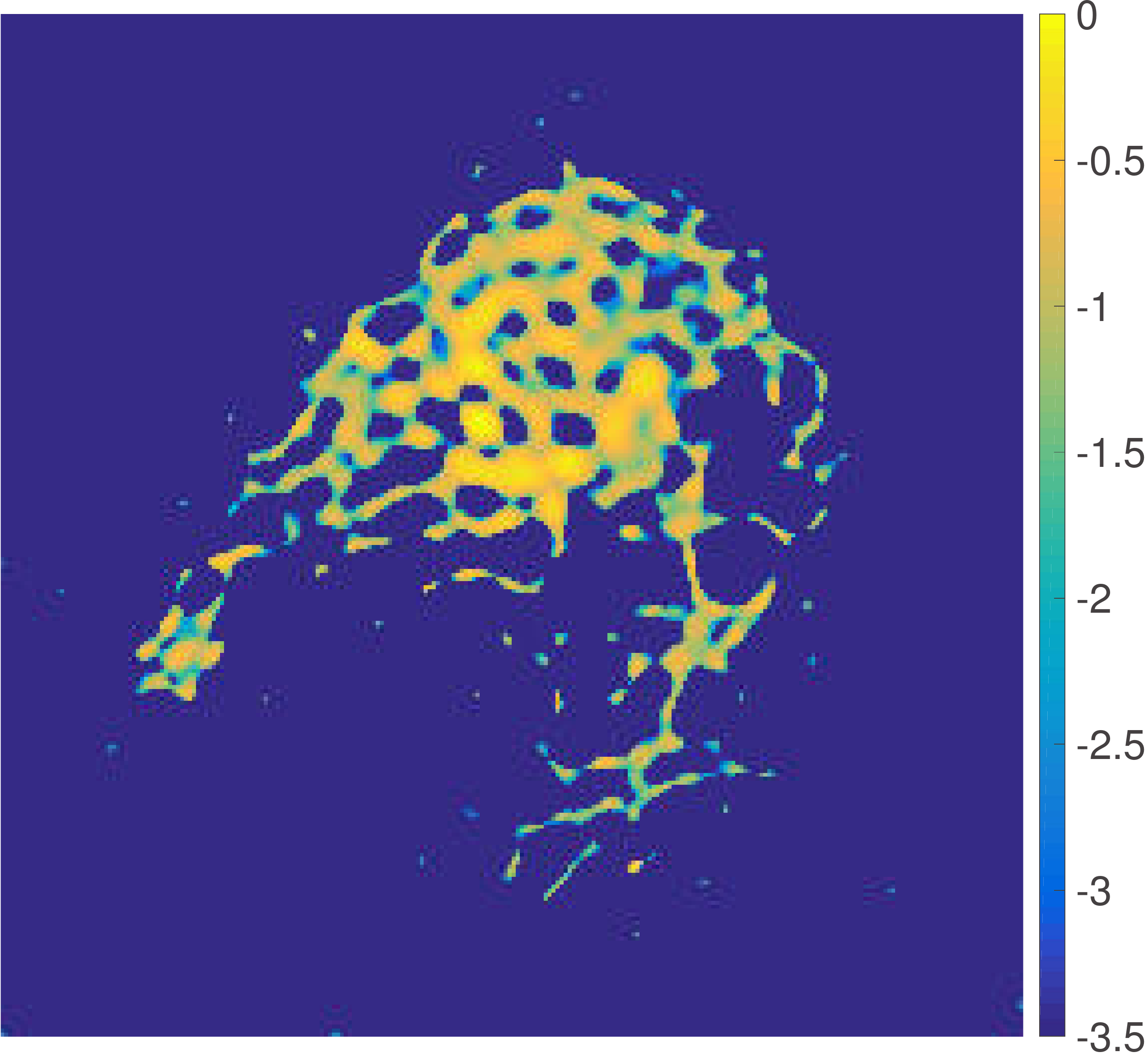}\hspace{2pt}
		\includegraphics[trim={0px 0px 0px 0px}, clip, height=0.21\linewidth]{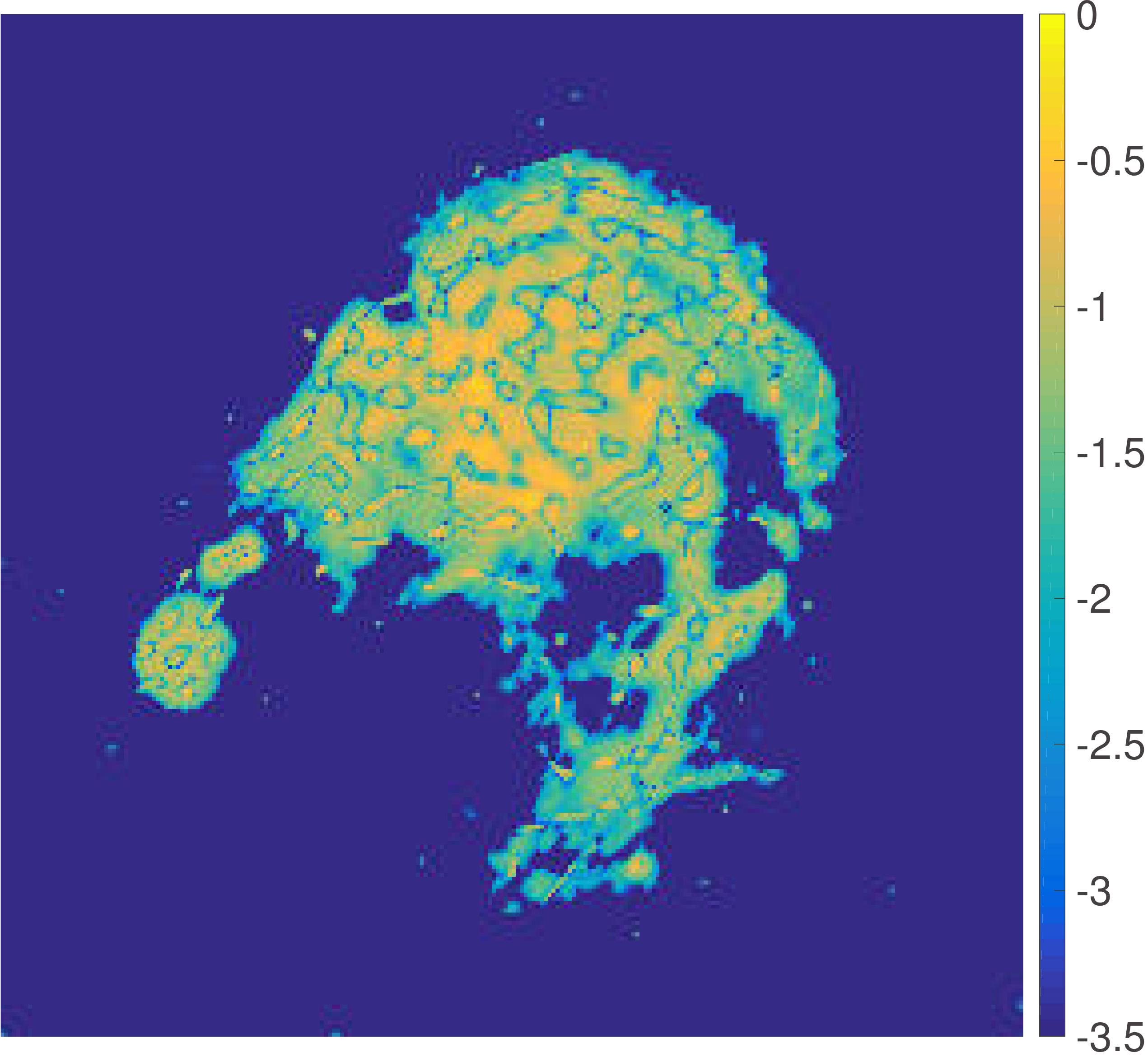}\hspace{2pt}
		\includegraphics[trim={0px 0px 0px 0px}, clip, height=0.21\linewidth]{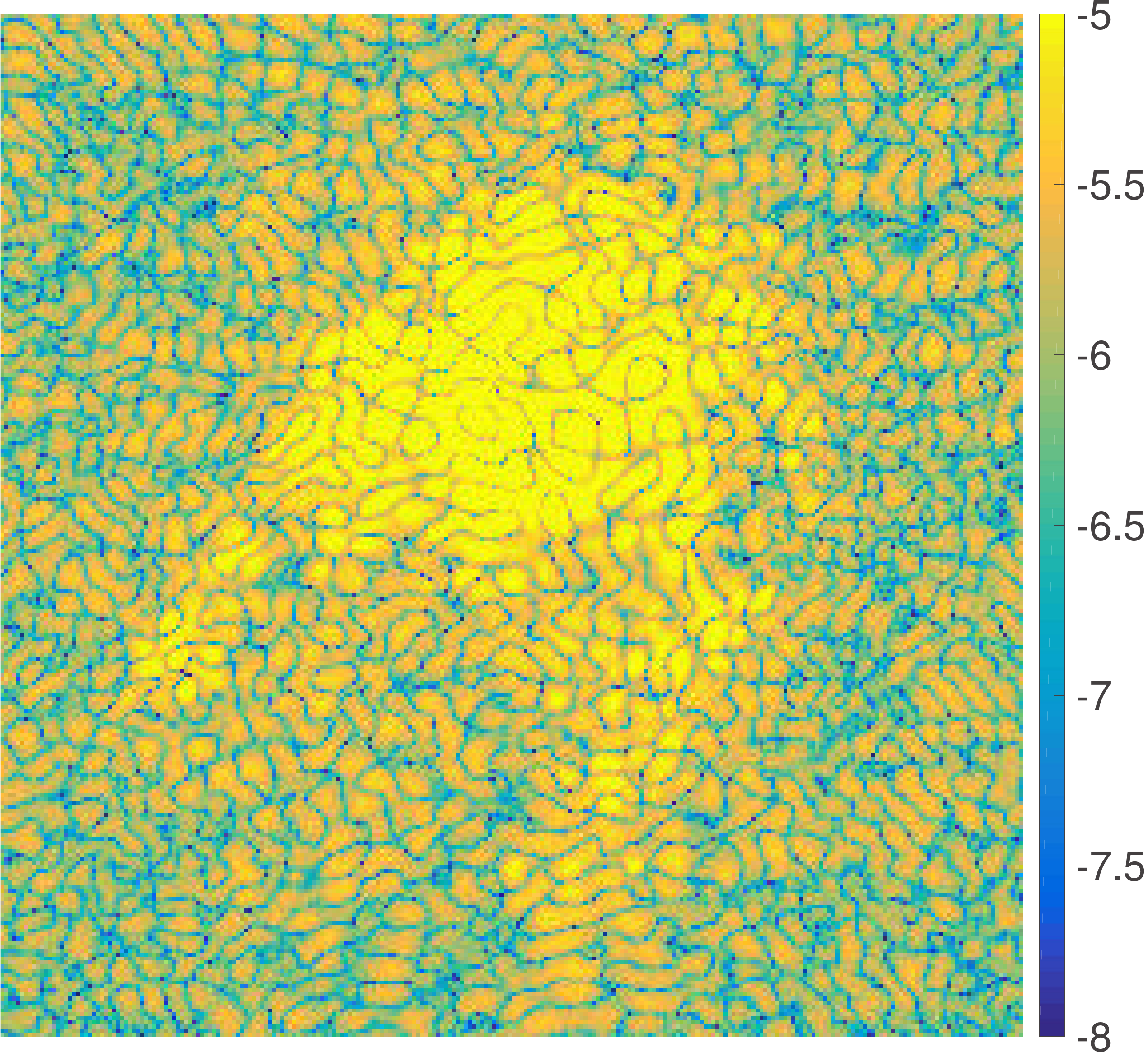}
		\caption{M31}
		\label{fig:m31_discardeigenvalues_comp}
	\end{subfigure}
	\begin{subfigure}{1\linewidth}
		\centering		
		\includegraphics[trim={0px 0px 0px 0px}, clip, height=0.21\linewidth]{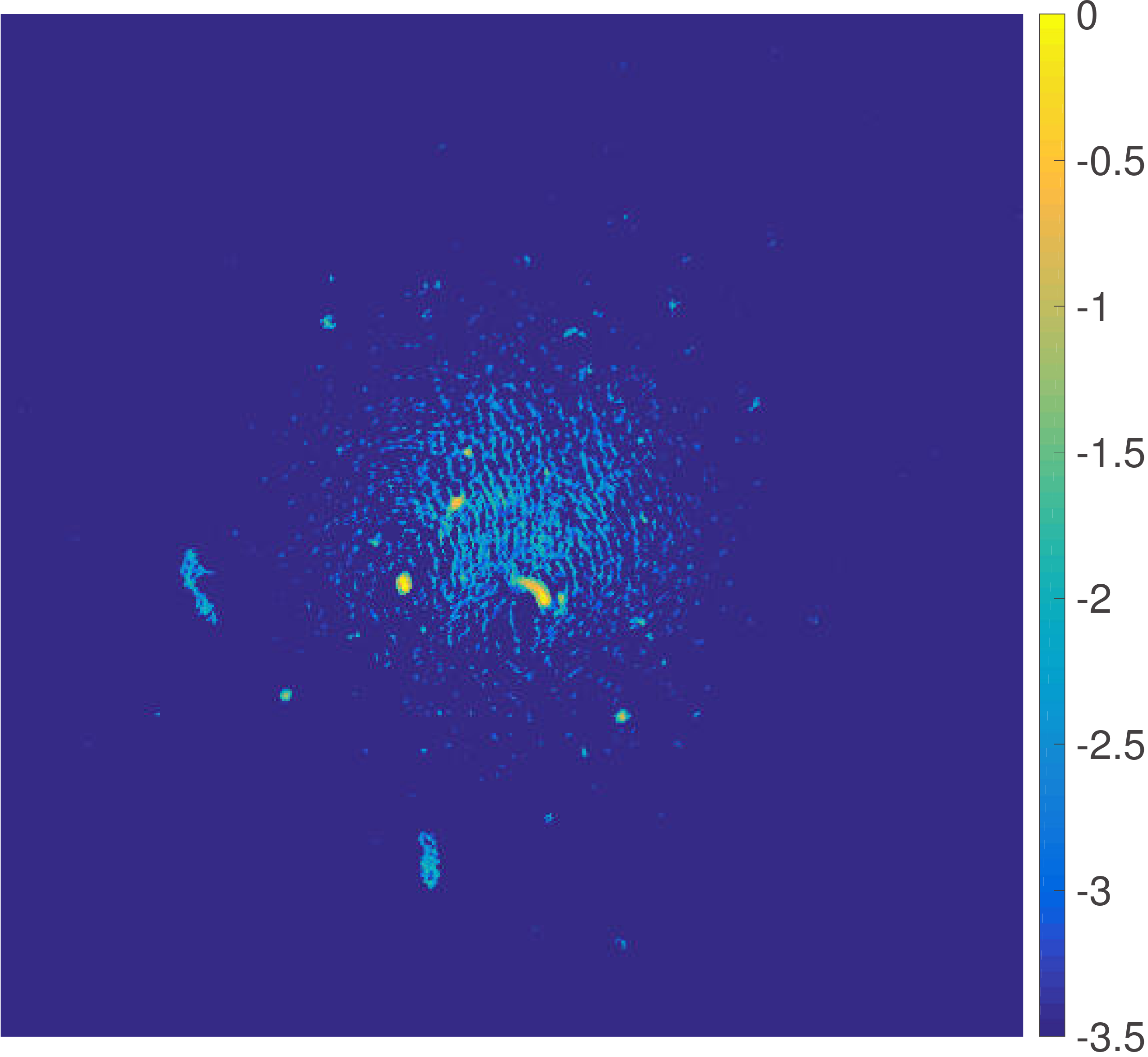}\hspace{2pt}
		\includegraphics[trim={0px 0px 0px 0px}, clip, height=0.21\linewidth]{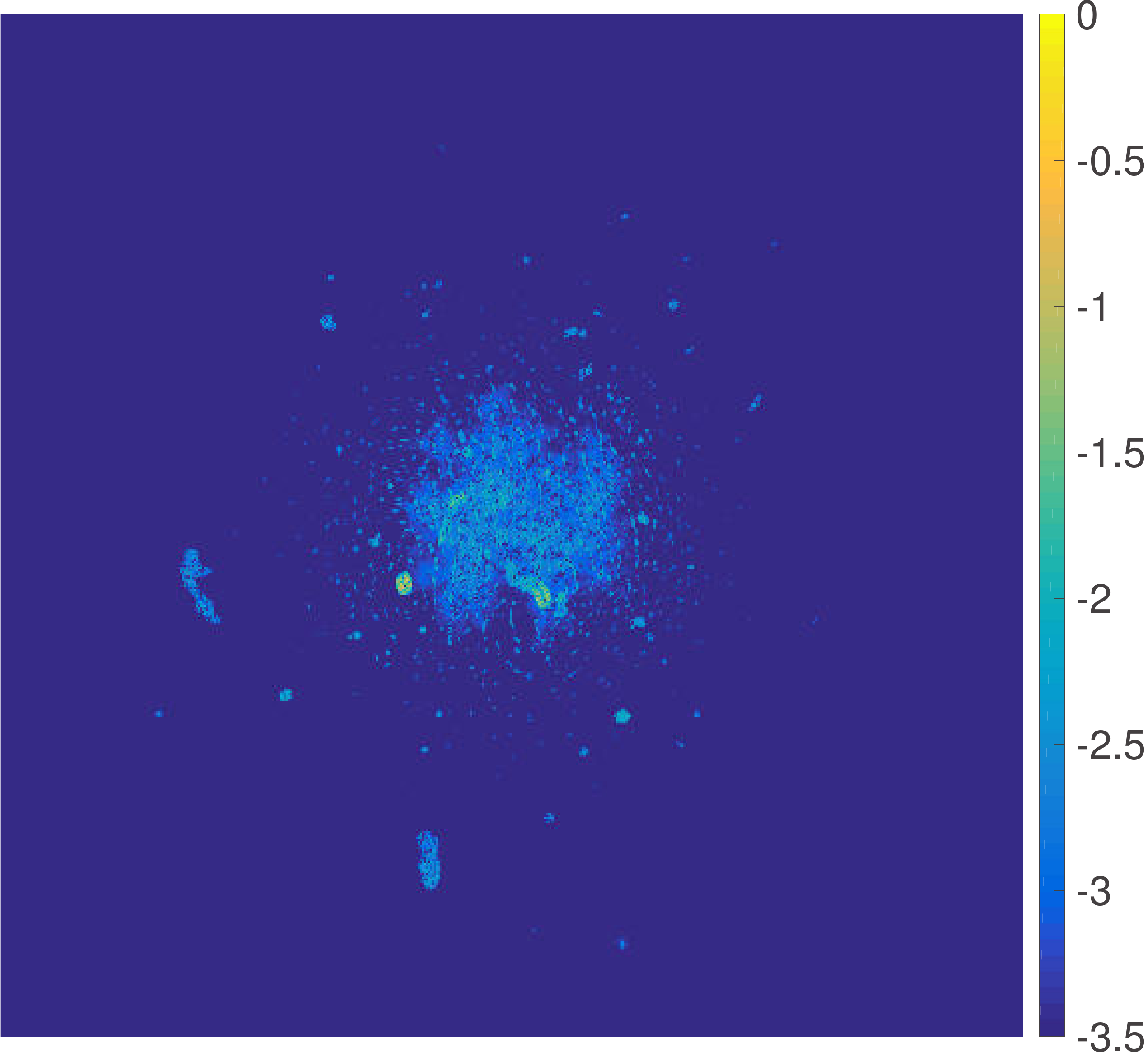}\hspace{2pt}
		\includegraphics[trim={0px 0px 0px 0px}, clip, height=0.21\linewidth]{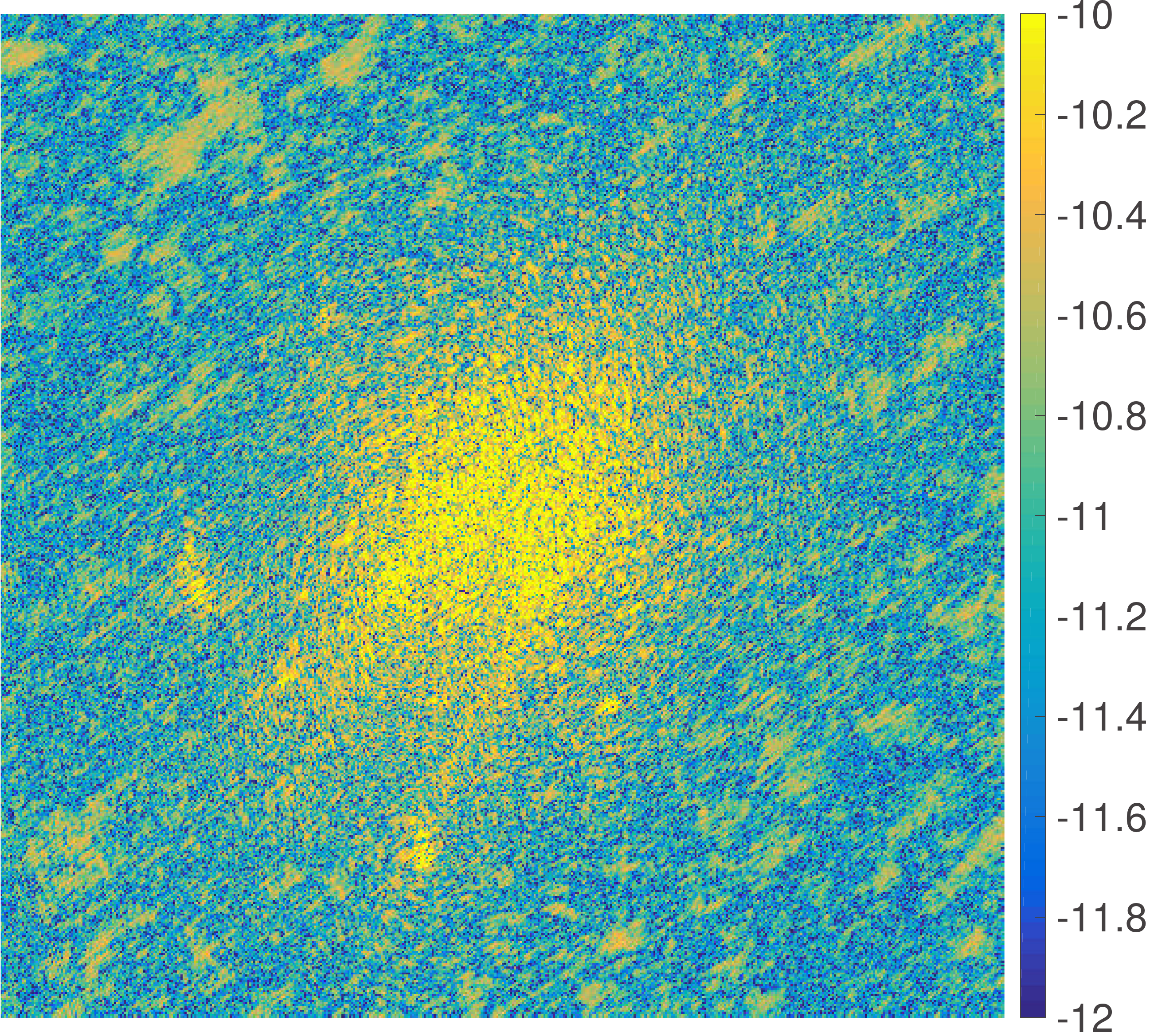} \\
		\vspace{3pt}		
		\includegraphics[trim={0px 0px 0px 0px}, clip, height=0.21\linewidth]{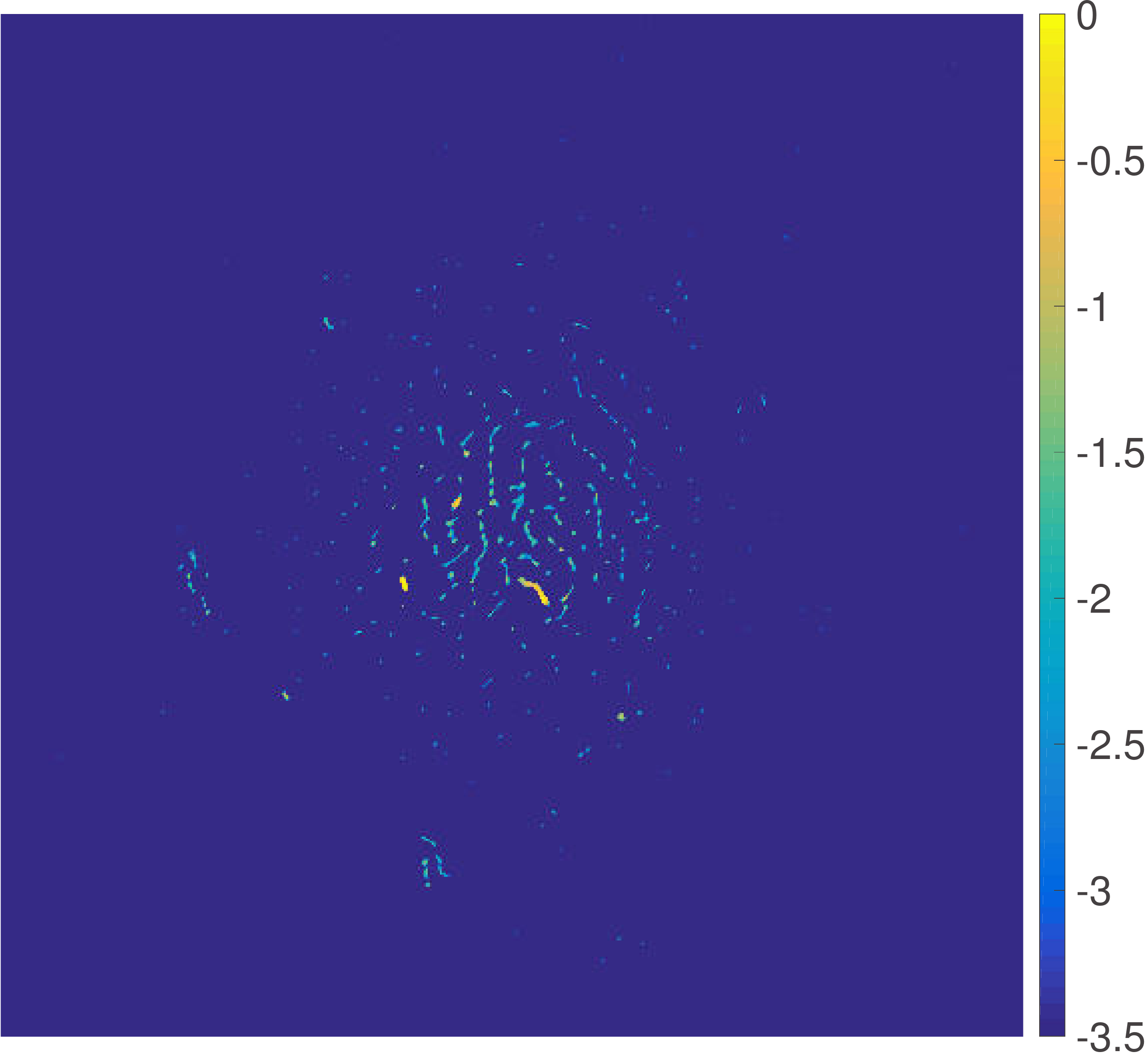}\hspace{2pt}
		\includegraphics[trim={0px 0px 0px 0px}, clip, height=0.21\linewidth]{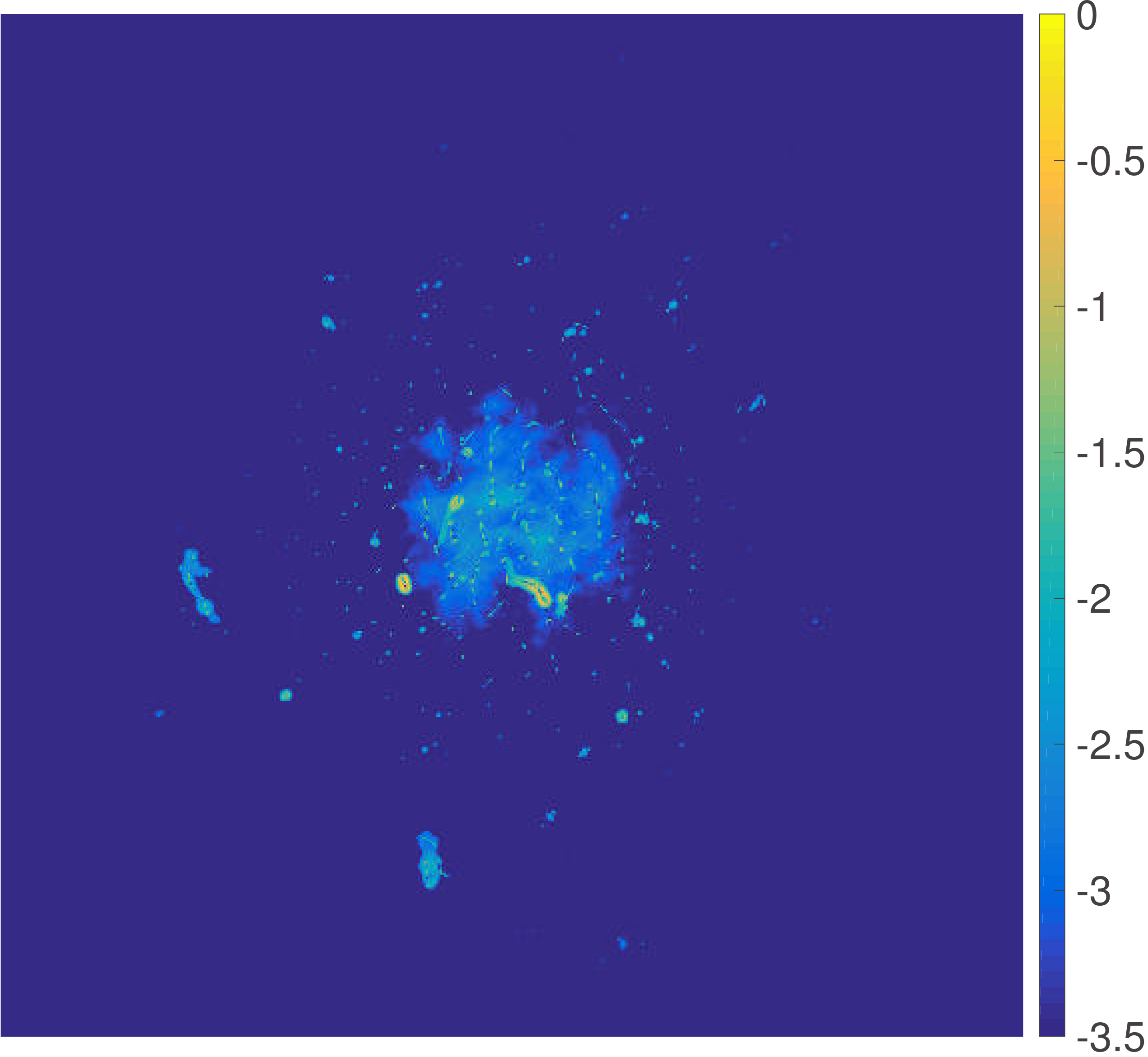}\hspace{2pt}
		\includegraphics[trim={0px 0px 0px 0px}, clip, height=0.21\linewidth]{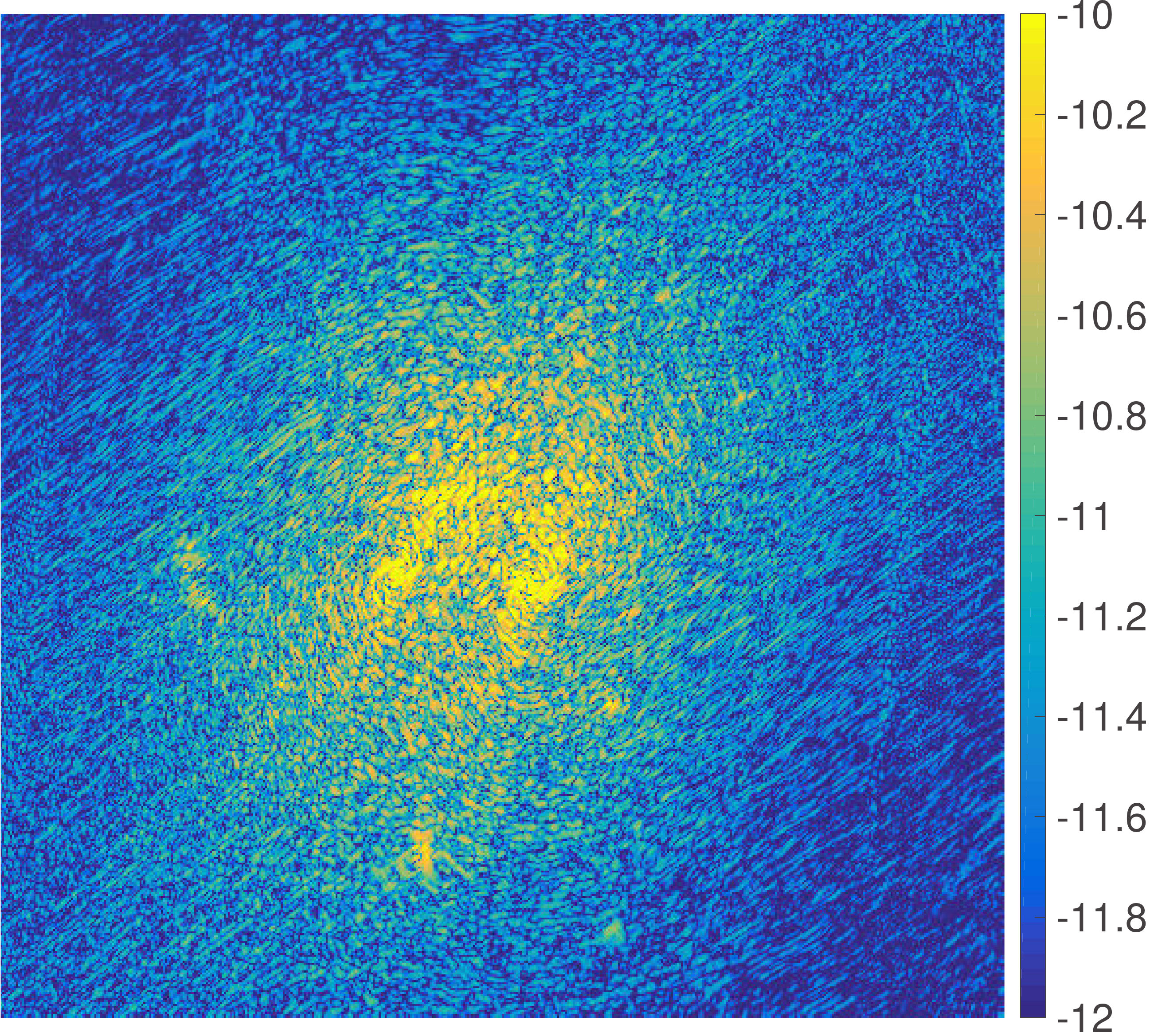}
		\caption{Galaxy cluster}
		\label{fig:galaxycluster_discardeigenvalues_comp}
	\end{subfigure}
	\begin{subfigure}{1\linewidth}
		\centering		
		\includegraphics[trim={0px 0px 0px 0px}, clip, height=0.14\linewidth]{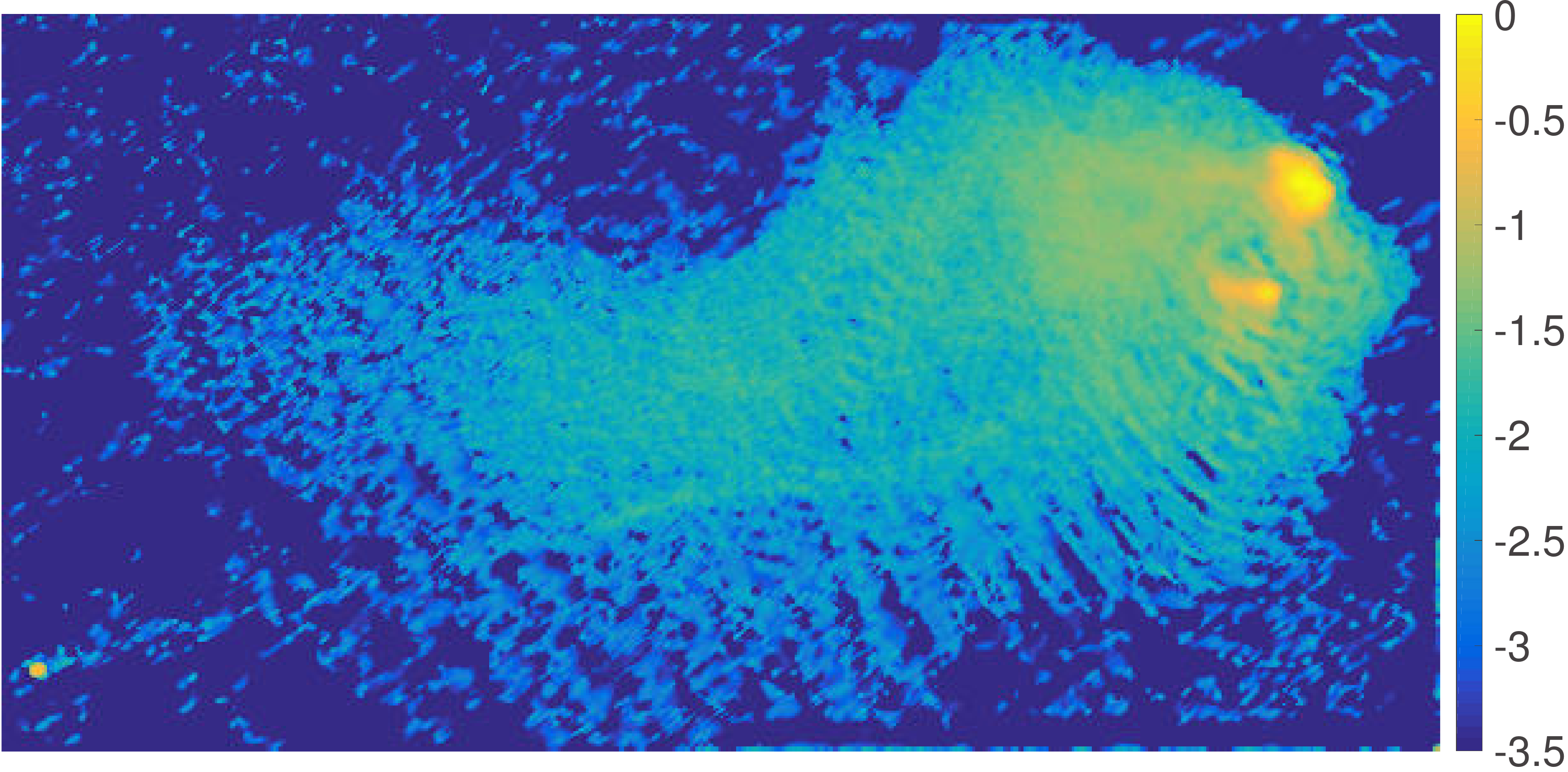}\hspace{2pt}
		\includegraphics[trim={0px 0px 0px 0px}, clip, height=0.14\linewidth]{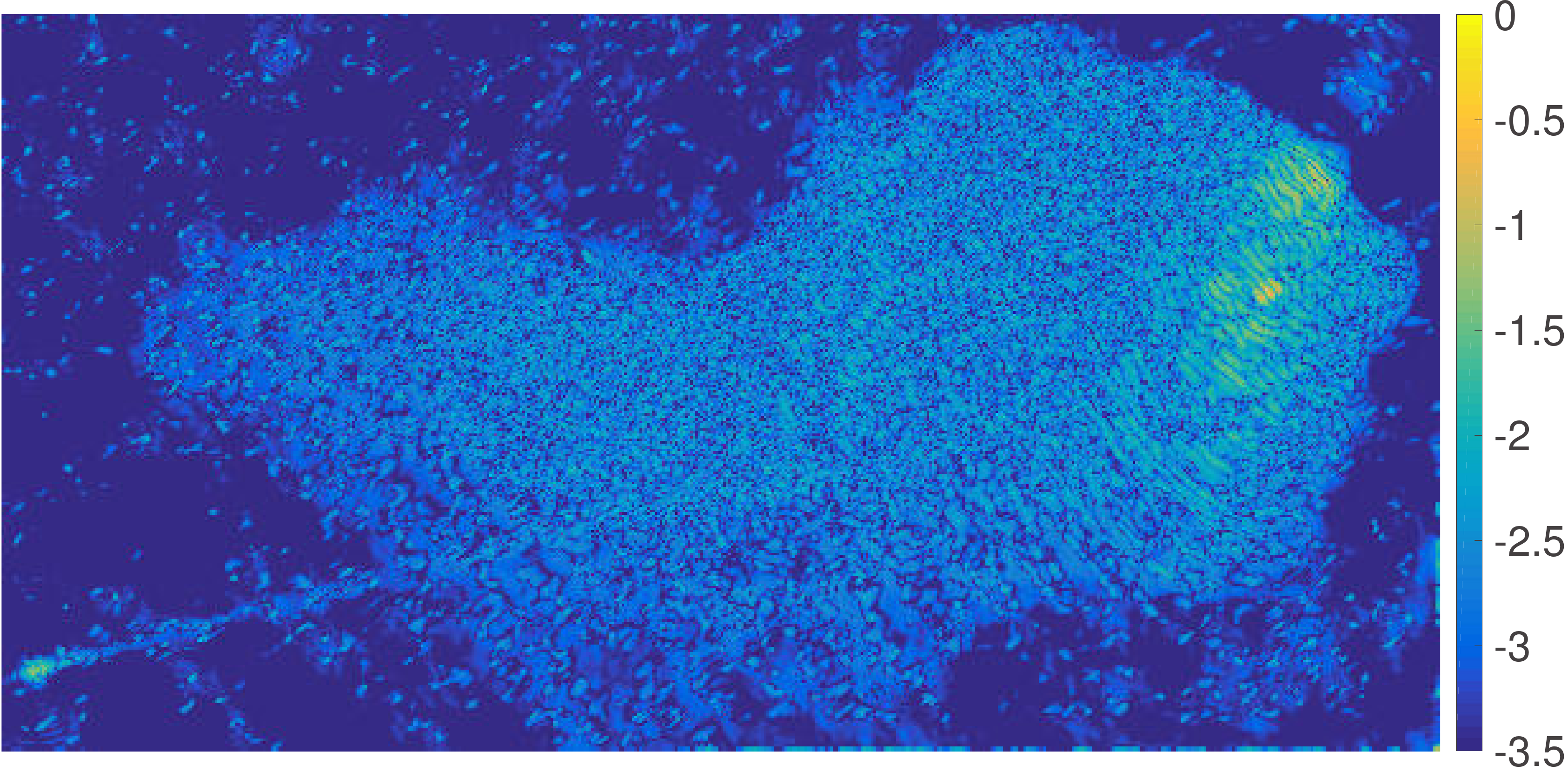}\hspace{2pt}
		\includegraphics[trim={0px 0px 0px 0px}, clip, height=0.14\linewidth]{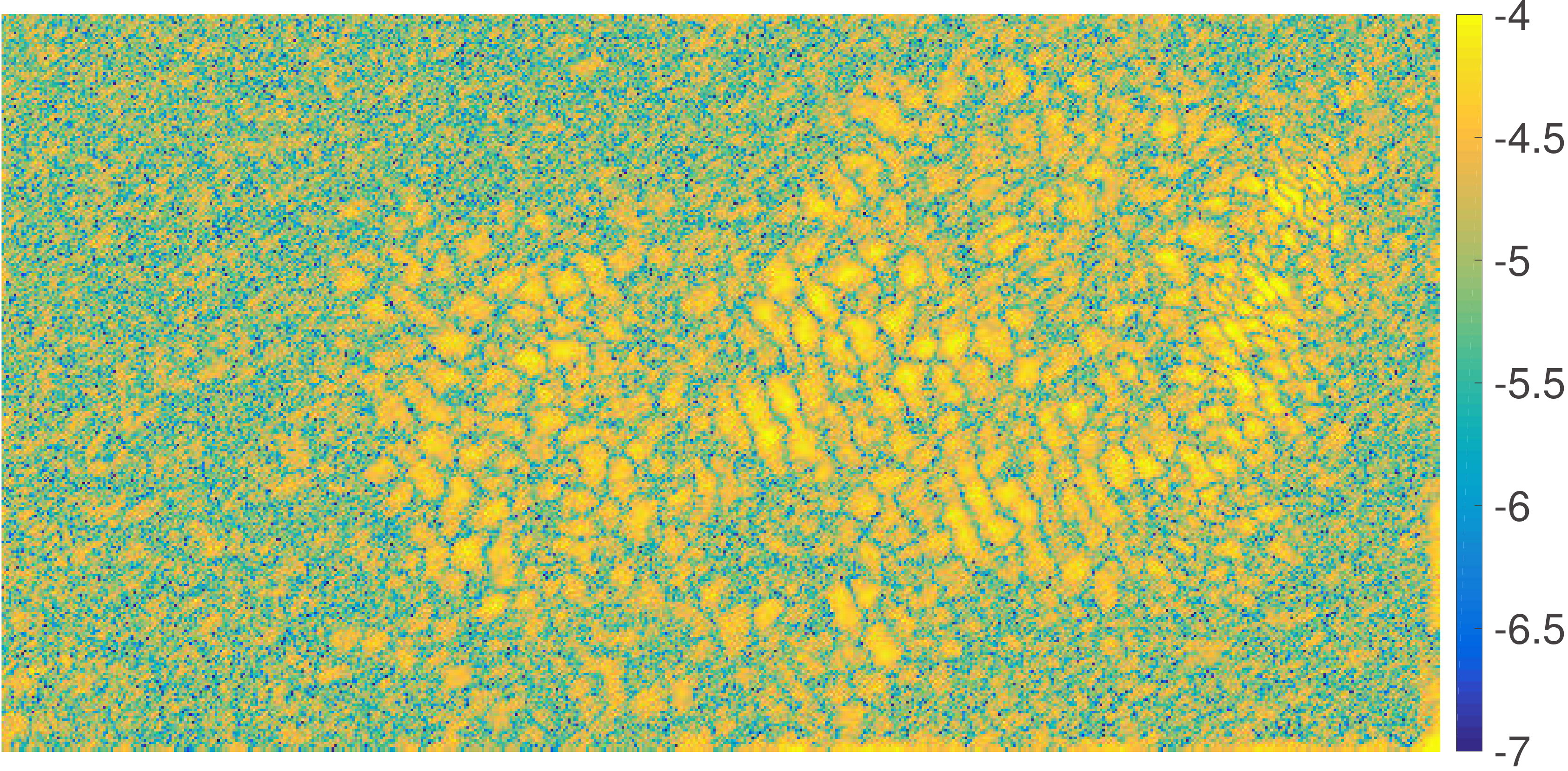}
		\vspace{3pt}		
		\includegraphics[trim={0px 0px 0px 0px}, clip, height=0.14\linewidth]{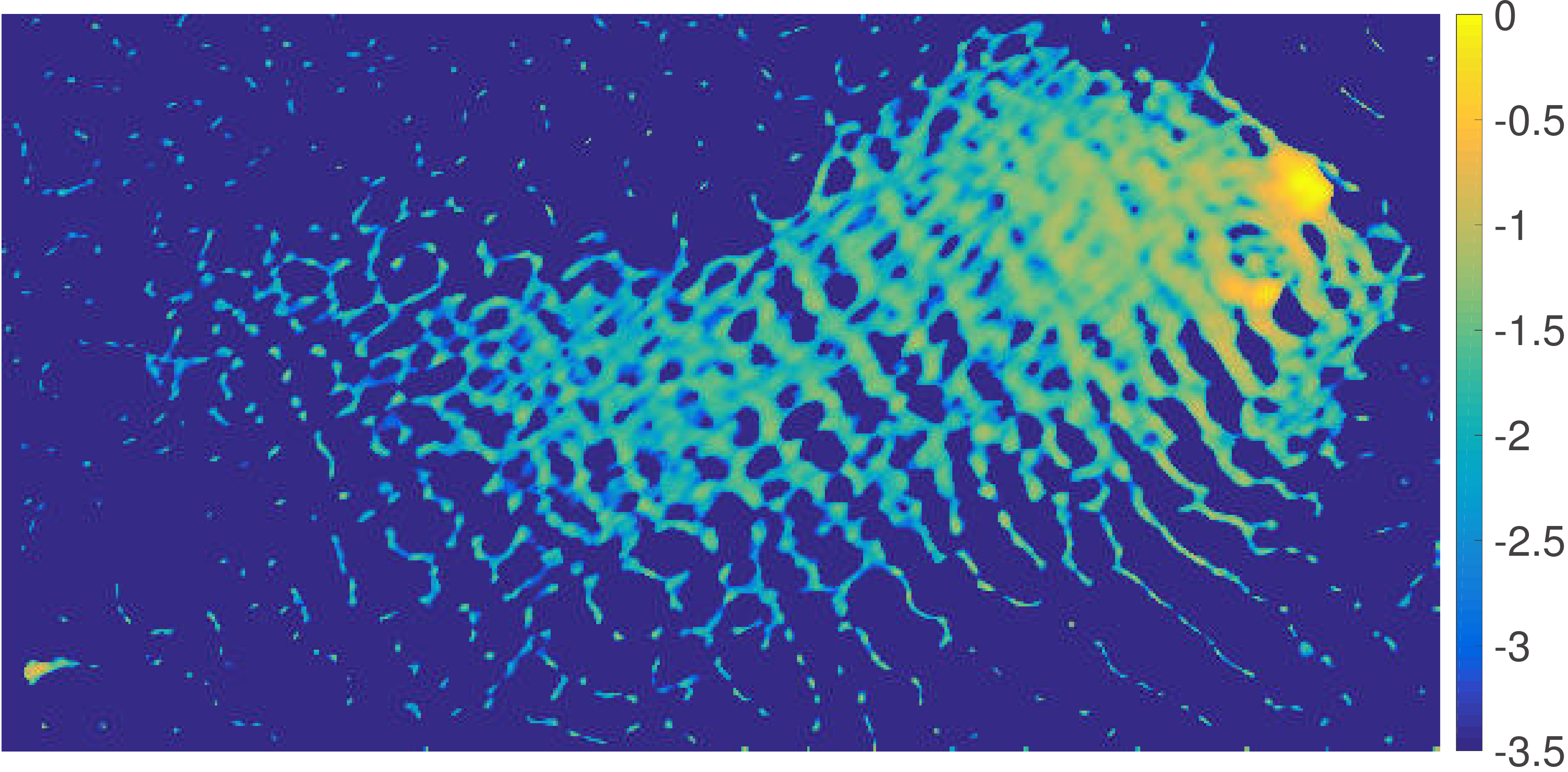}\hspace{2pt}
		\includegraphics[trim={0px 0px 0px 0px}, clip, height=0.14\linewidth]{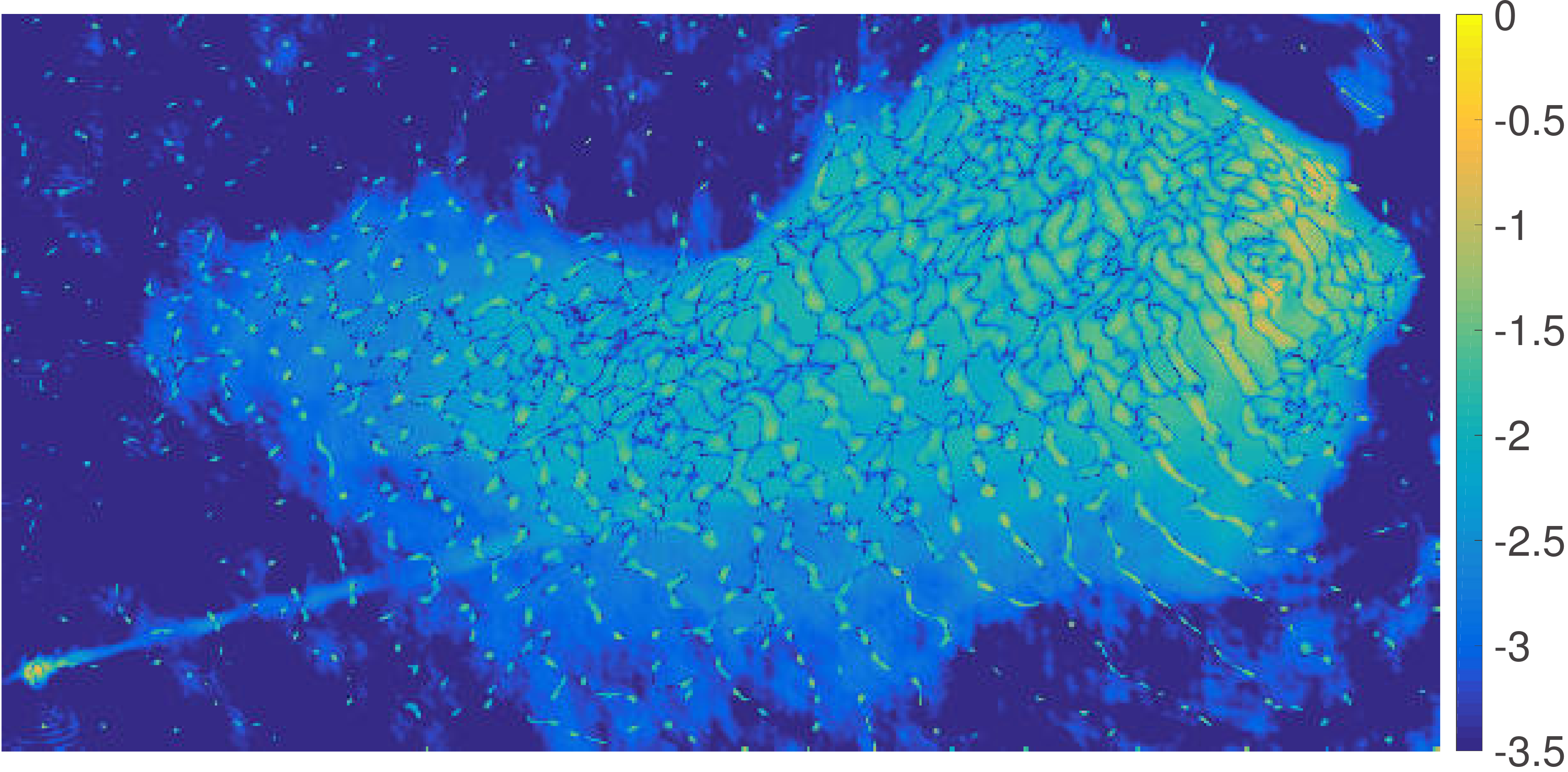}\hspace{2pt}
		\includegraphics[trim={0px 0px 0px 0px}, clip, height=0.14\linewidth]{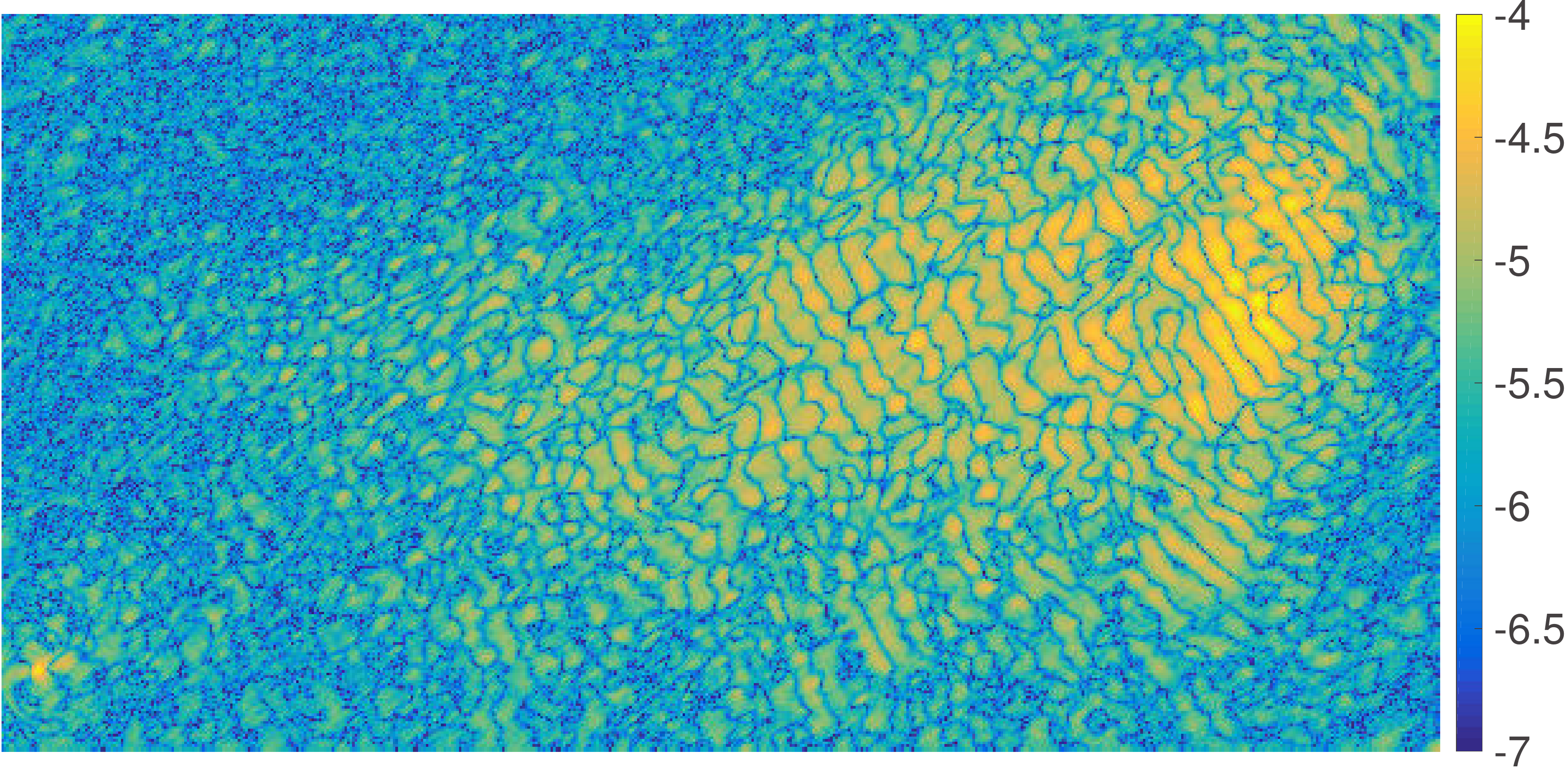}
		\caption{Cygnus A}
		\label{fig:cyga_discardeigenvalues_comp}
	\end{subfigure}
	\caption{Visual comparison of image reconstruction after data reduction to 5 per cent of image size. From left to right: reconstructed, error and residual images in $\log_{10}$ scale. (a) M31, initial data size = 50N. (b) Galaxy cluster image, initial data size = 50N. (c) Cygnus A (partial zoom on top right), initial data size = 25N. For each test image, the top row shows reconstruction from `reduced' visibilities after dimensionality reduction with $\RFPhit$ to $0.05N$, and the bottom row shows reconstruction from `reduced' visibilities after dimensionality reduction with $\RGt$ to $0.05N$.}
	\label{fig:discardeigenvalues_comp}
\end{figure*}

The $\Nz$ and $\nn$ values marked in Fig.~\ref{fig:snrs_discardeigenvalues} were computed with $\|\boldsymbol{x}\|_2$ known a priori from the test images.
The marked values correspond to $\gamma=1$ as given in Eq.~(\ref{eq:diagthreshold}) and Eq.~(\ref{eq:diagthresholdgt}), and provide an indication of a possible estimation of the appropriate final reduced data size, based on our noise-dependent thresholding considerations. 
The results indicate that a threshold value corresponding to $\gamma > 1$ can probably be safely considered for more aggressive reduction with no significant cost in reconstruction quality.

As mentioned in Section~\ref{subsec:holeycoverage}, the prior knowledge of $\|\boldsymbol{x}\|_2$ required to compute the thresholds is, in general, not available from interferometric data.
It could, however, be approximated by using the dirty image.
In the case of $\RFPhit$, simple simulations on the test images with known $\boldsymbol{x}$ suggested that for the approximations given by Eq.~(\ref{eq:approxfzero}), the use of the dirty image $(\PPhi'_{\mathrm{sing}})^{\dagger}\RFPhit\boldsymbol{y}$ instead of $\boldsymbol{x}$ leads to a discrepancy of a few orders of magnitude in the estimation of $\|\boldsymbol{x}\|_2$.
This is admittedly a loose bound, and the corresponding threshold was found to be suboptimal with respect to the discarding of data dimensions.
Further investigation is needed to check for better estimations of $\|\boldsymbol{x}\|_2$ from the data.

\section{Conclusions}
\label{sec:conclusions}
Data dimensionality reduction in radio interferometry is a key challenge for the scalability of imaging methods to the big data setting of the next-generation telescopes. 
We revisited the concept of dimensionality reduction of radio-interferometric data from a compressed sensing perspective. 
The proposed post-gridding linear data embedding approach consists of projecting the data, assumed to be of size much larger than the image size $M\gg N$, to the space spanned by the $\Nz$ left singular vectors of the measurement operator, thus preserving its null space. 
In the {\revised absence of DDEs and calibration errors}, we showed that this dimensionality reduction approach consists of first mapping gridded visibilities back to image space, i.e., computing the dirty image, and then performing a weighted subsampled {\revised discrete Fourier transform} to obtain the final reduced data vector with dimension below image size. 
The Fourier approximation model for the right singular vectors ensures a fast implementation of the full measurement operator after dimensionality reduction.
This dimensionality reduction also preserves the i.i.d. Gaussian properties of the original measurement noise thus making it directly suitable for use with convex optimization algorithms with an $\ell_2$-norm data fidelity term. 
The number of significant singular vectors can be conservatively evaluated by retaining all non-zero singular values, or for further dimensionality reduction to $\Nz \ll N$, by retaining only singular values above a noise-based threshold, effectively introducing a low-rank approximation of the original measurement operator. 
This is in contrast with current gridding-based imaging in radio interferometry, which reduces data on the oversampled discrete Fourier grid of size $\nn$ or to the dirty image of size $N$.

We show with realistic data simulated using SKA-like $uv$ coverages and using the SARA convex optimization method, that reconstruction quality is at least as good as with using the complete visibility data set of size $M$, while {\revised being computationally less expensive, both by having a smaller memory footprint thanks to a reduced data size, and through lower running time per iteration of the imaging algorithm.} 
It is also similar in reconstruction quality to results obtained with gridded visibilities or from the dirty image, but again enabling significantly more reduction below image size.

Another contribution from this work is that dimensionality reduction below $\nn$ can also be achieved from gridded visibilities by discarding those visibilities below a noise-dependent threshold. 
This reduction by thresholding is however significantly less optimal when applied on gridded visibilities, than on the singular value decomposition.

Our proposed dimensionality reduction method is available as \textsc{matlab} code on GitHub at \url{http://basp-group.github.io/fourierdimredn}. 
Further work integrating this method in the \textsc{purify} software is foreseen as part of the research towards scalable HPC-ready algorithms for radio-interferometric imaging. 
{\revised As the current proposed method assumes correctly calibrated data with negligible issues arising from imperfections in data acquisition, future work will also include extended testing with real data, including testing the robustness of the proposed method to w-term effects and calibration errors in particular.}

\section*{Acknowledgements}
We would like to thank Dr~Arwa Dabbech and Dr~Alexandru Onose for discussions on \textsc{clean} variants and the tool chain simulating $uv$ coverages, {\revised Dr~Andr\'{e} Offringa for help and discussions on \textsc{ms-clean} using \textsc{wsclean},} and  Dr~Federica Govoni and Dr~Matteo Murgia for providing the simulated galaxy cluster image.
{\revised We would also like to thank the reviewer for their valuable comments and suggestions.}
The work presented in this article was supported by the Swiss National Science Foundation (SNSF) under grant 200020-146594.



\bibliographystyle{mnras}
\bibliography{biblio} 

\begin{thebibliography}{}
\makeatletter
\relax
\def\mn@urlcharsother{\let\do\@makeother \do\$\do\&\do\#\do\^\do\_\do\%\do\~}
\def\mn@doi{\begingroup\mn@urlcharsother \@ifnextchar [ {\mn@doi@}
  {\mn@doi@[]}}
\def\mn@doi@[#1]#2{\def\@tempa{#1}\ifx\@tempa\@empty \href
  {http://dx.doi.org/#2} {doi:#2}\else \href {http://dx.doi.org/#2} {#1}\fi
  \endgroup}
\def\mn@eprint#1#2{\mn@eprint@#1:#2::\@nil}
\def\mn@eprint@arXiv#1{\href {http://arxiv.org/abs/#1} {{\tt arXiv:#1}}}
\def\mn@eprint@dblp#1{\href {http://dblp.uni-trier.de/rec/bibtex/#1.xml}
  {dblp:#1}}
\def\mn@eprint@#1:#2:#3:#4\@nil{\def\@tempa {#1}\def\@tempb {#2}\def\@tempc
  {#3}\ifx \@tempc \@empty \let \@tempc \@tempb \let \@tempb \@tempa \fi \ifx
  \@tempb \@empty \def\@tempb {arXiv}\fi \@ifundefined
  {mn@eprint@\@tempb}{\@tempb:\@tempc}{\expandafter \expandafter \csname
  mn@eprint@\@tempb\endcsname \expandafter{\@tempc}}}

\bibitem[\protect\citeauthoryear{Bandeira, Dobriban, Mixon  \& Sawin}{Bandeira
  et~al.}{2012}]{bandeira_certifying_2012}
Bandeira A.~S.,  Dobriban E.,  Mixon D.~G.,   Sawin W.~F.,  2012,
  arXiv:1204.1580 [cs, math]

\bibitem[\protect\citeauthoryear{Baraniuk \& Wakin}{Baraniuk \&
  Wakin}{2007}]{baraniuk_random_2007}
Baraniuk R.~G.,  Wakin M.~B.,  2007, Foundations of Computational Mathematics,
  9, 51

\bibitem[\protect\citeauthoryear{Baraniuk, Davenport, DeVore  \&
  Wakin}{Baraniuk et~al.}{2008}]{baraniuk_simple_2008}
Baraniuk R.,  Davenport M.,  DeVore R.,   Wakin M.,  2008, Constructive
  Approximation, 28, 253

\bibitem[\protect\citeauthoryear{Bingham \& Mannila}{Bingham \&
  Mannila}{2001}]{bingham_random_2001}
Bingham E.,  Mannila H.,  2001, in Proc. {Seventh} {ACM} {SIGKDD}
  {International} {Conference} on {Knowledge} {Discovery} and {Data} {Mining}.
  {KDD} '01.
ACM, New York, NY, USA, pp 245--250

\bibitem[\protect\citeauthoryear{Broekema, Nieuwpoort  \& Bal}{Broekema
  et~al.}{2015}]{broekema_square_2015}
Broekema P.~C.,  Nieuwpoort R. V.~v.,   Bal H.~E.,  2015, Journal of
  Instrumentation, 10, C07004

\bibitem[\protect\citeauthoryear{Cand{\`e}s \& Romberg}{Cand{\`e}s \&
  Romberg}{2007}]{candes_sparsity_2007}
Cand{\`e}s E.,  Romberg J.,  2007, Inverse Problems, 23, 969

\bibitem[\protect\citeauthoryear{Cand{\`e}s, Romberg  \& Tao}{Cand{\`e}s
  et~al.}{2006a}]{candes_robust_2006}
Cand{\`e}s E.,  Romberg J.,   Tao T.,  2006a, IEEE Transactions on Information
  Theory, 52, 489

\bibitem[\protect\citeauthoryear{Cand{\`e}s, Romberg  \& Tao}{Cand{\`e}s
  et~al.}{2006b}]{candes_stable_2006}
Cand{\`e}s E.~J.,  Romberg J.~K.,   Tao T.,  2006b, Communications on Pure and
  Applied Mathematics, 59, 1207

\bibitem[\protect\citeauthoryear{Cand{\`e}s, Eldar, Needell  \&
  Randall}{Cand{\`e}s et~al.}{2011}]{candes_compressed_2011}
Cand{\`e}s E.~J.,  Eldar Y.~C.,  Needell D.,   Randall P.,  2011, Applied and
  Computational Harmonic Analysis, 31, 59

\bibitem[\protect\citeauthoryear{Carrillo, McEwen  \& Wiaux}{Carrillo
  et~al.}{2012}]{carrillo_sparsity_2012}
Carrillo R.~E.,  McEwen J.~D.,   Wiaux Y.,  2012, Monthly Notices of the Royal
  Astronomical Society, 426, 1223

\bibitem[\protect\citeauthoryear{Carrillo, McEwen  \& Wiaux}{Carrillo
  et~al.}{2014}]{carrillo_purify:_2014}
Carrillo R.~E.,  McEwen J.~D.,   Wiaux Y.,  2014, Monthly Notices of the Royal
  Astronomical Society, 439, 3591

\bibitem[\protect\citeauthoryear{Carrillo, Kartik, Thiran  \& Wiaux}{Carrillo
  et~al.}{2015}]{carrillo_scalable_2015}
Carrillo R.,  Kartik V.,  Thiran J.-P.,   Wiaux Y.,  2015, in Signal Processing
  with Adaptive Sparse Structured Representations (SPARS) 2015. Cambridge, UK

\bibitem[\protect\citeauthoryear{Combettes \& Pesquet}{Combettes \&
  Pesquet}{2011}]{combettes_proximal_2011}
Combettes P.~L.,  Pesquet J.-C.,  2011, in {Optimization} and {Its}
  {Applications}, Fixed-{Point} {Algorithms} for {Inverse} {Problems} in
  {Science} and {Engineering}.
Springer New York, pp 185--212

\bibitem[\protect\citeauthoryear{Cornwell}{Cornwell}{2008}]{cornwell_multiscale_2008}
Cornwell T.~J.,  2008, IEEE Journal of Selected Topics in Signal Processing, 2,
  793

\bibitem[\protect\citeauthoryear{Cornwell, Voronkov  \& Humphreys}{Cornwell
  et~al.}{2012}]{cornwell_wide_2012}
Cornwell T.~J.,  Voronkov M.~A.,   Humphreys B.,  2012, in International
  Society for Optics and Photonics. pp 85000L--85000L--12

\bibitem[\protect\citeauthoryear{Dabbech, Ferrari, Mary, Slezak, Smirnov  \&
  Kenyon}{Dabbech et~al.}{2015}]{dabbech_moresane:_2015}
Dabbech A.,  Ferrari C.,  Mary D.,  Slezak E.,  Smirnov O.,   Kenyon J.~S.,
  2015, Astronomy \& Astrophysics, 576, A7

\bibitem[\protect\citeauthoryear{Donoho}{Donoho}{2006}]{donoho_compressed_2006}
Donoho D.~L.,  2006, IEEE Trans. Inform. Theory

\bibitem[\protect\citeauthoryear{Ferrari, Mary, Flamary  \& Richard}{Ferrari
  et~al.}{2014}]{ferrari_distributed_2014}
Ferrari A.,  Mary D.,  Flamary R.,   Richard C.,  2014, in IEEE Sensor Array
  and Multichannel Signal Processing Workshop (SAM). A Coru\~{n}a, Spain, pp
  389--392

\bibitem[\protect\citeauthoryear{Ferrari, Deguignet, Ferrari, Mary, Schutz  \&
  Smirnov}{Ferrari et~al.}{2015}]{ferrari_multi-frequency_2015}
Ferrari A.,  Deguignet J.,  Ferrari C.,  Mary D.,  Schutz A.,   Smirnov O.,
  2015, arXiv:1504.06847 [astro-ph]

\bibitem[\protect\citeauthoryear{Fessler \& Sutton}{Fessler \&
  Sutton}{2003}]{fessler_nonuniform_2003}
Fessler J.~A.,  Sutton B.~P.,  2003, IEEE Transactions on Signal Processing,
  51, 560

\bibitem[\protect\citeauthoryear{Fornasier \& Rauhut}{Fornasier \&
  Rauhut}{2011}]{fornasier_compressive_2011}
Fornasier M.,  Rauhut H.,  2011, in , Handbook of {Mathematical} {Methods} in
  {Imaging}.
Springer New York, pp 187--228

\bibitem[\protect\citeauthoryear{Foucart \& Rauhut}{Foucart \&
  Rauhut}{2013}]{foucart_mathematical_2013}
Foucart S.,  Rauhut H.,  2013, A {Mathematical} {Introduction} to {Compressive}
  {Sensing}.
Birkh\"{a}user Basel

\bibitem[\protect\citeauthoryear{Garsden et~al.,}{Garsden
  et~al.}{2015}]{garsden_lofar_2015}
Garsden H.,  et~al., 2015, Astronomy \& Astrophysics, 575, A90

\bibitem[\protect\citeauthoryear{Golub \& Loan}{Golub \&
  Loan}{1996}]{golub_matrix_1996}
Golub G.~H.,  Loan C. F.~V.,  1996, {Matrix {Computations}}.
JHU Press

\bibitem[\protect\citeauthoryear{Hegde, Wakin  \& Baraniuk}{Hegde
  et~al.}{2007}]{hegde_random_2007}
Hegde C.,  Wakin M.,   Baraniuk R.,  2007, Neural Information Processing
  Systems, pp 641--648

\bibitem[\protect\citeauthoryear{H\"{o}gbom}{H\"{o}gbom}{1974}]{hogbom_aperture_1974}
H\"{o}gbom J.~A.,  1974, Astronomy and Astrophysics Supplement Series, 15, 417

\bibitem[\protect\citeauthoryear{Johnson \& Lindenstrauss}{Johnson \&
  Lindenstrauss}{1984}]{johnson_extensions_1984}
Johnson W.~B.,  Lindenstrauss J.,  1984, Contemporary Mathematics, 26, 189

\bibitem[\protect\citeauthoryear{Kartik, Carrillo  \& Wiaux}{Kartik
  et~al.}{2015}]{kartik_dimension_2015}
Kartik V.,  Carrillo R.,   Wiaux Y.,  2015, in Proc. International BASP
  Frontiers Workshop 2015. Villars, Switzerland

\bibitem[\protect\citeauthoryear{Krahmer \& Ward}{Krahmer \&
  Ward}{2010}]{krahmer_new_2010}
Krahmer F.,  Ward R.,  2010, arXiv:1009.0744 [cs, math]

\bibitem[\protect\citeauthoryear{Li, Hastie  \& Church}{Li
  et~al.}{2006}]{li_very_2006}
Li P.,  Hastie T.~J.,   Church K.~W.,  2006, in Proc. 12th {ACM} {SIGKDD}
  {International} {Conference} on {Knowledge} {Discovery} and {Data} {Mining}.
  {KDD} '06.
ACM, New York, NY, USA, pp 287--296

\bibitem[\protect\citeauthoryear{Li, Cornwell  \& de Hoog}{Li
  et~al.}{2011}]{li_application_2011}
Li F.,  Cornwell T.~J.,   de Hoog F.,  2011, Astronomy \& Astrophysics, 528,
  A31

\bibitem[\protect\citeauthoryear{Mahoney}{Mahoney}{2011}]{mahoney_randomized_2011}
Mahoney M.~W.,  2011, Found. Trends Mach. Learn., 3, 123

\bibitem[\protect\citeauthoryear{Mallat \& Zhang}{Mallat \&
  Zhang}{1993}]{mallat_matching_1993}
Mallat S.~G.,  Zhang Z.,  1993, IEEE Transactions on Signal Processing, 41,
  3397

\bibitem[\protect\citeauthoryear{McEwen \& Wiaux}{McEwen \&
  Wiaux}{2011}]{mcewen_compressed_2011}
McEwen J.~D.,  Wiaux Y.,  2011, Monthly Notices of the Royal Astronomical
  Society, 413, 1318

\bibitem[\protect\citeauthoryear{{Murgia}, {Govoni}, {Feretti}, {Giovannini},
  {Dallacasa}, {Fanti}, {Taylor}  \& {Dolag}}{{Murgia}
  et~al.}{2004}]{Murgia2004}
{Murgia} M.,  {Govoni} F.,  {Feretti} L.,  {Giovannini} G.,  {Dallacasa} D.,
  {Fanti} R.,  {Taylor} G.~B.,   {Dolag} K.,  2004, \aap, 424, 429

\bibitem[\protect\citeauthoryear{Nelson, Price  \& Wootters}{Nelson
  et~al.}{2014}]{nelson_new_2014}
Nelson J.,  Price E.,   Wootters M.,  2014, in Proc. {Twenty}-{Fifth} {Annual}
  {ACM}-{SIAM} {Symposium} on {Discrete} {Algorithms}. {SODA} '14.
SIAM, Portland, Oregon, pp 1515--1528

\bibitem[\protect\citeauthoryear{Offringa, McKinley, Hurley-Walker
  et~al.}{Offringa et~al.}{2014}]{offringa_wsclean_2014}
Offringa A.~R.,  McKinley B.,  Hurley-Walker  et~al., 2014, Monthly Notices of
  the Royal Astronomical Society, 444, 606

\bibitem[\protect\citeauthoryear{Onose, Carrillo, Repetti, McEwen, Thiran,
  Pesquet  \& Wiaux}{Onose et~al.}{2016}]{onose_scalable_2016}
Onose A.,  Carrillo R.~E.,  Repetti A.,  McEwen J.~D.,  Thiran J.-P.,  Pesquet
  J.-C.,   Wiaux Y.,  2016, arXiv:1601.04026 [astro-ph]

\bibitem[\protect\citeauthoryear{Rau, Bhatnagar, Voronkov  \& Cornwell}{Rau
  et~al.}{2009}]{rau_advances_2009}
Rau U.,  Bhatnagar S.,  Voronkov M.,   Cornwell T.,  2009, Proc. IEEE, 97, 1472

\bibitem[\protect\citeauthoryear{Sardarabadi, Leshem  \& van~der
  Veen}{Sardarabadi et~al.}{2016}]{sardarabadi_radio_2016}
Sardarabadi A.~M.,  Leshem A.,   van~der Veen A.-J.,  2016, Astronomy \&
  Astrophysics, 588, A95

\bibitem[\protect\citeauthoryear{Sault \& Wieringa}{Sault \&
  Wieringa}{1994}]{sault_mfclean_1994}
Sault R.~J.,  Wieringa M.~H.,  1994, \aaps, 108, 585

\bibitem[\protect\citeauthoryear{{Schwab}}{{Schwab}}{1984}]{schwab_relaxing_1984}
{Schwab} F.~R.,  1984, \aj, 89, 1076

\bibitem[\protect\citeauthoryear{Smirnov}{Smirnov}{2011}]{smirnov_revisiting_2011}
Smirnov O.~M.,  2011, Astronomy \& Astrophysics, 527, A106

\bibitem[\protect\citeauthoryear{Sullivan et~al.,}{Sullivan
  et~al.}{2012}]{sullivan_fast_2012}
Sullivan I.~S.,  et~al., 2012, The Astrophysical Journal, 759, 17

\bibitem[\protect\citeauthoryear{Tegmark}{Tegmark}{1997}]{tegmark_how_1997}
Tegmark M.,  1997, ApJ, 480, L87

\bibitem[\protect\citeauthoryear{Thompson, Moran  \& Jr}{Thompson
  et~al.}{2001}]{thompson_interferometry_2001}
Thompson A.~R.,  Moran J.~M.,   Jr G. W.~S.,  2001, {Interferometry and
  {Synthesis} in {Radio} {Astronomy}}.
Wiley, New York

\bibitem[\protect\citeauthoryear{Tropp}{Tropp}{2011}]{tropp_improved_2011}
Tropp J.~A.,  2011, Advances in Adaptive Data Analysis, 03, 115

\bibitem[\protect\citeauthoryear{Wiaux, Jacques, Puy, Scaife  \&
  Vandergheynst}{Wiaux et~al.}{2009a}]{wiaux_compressed_2009}
Wiaux Y.,  Jacques L.,  Puy G.,  Scaife A. M.~M.,   Vandergheynst P.,  2009a,
  Monthly Notices of the Royal Astronomical Society, 395, 1733

\bibitem[\protect\citeauthoryear{Wiaux, Puy, Boursier  \& Vandergheynst}{Wiaux
  et~al.}{2009b}]{wiaux_spread_2009}
Wiaux Y.,  Puy G.,  Boursier Y.,   Vandergheynst P.,  2009b, Monthly Notices of
  the Royal Astronomical Society, 400, 1029

\bibitem[\protect\citeauthoryear{Wijnholds \& van~der Veen}{Wijnholds \&
  van~der Veen}{2011}]{wijnholds_data_2011}
Wijnholds S.~J.,  van~der Veen A.~J.,  2011, in 2011 {IEEE} {International}
  {Conference} on {Acoustics}, {Speech} and {Signal} {Processing} ({ICASSP}).
  Prague, Czech Republic, pp 2704--2707

\bibitem[\protect\citeauthoryear{Wijnholds, van~der Veen, De~Stefani, La~Rosa
  \& Farina}{Wijnholds et~al.}{2014}]{wijnholds_signal_2014}
Wijnholds S.,  van~der Veen A.-J.,  De~Stefani F.,  La~Rosa E.,   Farina A.,
  2014, in 2014 {IEEE} {International} {Conference} on {Acoustics}, {Speech}
  and {Signal} {Processing} ({ICASSP}). Florence, Italy, pp 5382--5386

\bibitem[\protect\citeauthoryear{Wolz, McEwen, Abdalla, Carrillo  \&
  Wiaux}{Wolz et~al.}{2013}]{wolz_revisiting_2013}
Wolz L.,  McEwen J.~D.,  Abdalla F.~B.,  Carrillo R.~E.,   Wiaux Y.,  2013,
  Monthly Notices of the Royal Astronomical Society, 436, 1993

\bibitem[\protect\citeauthoryear{Woodruff}{Woodruff}{2014}]{woodruff_sketching_2014}
Woodruff D.~P.,  2014, Foundations and Trends in Theoretical Computer Science,
  10, 1

\bibitem[\protect\citeauthoryear{Yang \& Zhang}{Yang \&
  Zhang}{2011}]{yang_alternating_2011}
Yang J.,  Zhang Y.,  2011, SIAM Journal on Scientific Computing, 33, 250

\makeatother
\end{thebibliography}




%
%


\bsp	
\label{lastpage}
\end{document}